\documentclass[usegraphicx,usenatbib]{mn2e}
\usepackage{amssymb}
\usepackage{epsf,multirow}

\begin{document}

\title[Cosmological parameters from the SDSS-DR6 LRGs $\xi(s)$]
{
Cosmological parameter constraints from SDSS luminous red galaxies: 
a new treatment of large-scale clustering. 
}
\author[A.G. S\'anchez et al.]
{\parbox[t]{\textwidth}{
Ariel G. S\'anchez$^{1}$\thanks{E-mail: arielsan@mpe.mpg.de},
M. Crocce$^{2}$,
A. Cabr\'e$^{2}$,
C. M. Baugh$^{3}$ and
E. Gazta\~naga$^{2}$
}
\vspace*{6pt} \\ 
$^{1}$ Max-Planck-Institut f\"ur Extraterrestrische Physik, Giessenbachstrasse, 85748 Garching, Germany.\\
$^{2}$ Institut de Ci\`encies de l'Espai, CSIC/IEEC, Campus UAB, F. de Ci\`encies, Torre C5 par-2, Barcelona 08193, Spain.\\
$^{3}$ The Institute for Computational Cosmology, 
Department of Physics, University of Durham, South Road, Durham DH1
3LE, UK.\\
\\
}
\date{Submitted to MNRAS}
\maketitle

\begin{abstract}
We apply a new model for the spherically averaged correlation function at large pair 
separations to the measurement of the clustering of luminous red galaxies (LRGs) made from the SDSS by 
\citet{cabre2008}. Our model takes into account the form of the BAO peak and the 
large scale shape of the correlation function. We perform a Monte Carlo Markov chain 
analysis for different combinations of datasets and for different parameter sets. 
When used in combination with a compilation of the latest 
CMB measurements, the LRG clustering and the latest supernovae results give 
constraints on cosmological parameters which are 
comparable and in remarkably good agreement, 
resolving the tension reported in some studies. The best 
fitting model in the context of a flat, $\Lambda$-CDM cosmology is specified by 
$\Omega_{\rm m} = 0.261\pm0.013$, $\Omega_{\rm b} = 0.044 \pm 0.001$, 
$n_{\rm s} = 0.96 \pm 0.01$, $H_{0}=71.6 \pm 1.2 \, {\rm km} \,{\rm s}^{-1}\,{\rm Mpc}^{-1}$
and $\sigma_{8} = 0.80 \pm 0.02$. If we allow the time-independent dark energy equation of state 
parameter to vary, we find results consistent with a cosmological constant at 
the $5\%$ level using all data sets: $w_{\rm DE} = -0.97 \pm 0.05$. 
The large scale structure measurements by themselves can constrain the dark energy 
equation of state parameter to $w_{\rm DE}=-1.05^{+0.16}_{-0.15}$, independently 
of CMB or supernovae data.
 We do not find convincing evidence for an evolving equation of state.
We provide a set of ``extended distance priors'' that contain the most relevant
information from the CMB power spectrum and the shape of the LRG correlation
function which can be used to constrain dark energy models and
spatial curvature. Our model should provide an accurate description
of the clustering even in much larger, forthcoming surveys,
such as those planned with NASA's JDEM or ESA's Euclid mission.
\end{abstract}


\section{Introduction}
\label{sec:intro}

The acoustic peaks imprinted on the temperature power spectrum of 
the cosmic microwave background (CMB) have now been measured with 
impressive precision by a number of experiments \citep{lee2001,
bennet03,hinshaw03,hinshaw07,hinshaw08,jones2006,reichardt2008}. 
These observations 
place tight constraints on the values of many of the fundamental 
cosmological parameters. With the fifth year of integration from the 
WMAP satellite there is seemingly little room left for any deviation 
from the basic $\Lambda$CDM model \citep{dunkley2008,komatsu08}. 
However, degeneracies exist between some parameters which cannot be 
broken by CMB data alone (e.g. \citep{bond1997,efstathiou1999,
bridle2003}. Perhaps the two most important examples are the curvature 
of the Universe and the equation of state of the dark energy, 
$w_{\rm DE}= P_{\rm DE}/\rho_{\rm DE}$, where $P_{\rm DE}$ is 
the pressure of the dark energy and $\rho_{\rm DE}$ is its density. 
Meaningful constraints cannot be obtained on these parameters 
using CMB data in isolation. 

The full potential of the CMB measurements is realized when these data 
are combined with other observations, such as the Hubble diagram of type Ia 
supernovae and the large scale structure of the Universe as traced by 
galaxies \citep{riess1998,perlmutter1999,efstathiou02,percival02,spergel03,riess2004,tegmark04,
seljak05,sanchez06,astier06,seljak06,wang06b,wood-vasey07,spergel07,komatsu08,okumura08,xia08,ferramacho08}.
These complementary data sets come from the late Universe 
compared with the CMB data, and the interpretation of the observations 
is more complicated and controversial. 

Type Ia supernovae (SNe) have been proposed as standard candles which can 
probe the luminosity distance - redshift relation. 
The first strong evidence in support of a cosmological constant 
came from combining the SNe data with CMB measurements \citep{riess1998,
perlmutter1999}. The type Ia Hubble diagram has come under intense scrutiny 
to uncover any hint of non-standardness arising from the nature of the host 
galaxy, possible evolution with redshift or variation in dust extinction 
\citep{sullivan03,gallagher05,ellis08,howell08}. A recent joint analysis 
of SNe from different datasets suggests that the systematic 
error on the equation of state parameter from a joint CMB and SNe analysis 
is comparable to the size of the random error \citep{kowalski08}. 

The power spectrum of galaxy clustering has also been used in 
combination with CMB data (e.g. Percival et~al. 2002). According to the standard 
lore, the galaxy power spectrum on large scales is expected to 
have a simple relation to the underlying dark matter spectrum. 
Moreover, the shape of the spectrum is believed to closely follow 
that expected in linear perturbation theory for the matter, which 
can be readily computed given a set of values for the cosmological 
parameters. However, with the availability of improving measurements 
and more refined modelling of the galaxy power spectrum it has become clear that this simple 
picture is no longer sufficiently accurate to describe the data. 
Additional levels of modelling of the deviations from linear theory 
have to be incorporated into the analysis. These include empirical 
models of the nonlinear distortion of the power spectrum \citep{smith03}
and possible scale dependent biases between the clustering of galaxies and mass  
\citep{cole05,hamann08,cresswell09}. Recent studies have cast doubt on the 
accuracy of these prescriptions \citep{sanchez08,reid08}. In principle, 
if the simple models described the form of the observed power spectrum, then 
the power spectra measured from different galaxy samples should yield 
equivalent constraints on cosmological parameters. However, \citet{sanchez08}
found a fundamental difference in the shapes of the galaxy power 
spectra measured from the two-degree galaxy redshift survey and the 
Sloan Digital Sky Survey, after attempting to correct the measured spectra. 
These authors found that the red selection of the SDSS galaxies results 
in a strong scale dependent bias \citep[see also][]{swanson08}. 
Similar scale dependent effects have been seen in the power spectrum 
of dark matter haloes and galaxies modelled in simulations 
\citep{smith07,angulo08}. 

Recent analyses have not attempted to model the overall shape or amplitude 
of the power spectrum, as a consequence of the difficulties described above 
in interpreting the results from different samples. Instead, attention has 
shifted to a pattern of oscillatory features called the baryonic acoustic 
oscillations (BAO), which are imprinted on the matter power spectrum. 
These features have been advocated as a standard ruler which can be used 
to measure the distance-redshift relation, and hence constrain the dark 
energy equation of state  \citep{blake03,hu03,linder03,seo03,wang06,guzik2007,
seo2007,seo2008}. The BAO arise from oscillations in the baryon-photon fluid 
prior to matter-radiation decoupling. This phenomenon gives rise to the 
peaks seen in the power spectrum of temperature fluctuations in the CMB. 
In the matter power spectrum, the oscillations have a much smaller amplitude 
as baryons only account for around 20 per cent of the total matter density 
of the Universe. Furthermore, the oscillations in the matter spectrum are 
out of phase with those in the CMB \citep{sugiyama95,EH98,EH99,meiksin1999}. 
The oscillation scale is related to the size of the sound horizon at 
recombination, which can be measured with high accuracy from the CMB 
\citep{komatsu08}. The apparent size of the BAO ruler depends upon the 
parameters $w_{\rm DE}$ and $\Omega_k$, as these determine the angular 
diameter distance out to a given redshift. In practice, the BAO are not 
precisely a standard ruler at the level of precision demanded for 
their interpretation in future surveys. Nevertheless, by modelling the 
appearance of the BAO accurately, they are still a valuable probe 
of cosmological parameters \citep{sanchez08b, smith08}. 
Careful simulation work and modelling has shown 
that techniques can be developed which can overcome long-wavelength 
gradients in the power spectrum to yield robust constraints on the BAO 
using the galaxy power spectrum \citep{percival07b,smith07,angulo08,crocce08,takahashi08,seo2008}.  

The BAO signal has been detected in both the 2dFGRS and SDSS surveys 
\citep{cole05,eisenstein05}. The most powerful BAO 
measurements currently come from samples of luminous red galaxies (LRGs) 
\citep{cole05,hutsi06,padmanabhan07,percival07b,percival07c,okumura08,
cabre2008,gaztanaga08b,martinez2008}.
Detections at a lower significance have also been reported using galaxy clusters 
\citep{hutsi07,estrada08}. 
The imprint of these features 
has even been found in 
the three point function of LRGs \citep{gaztanaga08a}. Despite this rapid 
progress, the conclusions drawn from measurements of the BAO remain unclear. 
For example, \citet{percival07c} analysed the BAO signal in a joint galaxy 
sample drawn from the Sloan Digital Sky Survey (SDSS) data release five 
(DR5) and the two-degree Field Galaxy Redshift Survey (2dFGRS). 
Ignoring the information from the amplitude and long-wavelength shape of 
the power spectrum, and simply isolating the BAO, the results of 
Percival et~al. showed a 2.4$\sigma$ discrepancy with the distance 
measurements inferred from the supernovae type Ia (SN) data by 
\citet{astier06}, signaling a possible problem in the modelling 
of the BAO data, or a challenge to the $\Lambda$CDM model.

Much theoretical work has been devoted to uncovering scale dependent 
effects in the BAO and in improving the modelling of the signal in 
galaxy surveys \citep{angulo05,huff07,smith07,angulo08,smith08,crocce08,seo2008,desjacques08}.
In order to realize the full potential of the BAO technique as a cosmological probe 
it is essential to quantify any systematics in the signal and to 
understand how the measurements relate to cosmological parameters. 
First, let us debunk some possible preconceptions about BAO. As we have 
remarked above, the BAO are not precisely a standard ruler. In the 
correlation function, the Fourier transform of the power spectrum, 
the BAO appear as a broad peak at large pair separations \citep{matsubara04}.   
\citet{sanchez08b} showed that even in linear perturbation theory, the 
maximum of the BAO peak in correlation function does not coincide with the sound 
horizon scale at the percent level of accuracy required to fully exploit the 
measurements expected from forthcoming galaxy surveys. Furthermore, 
\citet{smith08} and \citet{crocce08} have shown that both large volume 
numerical simulations and theoretical predictions based on renormalized 
perturbation theory (RPT) indicate that the BAO peak in the correlation 
function is shifted and distorted in a non-trivial manner relative to the prediction of linear 
perturbation theory. If unaccounted for, these shifts bias the constraints 
obtained by using the BAO measurements as a standard ruler. 

Careful modelling of the correlation function is therefore required 
to extract the cosmological information encoded in large scale structure.
\citet{sanchez08b} argued that the correlation function is less affected 
by scale dependent effects than the power spectrum and that a simple model 
for the correlation function proposed by \citet{crocce08}, based on 
RPT \citep{crocce06,crocce06b}, gives an essentially unbiased 
measurement of the dark energy equation of state. This means that 
information from the large scale shape of the correlation function, 
in addition to the form of the BAO peak, can be used to provide 
robust constraints on cosmological parameters. The correlation function 
therefore provides a better constraint on the distance scale than the 
more conservative, ``BAO only'' approach required when using the power spectrum 
(i.e. which requires the long wavelength shape information to be discarded 
along with the amplitude).

In this paper we apply this new model to the shape of the redshift space 
correlation function, $\xi(s)$, of the SDSS Data release 6 (DR6) 
LRGs measured by \citet{cabre2008}. The LRG sample analysed by \citet{cabre2008} is twice the 
size of the sample used to make the first detection of the BAO by \citet{eisenstein05}, 
and has also yielded the first measurement of the radial BAO signal, which 
constrains the Hubble parameter \citep{gaztanaga08b}. 
We combine the LRG clustering information with the latest measurements 
of CMB and SNe data. We focus on the constraints on the dark energy 
equation of state and the curvature of the Universe, which are the 
parameters where the extra information from the shape of $\xi(s)$ can 
dramatically improve upon the CMB only constraints. We also pay special 
attention to the consistency of the results obtained with different 
dataset combinations.

The outline of the paper is as follows. In Section~\ref{sec:data}, we 
describe the data used in our parameter estimation. In 
Section~\ref{sec:method} we describe the details of our modelling of 
the shape of the redshift space correlation function and compare it 
with measurements in N-body simulations. We also set out the 
different parameter spaces that we study and describe our methodology 
for parameter estimation. In Section~\ref{sec:test} we 
assess the impact of the details of the parameter estimation technique on the obtained 
constraints on cosmological parameters. Section~\ref{sec:results} presents 
our main results for the parameter constraints obtained by comparing 
theoretical models to the CMB data and the correlation function of the 
SDSS-DR6. 
In Section~\ref{sec:distances} we focus on the constraints on certain 
distance combinations obtained from the shape of $\xi(s)$. 
We summarize our conclusions in Section~\ref{sec:conclusions}.
Appendix A gives the theoretical motivation for the model we use to describe 
the correlation function and Appendix B gives the covariance matrix for 
the distance constraints. 

\section{The Datasets}
\label{sec:data}
Here we describe the different datasets that we use to 
constrain cosmological parameters. The modelling of the 
correlation function of luminous red galaxies is described 
later on in Section~\ref{ssec:rpt}. The datasets described 
below are used in different combinations to check the 
consistency of the constraints returned. 

\subsection{The redshift space correlation function of SDSS-DR6 LRGs: the monopole}
\label{ssec:dr7}

Luminous red galaxies (LRGs) are an efficient tracer of the large scale 
structure of the Universe. These galaxies are selected by color and magnitude 
cuts designed to identify intrinsically red, bright galaxies using SDSS 
photometry (see Eisenstein et al 2001 for a complete description of the 
color cuts). LRGs can be seen out to higher redshifts than galaxies in a 
simple magnitude limited catalogue, and so map a larger volume of the 
Universe. LRGs have a low space density compared to $L_*$ galaxies, which 
means that fewer redshifts have to be measured to map out the same volume. 
The low space density, which translates into a higher shot noise, is 
compensated for by the stronger than average clustering of LRGs, which 
maintains the signal-to-noise of the correlation function at a level 
which can be measured.

\begin{figure}
\centering
\centerline{\includegraphics[width=\columnwidth]{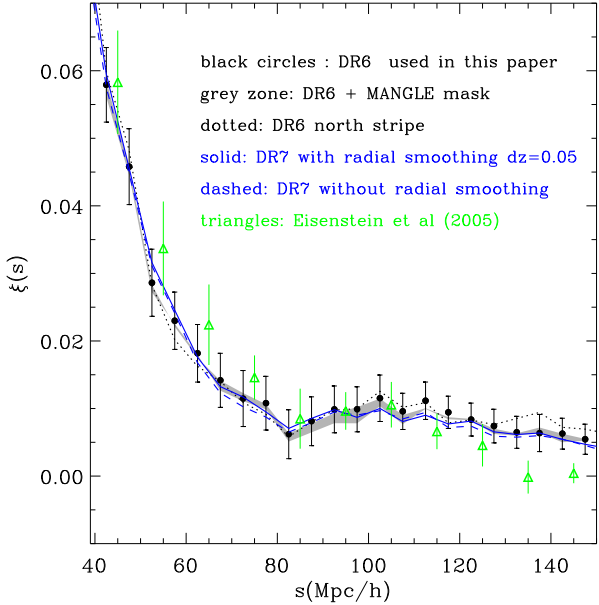}}
\caption{
The spherically averaged correlation function of LRGs in redshift space. 
Circles with errorbars show the correlation function used in this paper. 
The shaded region shows the dispersion in the correlation function obtained when 
using the new MANGLE mask of Swanson et al. (2008) with different completeness fractions.  
The dotted line shows the result for the north stripe of DR6. We have also calculated 
the correlation function for the new DR7 (solid line), estimated using a random catalog 
generated from a smoothed version of the selection function. 
The dashed line shows the estimate from DR7 without smoothing. 
The measurement from Eisenstein et~al. (2005) is shown using triangles; note these authors 
had fewer LRGs and used broader bins. 
}
\label{fig:corfunc}
\end{figure}

Here we use the measurement of the 2-point correlation function of LRGs 
in redshift space made by \cite{cabre2008} shown by the black points 
in Fig.~\ref{fig:corfunc}. These authors studied the clustering of LRGs 
in Data Release 6 (DR6) of the Sloan Digital Sky Survey (SDSS), which has 75\,000 
LRG galaxies spanning a volume of $1\,h^{-3}\,{\rm Gpc}^3$ over the redshift 
interval $0.15<z<0.47$. The comparison between DR6 and the result of 
\citet{eisenstein05} is plotted in Fig.B4 of \citet{cabre2008} and Fig.~\ref{fig:corfunc}; 
the two results are in good agreement with the DR6 result showing an improvement 
of a factor of about $\sqrt2$ in the size of the errors. \citet{cabre2008} carried 
out an extensive series of tests of their measurement of the LRG correlation function. 
They found that systematic uncertainties in the estimation of the radial selection 
have a small impact on the results (ie see their figure B2). The adoption of different 
weighting schemes also produces negligibly small changes (their Fig.B5 and B11). 
These careful tests indicate that this new estimation of the correlation function 
is robust and that any possible systematic effects are small compared to the error bars.

Here we recap two aspects of the analysis of \citet{cabre2008} which are particularly 
relevant to this paper: the treatment of the survey mask and the estimation of the 
covariance matrix of the correlation function.

An accurate knowledge of the angular and radial selection of a galaxy 
catalogue, including the redshift completeness as a function of 
magnitude and position on the sky, is an essential prerequisite for 
a measurement of clustering. This information allows the mean density 
of galaxies in the survey to be estimated. After the release of DR6, 
\cite{swanson08} provided this information in a readily usable form, translating 
the original mask files extracted from the NYU Value-Added Galaxy Catalog \citep{blanton2005}, 
from MANGLE into Healpix format \citep{gorski99}. \cite{cabre2008} describe 
how they constructed a survey ``mask" for LRGs and tested the impact of the mask on 
clustering measurements using mock catalogues. Using the same techniques, we have 
also examined the correlation function of LRGs in DR7, which has become available 
since the submission of \cite{cabre2008}. In Fig.~\ref{fig:corfunc}, we plot a summary 
of the possible systematics in the estimation of the correlation function. The estimates 
corresponding to different versions of the survey mask and different ways of generating 
a catalogue of random points are in remarkably good agreement with one another. Furthermore, 
the results from DR6 and DR7 are in excellent agreement; DR7 represents only a modest improvement 
over DR6. The results of \cite{eisenstein05} are plotted in Fig.~\ref{fig:corfunc} using 
triangles. Their estimate is consistent with that of \citet{cabre2008} within the errors 
(note that their binning is different, and also, perhaps, their normalisation, due to a small 
shift which could be attributed to systematics (see Section~\ref{sec:Kshift})).
Whilst there is a modest improvement in the parameter constraints on using the DR7 
measurement, in line with the incremental change in the number of LRGs and the solid angle 
covered, we have decided to retain the DR6 measurement in this paper, because 
the survey mask has been tested more extensively in that case. The DR6 measurement 
represents a factor of $\sim 2$ more volume than that covered by the LRG sample 
analyzed by Eisenstein et~al. (2005). 

\cite{cabre2008} constructed a covariance matrix for the LRG correlation 
function using mock catalogues drawn from the MareNostrum Institut de Ciencias del 
Espacio (MICE\footnote{http://www.ice.cat/mice}) N-body simulations 
\citep{fosalba08,crocce09}. The extremely large volume of this run (box size $7680\,h^{-1}\,{\rm Mpc}$) 
allowed 216 essentially independent LRG DR6 mocks to be extracted. 
\cite{cabre2008} investigated different methods to estimate the error on 
the correlation function (see their Appendix A). They found that the jackknife error, 
an internal estimate made from the dataset itself (JK; see Norberg et~al. 2008), 
gave a reasonable match to the diagonal elements of the covariance matrix obtained 
directly from the 216 mock catalogues. However, the JK estimate of the off-diagonal 
elements of the covariance matrix is noisier than that obtained from the mocks. 
In this analysis we construct the full covariance matrix of the measurement from the correlation
matrix estimated from mock catalogues of dark matter haloes with similar clustering and abundance
to the LRGs, rescaled by the JK estimate of the variance from the data, which has the advantage of
being independent of the cosmological model adopted in the N-body simulation.

\subsubsection{The implications of a possible constant systematic shift 
in clustering amplitude}
\label{sec:Kshift}
Small systematic effects including, for example, the integral constraint, 
calibration errors or evolutionary effects, will, if unaccounted for, 
appear as an additive term in the measured correlation function: 
$\xi(s)= \xi_{\rm true}(s)+ \xi_{\rm sys}(s)$, where $ \xi_{\rm sys}(s)$ stands 
for the systematic error. The simplest model for $ \xi_{\rm sys}(s)$ is a constant
shift, $ \xi_{\rm sys}(s)=K$, which we label here as the {\it K-shift}.
Systematic shifts include both effects which are unaccounted for in the estimate of
the radial selection function and angular calibration errors which can move galaxies in or out of the 
galaxy sample. These can introduce spurious fluctuations in the observed density.
A 1\% variation in $(r-i)$ color or a 3\% shift in $r$ magnitude can introduce a 10\% modulation
in the LRG target number density \citep{eisenstein01}. However, \citet{hogg2005} found that the
large scale density variations in the final LRG sample are completely consistent with
the predictions of biased $\Lambda$CDM models, showing that when averaged over large angular
scales, the fluctuations in the photometric calibration of the SDSS do not affect significantly
the uniformity of the LRG sample. Even a spurious number density fluctuation as large as 
$\delta\sim5\%$ can only produce a shift $K= \delta^2<0.0025$.
As illustrated in Fig.\ref{fig:corfunc} the potential systematics 
that we are able to check seem to produce shifts of $|K|<0.001$.

These are small changes, but could our analysis below
be affected by such a systematic effect, if present?
We have checked this explicitly 
by repeating our analysis allowing for a constant additive term $K$ as an extra free parameter,
which we marginalize over. We imposed a prior on this K-shift as large as $|K|=0.01$. 
This corresponds to 10\% density fluctuations, well above the expected systematics in the LRG sample.
We find that even with this wide prior in $K$, the marginalization over the additive K-shift does not
change the obtained errors of the cosmological parameters and only changes the mean values by
less than 0.5$\sigma$ in the most extreme cases. This small difference does not justify the inclusion
of this extra parameter. But note that if we want to address the question of what is the {\it absolute} 
goodness of fit, then it might be important to consider such a K-shift 
as an additional degree of freedom. We will come back to this question 
in Sec.~\ref{sec:conclusions}.

\subsection{The redshift space correlation function of SDSS-DR6 LRGs:
 the radial BAO peak}
\label{ssec:rbao}

The main piece of clustering information we shall use in this paper is
the monopole or spherical average of the two-point redshift space
correlation function described in the previous subsection. However,
there is also useful cosmological information in the form of the
correlation function split into bins of pair separation parallel
($\pi$) and perpendicular ($\sigma$) to the line of sight,
$\xi(\sigma,\pi)$.  \citet{gaztanaga08b} found a significant detection
of a peak along the line-of-sight direction, the position and shape of
which is consistent with it being the baryonic acoustic peak.

\citet{gaztanaga08b} measured the position of the radial BAO peak
(rBAO) using the full LRG DR6 sample, and two sub-samples at low,
$z=0.15-0.30$, and high redshifts, $z=0.40-0.47$ (see their Table
II). In this paper we make use of these two last measurements, which
can be treated as independent due to the large separation between 
these redshift intervals. When combined with a measurement of the sound
horizon scale from the CMB, the position of the BAO feature can be
used as a standard ruler to obtain a constraint on the value of the
Hubble parameter, $H(z)$, as a function of redshift
\citep{gaztanaga08b}. 
Here, instead of calibrating the radial BAO distance $r_{\rm BAO}(z)$ 
with the CMB, we follow the same approach as \citet{gaztanaga08c} and use the
measurement of the dimensionless redshift scale
\begin{equation}
 \Delta z_{\rm BAO}(z) = r_{\rm BAO}(z)\frac{H_{\rm fid}(z)}{c}.
\label{eq:dimensionless}
\end{equation}
In this equation $H_{\rm fid}(z)$ is the Hubble constant at the redshift of the measurement in the
fiducial cosmology assumed by \citet{gaztanaga08b} to convert the observed galaxy redshifts into 
comoving distances. Note that the factor $H_{\rm fid}(z)/c$ in Eq.~(\ref{eq:dimensionless})
ensures that $\Delta z_{\rm BAO}$ does not depend on the fiducial cosmology assumed to obtain the
rBAO measurement.
Section~\ref{ssec:rbao_method} describes the details of our modelling
of these measurements.

\begin{figure}
\centering
\centerline{\includegraphics[width=\columnwidth]{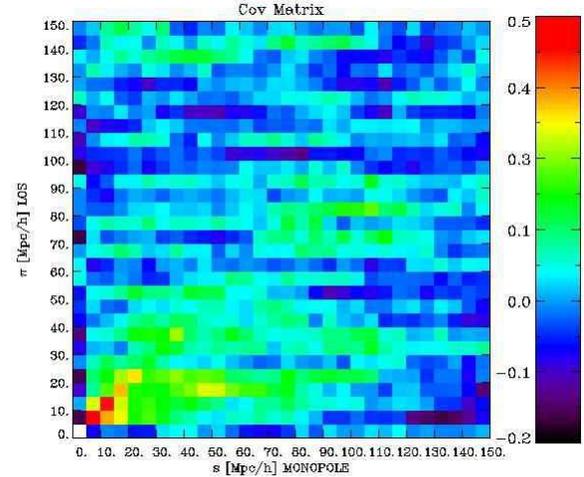}}
\caption{
Normalized covariance matrix estimated from 216 mocks LRG samples.
Here we test the covariance between the monopole correlation (horizontal axis)
and the radial (LOS) correlation (vertical axis). The correlation is only
important  on small scales and is negligible (less than 10\%) for the scales
of interest here ($>40\,h^{-1}\,{\rm Mpc}$).
}
\label{fig:covrBAO}
\end{figure}

Because of the narrow range of $\sigma$ values ($0.5<\sigma<5.5\,h^{-1}\,{\rm Mpc}$)
used in the radial measurement, the rBAO results are essentially independent
from the monopole of the two-point function; fewer than 1\% of the bins considered
in the spherically averaged correlation function are used in the estimate of the
radial correlation. For this reason we treat these two datasets as independent.
We tested this explicitly using the mock LRG catalogues from
\cite{cabre2008} (the same set used to estimate the covarince
matrix in the monopole), to evaluate the covariance of the
radial and the monopole correlations. Fig.~\ref{fig:covrBAO} shows
the normalized covariance:

\begin{equation}\label{eq:covMC} 
C_{ij} =  \frac{1}{M}
\sum_{k=1}^{M} \frac{[\xi_{\rm m}(i)^{k}-\widehat\xi_{\rm m}(i)]}{\sigma_{\rm m}(i)} 
\frac{[\xi_{\rm r}(j)^{k}-\widehat\xi_{\rm r}(j)]}{\sigma_{\rm r}(j)}
\end{equation}
where $\xi_{\rm m}(i)^{k}$ and $\xi_{\rm r}(j)^{k}$ are the monopole and 
radial correlation in the $i$-th and $j$-th bins respectivelly 
measured in the $k$-th mock catalogue $(k=1,...M)$ and 
$\widehat\xi$ and $\sigma$ are the corresponding mean
and rms fluctuations over the $M$ realizations.  As shown in Fig.\ref{fig:covrBAO}
the covariance on the scales of interest in our analysis (larger than 
$40 \,h^{-1}\,{\rm Mpc}$) is quite small (less than the noise in the estimation, of
about 10\%) which shows that the two datasets are indeed independent in practice. 
This implies that the rBAO measurements obtained from the radial correlation functions are
independent from the monopole $\xi(s)$.

\subsection{CMB data: temperature and polarization power spectra}
\label{ssec:cmb}

The accuracy of recent observations of the CMB mean that this is the 
single most powerful dataset for constraining the values of cosmological 
parameters.
The compilation of CMB measurements we use includes the 
temperature power spectrum in the range $2 \leq \ell \leq 1000$ and  
the temperature-polarization power spectrum for $2 \leq \ell \leq 450$
of the first five years of observations of the WMAP satellite \citep{hinshaw08, nolta2008};
the bandpower temperature  spectrum from the 2008 observations of the Arcminute
Cosmology Bolometer Array Receiver \citep[][ACBAR,]{kuo2007,reichardt2008} over the spherical 
harmonic range $910 < \ell < 1850$; temperature and polarization data from 
the Cosmic Background Imager \citep[][CBI,]{readhead2004} with $855 < \ell < 1700$;
observations from the 2003 flight of BOOMERANG \citep{ruhl2003,jones2006,montroy2006,
piacentini2006} in the range $925 < \ell < 1400$ and the recent results for 
the temperature and polarization power spectra measurements from QUaD \citep{ade08}
over the range $893 < \ell < 1864$. These measurements of the power spectrum of 
temperature fluctuations in the CMB cover the spherical harmonic range $2<\ell<1800$. 
Following \citet{dunkley2008}, in order to avoid cross-correlations with the WMAP data, 
we use only the bandpowers of the small scale CMB experiments that do not overlap 
with the signal-dominated WMAP data. Note that the QUaD measurements were not 
available at the time of the ``WMAP+CMB'' analysis carried out by Komatsu et~al. 
(2008).

\subsection{The Hubble diagram of type Ia supernovae}
\label{ssec:sn}

We also consider the constraints provided by the Hubble diagram of 
type Ia supernovae (SN) as provided by the UNION sample from \citep{kowalski08}. 
This compilation is drawn from 13 independent datasets processed using the 
{\sc SALT} light curve fitter \citep{guy2005} and analysed in a uniform way. 
The sample contains a set of 57 low-redshift SN, the recent samples from the 
SuperNova Legacy Survey \citep[SNLS,]{astier06} and the Equation of State 
SupErNovae trace Cosmic Expansion \citep[ESSENCE,][]{miknaitis2007}, the 
high-redshift sample from the Hubble Space Telescope \citep{riess2004,riess2007}, 
as well older datasets \citep{riess1998,perlmutter1999,tonry2003,barris2004}.
The final sample is the largest available to date, comprising 307 SN which pass 
the selection criteria.

\citet{kowalski08} suggest a way to include the effect of systematic errors when 
fitting the SN data. This estimation and those from other authors \citep{astier06,wv07,hicken09}, predict
different systematic errors in the estimated constraints on the dark energy equation
of state, with values ranging from 5\% to more than 10\%. These approaches differ in 
the choice of which potential sources of systematic errors are taken into account and the estimation of
their likely magnitude. It is our understanding that the community has not reached a
consensus about the correct way to estimate the effect of the systematic errors in the analysis of the
SN data. For this reason we follow other authors \citep[e.g.][]{komatsu08} and do not include the systematic
errors in our constraints on cosmological parameters. However, the inclusion of the systematic errors 
in the SN data has important implications for the derived values of the dark energy equation of state 
(see Section~\ref{ssec:wde}). This should be borne in mind when comparing constraints
obtained using SN data with those obtained using other datasets.

\section{Methodology}
\label{sec:method}

In this section we summarize the approach used to obtain constraints on cosmological 
parameters. We start by describing he different parameter sets that we consider.
The parametric model we use to describe the shape of the correlation function in
redshift space is presented \ref{ssec:rpt}. The theoretical motivation for this
parametric form can be found in Appendix~\ref{ssec:shape}. 
In Section~\ref{ssec:rpt}, we also compare the model with measurements made using 
numerical simulations. Section~\ref{ssec:rbao_method} describes the model we
implement to describe the radial BAO measurements. The methodology we follow to explore and 
constrain the parameter spaces is discussed in Section~\ref{ssec:mechanics}.

\subsection{The parameter space}
\label{ssec:param}

In this paper we make the basic assumption that the primordial
density fluctuations were adiabatic, Gaussian and had a power-law
spectrum of Fourier amplitudes, with a negligible contribution 
from tensor modes. From the analysis of the fifth year of
WMAP data, \citet{komatsu08} did not detect any deviation 
from these hypotheses at the 99\% confidence limit (CL).
Within this framework, a cosmological model can be defined by 
specifying the values of the following eight parameters:
\begin{equation}
\mathbf{P\equiv (}\Omega_{k},\omega_{\rm dm},\omega_{\rm b},\tau ,n_{\rm s},A_{\rm s},\Theta,
w_{\rm DE}).
\label{eq:param}
\end{equation}
We now go through the parameters in the above list, defining each one and 
also explaining how the values of other parameters are obtained, which we 
will refer to as derived parameters. 

The homogeneous background cosmology is described through the various 
contributions to the mass-energy density.
These are, in units of the critical density: $\Omega_{k}$, which describes 
the curvature of the universe; $\omega_{\rm dm}\equiv \Omega_{\rm dm}h^{2}$, 
the density of the dark matter (assumed cold, where $h$ is Hubble's constant 
in units of $100 \,{\rm km s}^{-1}{\rm Mpc}^{-1}$) and $\omega_{\rm b}\equiv 
\Omega_{\rm b}h^{2}$, the baryon density. We assume that massive neutrinos 
make no contribution to the mass budget.  
For most of this paper, we assume that the dark energy component has a constant 
equation of state independent of redshift, with the ratio of pressure
to density given by $w_{\rm DE}$. In section \ref{ssec:wa} we relax this assumption 
and analyse models allowing for a time variation in this parameter. In this case 
we use the standard linear parametrization given by \citep{chevalier2001,linder03}
\begin{equation}
w_{\rm DE}(a)=w_{\rm 0}+w_{a}(1-a),
\label{eq:wa}
\end{equation}
where $a$ is the expansion factor and $w_{\rm 0}$ and  $w_{a}$ are parameters.

The form of the initial fluctuations is described by two quantities; 
the scalar spectral index, $n_{\rm s}$ and the primordial amplitude of
the scalar fluctuations $A_{\rm s}$. These parameter values are quoted at 
the ``pivot'' scale wavenumber of $k=0.05\, {\rm Mpc}^{-1}$.

We assume that the reionization of the neutral intergalactic medium 
occurred instantaneously, with an optical depth given by $\tau$.
Finally, $\Theta$ gives the ratio of the sound horizon scale 
at the epoch of decoupling to the angular diameter distance 
to the corresponding redshift. 

There are further basic quantities whose values can be derived from those 
listed in the set of Eq.~(\ref{eq:param}):
\begin{equation}
\mathbf{P_{\rm derived} }\equiv (
\Omega_{\rm m},h,\Omega_{\rm DE},\sigma_{8},z_{\rm re},t_{0}).
\label{eq:paramderived}
\end{equation}
The matter density parameter is given by 
$\Omega_{\rm m} = \Omega_{\rm dm} + \Omega_{\rm b}$. 
The value of the Hubble constant is derived from 
$h = \sqrt{(\omega_{\rm dm}+\omega_{\rm b})/\Omega_{\rm m}}$.
The energy-density of the dark energy is set by 
$\Omega_{\rm DE} = 1 - \Omega_{\rm m} - \Omega_{k}$. 
The results for $A_{\rm s}$ can be translated into a constraint 
on $\sigma_{8}$, the {\it rms} linear
perturbation theory variance in spheres of radius $8~h^{-1}$Mpc,
using the matter fluctuation transfer function.
The redshift of reionization, $z_{\rm re}$, can be computed from 
the values of $\tau$, the Hubble constant and the matter and baryon 
densities \citep{tegmark1994}. The age of the universe is $t_{0}$.

The $\Lambda$CDM cosmological model is the simplest model 
which can account for the wide variety of cosmological observations 
available today. This model is characterized by six parameters: 
\begin{equation}
\mathbf{
P^{6}_{\rm varied} }\equiv(\omega_{\rm dm},\omega_{\rm b},n_{\rm s},
\tau,A_{\rm s},\Theta),  
\label{eq:param6}
\end{equation}
assuming $\Omega_{\rm k}=0$ and $w_{\rm DE}=-1$.
This parameter space is well constrained by the temperature and polarization 
power spectrum measurements from five years of integration of the WMAP satellite 
\citep{komatsu08}. In Section~\ref{ssec:lcdm} we analyse the impact of the 
measurement of the shape of $\xi(s)$ on the constraints in this parameter space.

Using the latest WMAP data, \citet{komatsu08} placed strong constraints on the 
possible deviations from the $\Lambda$CDM model, namely non-Gaussianity, the 
presence of isocurvature modes, deviations from a pure power law scalar primordial 
power spectrum, the presence of tensor modes, a non-negligible energy
component in the form of massive neutrinos and parity-violation interactions. 
However, there are two parameters that signal important deviations of the standard 
$\Lambda$CDM model that can not be tightly constrained from CMB data alone: the 
curvature of the Universe $\Omega_{k}$ and the dark energy equation of state $w_{\rm DE}$.
In order to assess the improvement on the constraints once 
the information on the shape of $\xi(s)$ is included in the analysis, we explore 
four parameter spaces which contain extensions of the simple $\Lambda$CDM set. 
First we extend the parameter set of Eq.~(\ref{eq:param6}) by adding a constant 
dark energy equation of state
\begin{equation}
\mathbf{
P^{6+{\rm w_{DE}}}_{\rm varied} }\equiv(\omega_{\rm dm},\omega_{\rm b},n_{\rm s},\tau,A_{\rm s},\Theta,w_{\rm DE}),
\label{eq:paramwde}
\end{equation}
fixing $\Omega_k=0$. The results obtained in this case are 
shown in Section~\ref{ssec:wde}.

In Section~\ref{ssec:wa} we include the parametrization of Eq.~(\ref{eq:wa}) in 
our analysis and explore the extended parameter space
\begin{equation}
\mathbf{
P^{6+{\rm w(a)}}_{\rm varied} }\equiv(\omega_{\rm dm},\omega_{\rm b},n_{\rm s},
\tau,A_{\rm s},\Theta,w_{\rm 0},w_a),
\label{eq:paramwa}
\end{equation}
where we also implement the hypothesis of a flat universe with $\Omega_k=0$. 
We also analyse the effect of dropping this assumption.
First, in Section~\ref{ssec:omk}, we include $\Omega_k$ as a free parameter, with 
\begin{equation}
\mathbf{
P^{6+{\rm \Omega_k}}_{\rm varied} }\equiv(\omega_{\rm dm},\omega_{\rm b},
n_{\rm s},\tau,A_{\rm s},\Theta,\Omega_k),
\label{eq:paramomk}
\end{equation}
assuming that the dark energy is given by a cosmological constant (or vacuum energy) 
with $w_{\rm DE}=-1$. Finally, in Section~\ref{ssec:wok}, we also allow this 
parameter to vary freely and we explore the full parameter space of Eq.~(\ref{eq:param}).

\subsection{A physically motivated model for the correlation function}
\label{ssec:rpt}

In the linear perturbation theory regime (valid when the fluctuation amplitude is 
small, for instance at high redshift or on very large scales), the shape of the 
matter correlation function is well understood and can be readily obtained using 
linear Boltzmann solvers, such as CAMB \citep{camb}. The shape 
of the correlation function, in the context of a standard adiabatic CDM model, 
is sensitive to the values of the matter density, $\Omega_m h^2$, the baryon 
density $\Omega_b h^2$, the spectral tilt $n_{\rm s}$ and the density parameter 
of massive neutrinos. The evolution with redshift of the correlation function is 
well understood in the linear regime. In this case, each Fourier mode of the density 
field evolves independently of the others and the shape of $\xi$ is unaltered, 
although the overall amplitude changes with time. 

Unfortunately this simple behaviour is modified by a number of nonlinear phenomena, which affect 
different scales at different epochs. These include nonlinear effects generated by 
the latter stages of gravitational instability, redshift-space distortions caused 
by gravitationally induced motions and a possible non-trivial scale dependent bias 
relation between the distribution of galaxies and the underlying dark matter field
\citep[see, for example, the step-by-step illustration of these effects given by][]{angulo08}. 
Nonlinear growth results in cross-talk between different Fourier modes 
and introduces scale dependent patterns in the clustering, even on large scales. 
This is particularly noticeable in the appearance of the BAO bump, which is 
sensitive to the match between the amplitude and phases of the fluctuations around 
the peak scale. Due to the distortion of the Fourier modes from their original values, 
nonlinear growth causes the BAO bump to become smeared out and also to lose contrast. 
Such effects cannot be ignored when modelling low-redshift data.

\begin{figure}
\centering
\centerline{\includegraphics[width=\columnwidth]{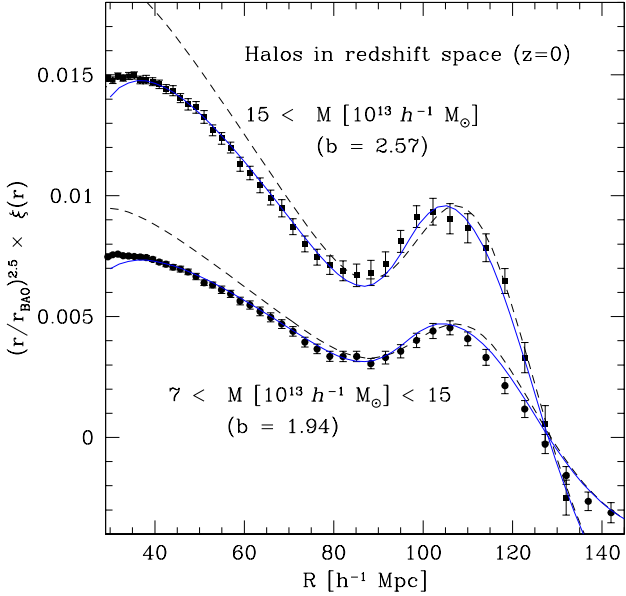}}
\caption{
The correlation function of dark matter halos in redshift space measured in 
an ensemble of 50 large-volume N-body simulations (with total volume of $\sim 105\,h^{-3}\,
{\rm Gpc}^3$). The error bars correspond to the error on the mean of the ensemble and 
are obtained from the scatter among the $50$ realizations. 
The best fitting parametric model used in this work, Eq.~(\ref{eq:xi_model}), is shown by 
the solid blue line (note the fit takes into account the covariance between the bins). 
The dashed line corresponds to setting $A_{\rm MC}=0$ and 
highlights the importance of this term in matching the shape of the correlation 
function at $r < 80\,h^{-1}\,{\rm Mpc}$. The scaling uses $r_{\rm BAO}=102 h^{-1}$Mpc.}
\label{fig:modelRZspace}
\end{figure}

On large scales, ($r> 30 \,h^{-1}\,{\rm Mpc}$), the correlation function 
falls sharply with increasing comoving pair separation, scaling roughly as a power 
law $\sim r^{-\gamma}$, with $\gamma \sim 2.5 $. At even larger separations, this 
behaviour is altered by the emergence of a bump known as the BAO peak (e.g. see Fig.~\ref{fig:modelRZspace}).  
This feature, first measured for LRGs by \citet{eisenstein05}, is centered 
at about $110\,h^{-1}\,{\rm Mpc}$ and has a width of $\sim 20\,h^{-1}\,{\rm Mpc}$.
The amplitude of the peak in the LRG correlation function corresponds to 
a $\sim 1\%$ excess in the number of LRG pairs above the number expected in a 
random distribution \citep[e.g.][]{cabre2008}.

To model the shape of the correlation function on large scales, we 
follow \citet{crocce08} and \citet{sanchez08b} and adopt the following parametrization:
\begin{equation}
 \xi_{\rm NL}(r) = b^2 \{ \xi_{\rm L}(r)\otimes {\rm e}^{-(k_{\star}r)^2} 
+ A_{\rm MC} \,\xi'_{\rm L}(r)\,\xi^{(1)}_{\rm L}(r) \}, 
\label{eq:xi_model}
\end{equation}
where $b$, $k_{\star}$ and $A_{\rm MC}$ are nuisance parameters, and the symbol $\otimes$ 
denotes a convolution. Here $\xi'_{\rm L}$ is the derivative of the linear correlation 
function and $\xi^{(1)}_{\rm L}(r)$ is defined by the integral: 
\begin{equation}
 \xi_{\rm L}^{(1)}(r) \equiv \hat{r} \cdot \nabla^{-1}\xi_{\rm L}(r)
=4\pi\,\int P_{\rm L}(k)\,j_1(kr)k\,{\rm d}k ,
\label{eq:xi1}
\end{equation}
with $j_{\rm 1}(y)$ denoting the spherical Bessel function of order one. 
The model in Eq.~\ref{eq:xi_model} is primarily motivated by RPT, where the matter power spectrum is written
as $P=G^2P_L+P_{MC}$, with $G$ a {\it nonlinear} growth factor and $P_{\rm MC}$ being the power generated by
mode-coupling. To a very good approximation $G$ is of Gaussian form, while at large scales the leading order
contribution of $P_{\rm MC}$ in real space is $\sim \xi^{(1)}_L \xi^{\prime}_L$ (see Appendix \ref{ssec:shape}
for a detailed discussion). The Gaussian degradation of the BAO information was also shown to be a good
description by \citet{eisenstein06}.

We now demonstrate how accurately the model of Eq.~(\ref{eq:xi_model}) can reproduce 
the spatial clustering of halo samples with comoving number densities similar to  
that of the LRGs in DR6. To this end we utilize an ensemble of $50$ realizations of 
collisionless dark matter N-body simulations. Each simulation contains $640^3$ 
particles in a comoving volume $V=L^3=(1280 \,h^{-1}\,{\rm Mpc})^3$. The 
cosmological parameters were set to $\Omega_m=0.27$, $\Omega_{\Lambda}=0.7$, 
$\Omega_b=0.046$ and $h=0.72$. The initial power spectrum had spectral index $n_s=1$ 
and was normalized to give $\sigma_8=0.9$ when linearly extrapolated to $z=0$. Halos were identified using the 
friends-of-friends algorithm with linking-length parameter $l=0.2$ \citep[see][for more details]{smith08,crocce08}.

Fig.~\ref{fig:modelRZspace} shows the 2-point correlation function measured 
in redshift-space at $z=0$ for two non-overlapping samples of dark matter haloes of masses 
$7\times 10^{13}<M [{\it h}^{-1}\,{\rm M}_{\odot}]<15\times 10^{13}$ 
and $15 \times 10^{13} <M [{\it h}^{-1}\,{\rm M}_{\odot}]$ (with number 
densities $\bar{n}[{\it h}^3\,{\rm Mpc}^{-3}]=1.88\times 10^{-5}$ and 
$3.46\times 10^{-5}$ respectively). 
Our redshift space measurements were done considering the
contribution of peculiar velocities along one dimension in the
simulation (i.e. recreating a plane-parallel configuration).

\citet{szapudi04} discussed in depth the issue of wide angle redshift space distortions.
He found that opening angles $\theta \lesssim 15^{\circ}-20^{\circ}$ would ensure the
validity of the plane-parallel approximation. For the redshift range of the LRG sample considered
in our analysis ($0.15 < z < 0.47$) and the largest pair-distance separation allowed 
($s_{\rm max}=150\,h^{-1}{\rm Mpc}$), the maximum opening angle is given by 
$\sin(\theta_{\rm max})=s_{\rm max}/D_{\rm A}(z=0.15)$, which implies $\theta_{\rm max} \approx 20^{\circ}$.
For the mean redshift of the sample the angle subtended by $s_{\rm max}$ is $\theta \approx 10^{\circ}$,
well within the validity of the plane-parallel approximation.

The error bars plotted show the error on the mean obtained 
from the ensemble of 50 simulations. Each simulation has a larger volume than the LRG sample we use
from DR6. Therefore these errors are $\approx \sqrt{2} \times \sqrt{50}= 10 $ times smaller than they
would be for the LRG sample. This means that the deviations between models (solid lines) and simulation
results (points) are not important for our purposes. 

The parametrization given in Eq.~(\ref{eq:xi_model}) corresponds to the solid 
blue line in Fig.~\ref{fig:modelRZspace}. The corresponding best-fit $\chi^2$ values 
for the nuisance parameters $(k_{\star}[{\it h}\,{\rm Mpc}^{-1}],b,A_{\rm MC})$ were $(0.142,1.94,3.23)$
and $(0.146,2.75,4.53)$ respectively. Clearly, the model can accurately describe the clustering of halos in redshift space.

\citet{sanchez08b} also compared Eq.~(\ref{eq:xi_model}) against measurements of the non-linear correlation function 
from a similar large ensemble of N-body simulations at various redshifts, and confirmed that this form gives an essentially
unbiased measurement of the dark energy equation of state using both, real and redshift space information.

The dashed line in Fig.~\ref{fig:modelRZspace} corresponds to Eq.~(\ref{eq:xi_model}) with $A_{\rm MC}=0$. From the plot
we can see that the inclusion of this term is the key to recovering the correct clustering shape at separations smaller than
the BAO bump. In addition it contributes slightly to the shape of the bump and alleviates a systematic effect related to the
position of the BAO peak \citep{crocce08} (but this is sub-dominant for the survey volume being considered). In subsequent
sections we check that this nuisance parameter is not degenerate with any of the cosmological parameters. 

On theoretical grounds, one expects the smoothing length $k_{\star}^{-1}$ to depend on cosmology (i.e. aside from galaxy
type or redshift) with, for example, a $10\%$ increase in $\Omega_{\rm m}$ expected to increase $k_{\star}$ by about $4\%$
\citep{crocce06b,matsubara08}. In view of this, we decided to consider $k_{\star}$ as a nuisance parameter at 
the expense of a possible increase in error bars. Note that this is at variance 
with the approach of \citet{eisenstein05}, \citet{tegmark2006} and \citet{percival07c} 
who kept this length fixed.

In summary, for each cosmological model we compute the linear correlation 
function $\xi_{\rm L}(r)$ using CAMB to generate the corresponding transfer 
function, and simulate nonlinear effects through Eq.~(\ref{eq:xi_model}) after 
computing $\xi_{\rm L}^{\prime}(r)$ and $\xi_{\rm L}^{(1)}(r)$ from 
Eq.~(\ref{eq:xi1}).

\subsection{A model for the radial acoustic scale}
\label{ssec:rbao_method}

Here we describe the simple model that we use to compute the dimensionless redshift radial 
acoustic scale $\Delta z_{\rm BAO}$ for a given choice of the cosmological parameters of Eq.~(\ref{eq:param}).

The value of $\Delta z_{\rm BAO}$ can be computed as,
\begin{equation}
\Delta z_{\rm BAO}(z)=\frac{H(z)r_{\rm s}(z_{\rm d})}{c},
\label{eq:deltaz}
\end{equation}
where $H(z)$ is the Hubble constant at the mean redshift of the measurements ($z_{\rm m}=0.24$
and 0.43), and $r_{\rm s}(z_{\rm d})$ is the comoving sound horizon at the drag epoch, which
is given by
\begin{equation}
 r_{\rm s}(z)=\frac{c}{\sqrt{3}}\int_0^{(1+z)^{-1}} \frac{{\rm d}a}{a^2H(a)\sqrt{1+Ra}},
\label{eq:soundh}
\end{equation}
where $R=3\Omega_{\rm b}/4\Omega_{\rm \gamma}$ and $\Omega_{\rm \gamma}=2.469\times10^{-5}\,h^{-2}$ 
for a CMB temperature $T_{\rm CMB}=2.725\,{\rm K}$. The value of $z_{\rm d}$ can be computed with high accuracy
from the values of $\omega_{\rm b}$ and $\omega_{\rm dm}$ using the fitting formulae of \citet{EH98}.

The radial BAO scale measurements can be used to place constraints on cosmological parameters independently of other datasets.
Being a purely geometrical test, as in the case of the SN data, the radial BAO measurements are not 
sensitive to all the cosmological parameters of Eq.~(\ref{eq:param}) since they contain no information about the
primordial spectrum of density fluctuations, that is $A_{\rm s}$ and $n_{\rm s}$, and of the optical depth to the
last scattering surface $\tau$.

\begin{table}
\begin{center}
\caption{ The parameter space probed in our analysis. We assume a flat 
prior in each case. The parameter spaces that we consider are set out 
in Section~\ref{ssec:param}. 
}
\end{center}
\begin{center}
\begin{tabular}[t]{cc}
\hline\hline
Parameter       & Allowed range \\ \hline\hline
$\omega_{\rm dm}$  & 0.01 -- 0.99   \\ 
$\omega_{\rm b}$   & 0.005 -- 0.1  \\ 
$\Theta  $      & 0.5 -- 10   \\ 
$\tau $             & 0 -- 0.8   \\ 
$n_{\rm s}$         & 0.5 -- 1.5    \\ 
$\ln(10^{10}A_{\rm s})$ & 2.7 -- 4.0 \\ 
$w_{\rm DE }$       & $-$2. -- 0  \\ 
$\Omega_{k}$   & $-$0.3 -- 0.3   \\ 
\hline\hline
\end{tabular}
\end{center}
\label{tab:priors}
\end{table}

\subsection{Practical issues when constraining parameters} 
\label{ssec:mechanics}

We use a Bayesian approach and explore the different parameter 
spaces defined in Section~\ref{ssec:param} 
using the Markov Chain Monte Carlo (MCMC) technique. Our results were generated with the  
publicly available {\sc CosmoMC} code of Lewis \& Bridle (2002). 
{\sc CosmoMC} uses the {\sc camb} package to compute power spectra for 
the CMB and matter fluctuations (Lewis, Challinor \& Lasenby 2000). 
We use a generalized version of {\sc camb} which supports a time-dependent 
dark energy equation of state \citep{fang2008}. For each parameter set considered, 
we ran twelve separate chains which were stopped when the Gelman and Rubin (1992) 
criteria reached $R<1.02$. We implemented flat priors on our base parameters. 
Table~1 summarizes the ranges considered for different 
cosmological parameters in the cases where their values are allowed to vary.

In order to establish the link between a given cosmological model and 
the datasets described in Section~\ref{sec:data} it is necessary to 
include a small set of extra parameters given by
\begin{equation}
\mathbf{P_{\rm extra} }\equiv (b,k_{\star},A_{\rm MC},A_{\rm SZ}),
\label{eq:paramextra}
\end{equation}
to the parameter sets described in Section~\ref{ssec:param}.
The bias factor $b$, describes the difference in amplitude between the galaxy
correlation function and that of the underlying dark matter. 
The values of $k_{\star}$ and $A_{\rm MC}$ from Eq.~(\ref{eq:xi_model}) are 
also included as free parameters in our parameter space. $A_{\rm SZ}$ gives 
the amplitude of the contribution from the Sunyaev-Zeldovich effect to the 
CMB angular power spectrum on small scales (high $\ell$). When quoting 
constraints on the parameters of Eq.~(\ref{eq:param}) and (\ref{eq:paramderived}), 
the values of these extra parameters are marginalized over. In the case of $b$, 
this is done using the analytic expression given in Appendix F of 
Lewis \& Bridle (2002), but the remaining parameters are included in our 
Monte Carlo analysis (see Section~\ref{ssec:mechanics}).

There is one important point that must be considered in order to make a 
comparison between the model of Eq.(\ref{eq:xi_model}) and the observational 
data of the LRG redshift-space correlation function.
When measuring $\xi(s)$, in order to map the observed galaxy redshifts and 
angular positions on the sky into distances, it is necessary to assume a 
fiducial cosmological model. This choice has an impact on the results obtained.

\begin{figure}
\centering
\centerline{\includegraphics[width=\columnwidth]{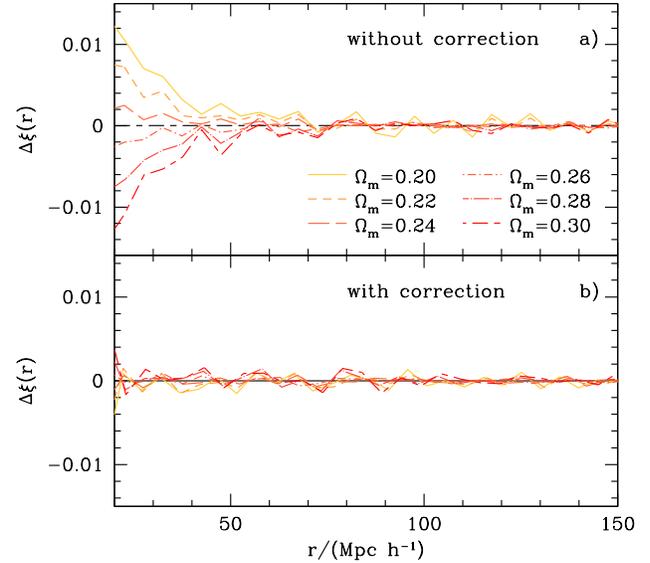}}
\caption{
The impact of a mismatch in cosmology on the form of the correlation function. 
The upper panel shows the difference between the correlation functions measured 
assuming different fiducial cosmologies ($\Lambda$CDM models with varying 
$\Omega_{\rm m}$) from that obtained assuming $\Omega_{\rm m}=0.25$. The lower 
panel shows the same comparison after applying the scale shift of Eq.~(\ref{eq:cor_fact}) 
to take into account the change in the value of $\Omega_{\rm m}$.
}
\label{fig:alpha}
\end{figure}

One possible way to deal with this is to re-compute the correlation function 
and its covariance matrix for the cosmology corresponding to each point in the 
Markov chains, and then use this measurement when computing the likelihood of 
the given cosmological model. This approach is infeasible since it would require 
an exceedingly large amount of computing time. Instead we follow an alternative 
approach by indirectly including the effect of the choice of the cosmology on 
the model correlation function. To do this we follow \citet{eisenstein05} and 
simply rescale the distances in the model correlation function by a factor
\begin{equation}
     \alpha = \frac{D_{{\rm V}}^{\rm model}}{D_{\rm V}^{{\rm fiducial}}},
\label{eq:cor_fact}
\end{equation}
where the effective distance $D_{{\rm V}}(z_{\rm m})$ to the mean redshift of the survey 
$z_{\rm m}=0.35$, is computed for each model as
\begin{equation}
 D_{\rm V}(z_{\rm m})=\left[ D_{\rm A}^2(z_{\rm m})\frac{cz}{H(z_{\rm m})}\right]^{1/3}.
\label{eq:dv}
\end{equation}
where $D_{\rm A}(z)$ is the comoving angular diameter distance given by
\begin{equation}
D_{\rm A}(z)=\frac{c}{H_0\sqrt{|\Omega_k}|}f_k \left( H_0\sqrt{|\Omega_k|}\int_0^z\frac{{\rm d}z'}{H(z')}\right),
\label{eq:dang}
\end{equation}
where
\begin{equation}
 f_k(x) = 
  \cases{ {\rm sinh}(x) &if$(\Omega_k>0)$, \cr
    x &if$(\Omega_k=0)$, \cr
    \sin(x) &if$(\Omega_k<0)$. \cr
  }
\end{equation}

The exponents within the square bracket on the right-hand 
side of Eq.~(\ref{eq:dv}) assume that the survey covers 
a wide solid angle, rather than a pencil-beam. 

We can test the effectiveness of the correction factor given by Eq.~(\ref{eq:cor_fact}) 
by applying it to different estimates of the correlation function computed with 
varying choices of the fiducial cosmology. \citet{cabre2008} computed the 
redshift space correlation function of SDSS-DR6 LRGs assuming different flat 
fiducial cosmologies with values of $\Omega_m$ ranging from 0.2 to 0.3. The upper 
panel of Fig.~\ref{fig:alpha} shows the difference of these estimates from the one 
obtained for $\Omega_{\rm m}=0.25$. It is clear that the choice of the cosmological 
model affects the shape of the correlation function. The effect of this choice is 
particularly important on small scales ($r<70\,h^{-1}\,{\rm Mpc}$), but also 
introduces a small distortion to the shape of the acoustic peak, shifting its 
position towards larger (smaller) scales for smaller (larger) values of $\Omega_{\rm m}$.
If unaccounted for, this difference would bias the constraints obtained on 
the cosmological parameters. The lower panel shows the same quantity once 
the correction factor of Eq.~(\ref{eq:cor_fact}) has been applied. Clearly, this 
simple correction is able to account for the choice of the fiducial cosmological 
model. We use this correction factor to translate the model correlation function to the
fiducial cosmology used by \citet{cabre2008} to estimate the LRG $\xi(s)$ (a flat $\Lambda$CDM
mode with $\Omega_{\rm m}=0.25$). We then compute the likelihood 
of the model assuming the Gaussian form ${\cal L}\propto\exp(-\chi^2/2)$.

\section{Testing the model of $\xi(s)$}
\label{sec:test}

In this section we analyse the sensitivity of the parameter constraints 
to the details of the procedure we follow to compute the likelihood of a given 
model. For the purpose of this exercise we use the information contained 
in the correlation function of LRGs combined with the latest results 
from the WMAP satellite alone to constrain the parameter set of 
Eq.~(\ref{eq:paramwde}), and assess the impact on the results of 
varying choices in our analysis procedure.

\begin{figure}
\centering
\centerline{\includegraphics[width=\columnwidth]{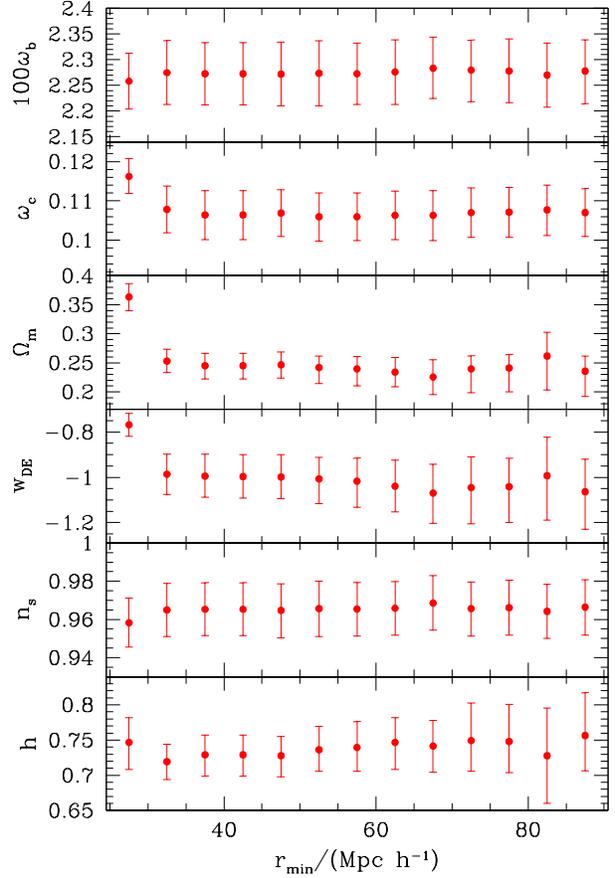}}
\caption{
The dependence of the constraints on cosmological parameters on the minimum pair 
separation, $r_{\rm min}$, included in the correlation function measurement.
The points show the mean value of the likelihood for each parameter and the 
error bars show the 68\% CL. 
}
\label{fig:scales}
\end{figure}

First we study the sensitivity of the results to the range of scales in $\xi(s)$
included in the analysis. \citet{sanchez08b} showed that the model of 
Eq.~(\ref{eq:xi_model}) gives an accurate description of the redshift-space 
halo correlation function measured from N-body simulations on scales in the 
range $60\,h^{-1}\,{\rm Mpc}\le r\le 180\,h^{-1} {\rm Mpc}$. In Section~\ref{ssec:rpt}, 
we showed that the second term in Eq.~(\ref{eq:xi_model}) also helps to reproduce
the results from N-body simulations down to scales of 
$40\,h^{-1}{\rm Mpc}$. Even though our measurement of $\xi(s)$ extends to larger scales, 
the measurement beyond $150\,h^{-1}\,{\rm Mpc}$ is noisy and does not have any effect on 
the constraints. Besides, \citet{cabre2008} showed that on these large scales, 
the measurement of the redshift space correlation function exhibits an excess in the 
amplitude with respect to the previous estimate of \citet{eisenstein05}. For this reason 
we have chosen to set a maximum distance $r_{\rm max}=150\,h^{-1}\,{\rm Mpc}$ in all 
subsequent analyses.

\begin{table*} 
\centering
 \begin{minipage}{172mm}
  \caption{
    The marginalized 68\% interval constraints on the cosmological parameters of 
    the $\Lambda$CDM model obtained using different combinations of the datasets 
    described in Section~\ref{sec:data}, as stated in the column headings.}
    \begin{tabular}{@{}lccccccc@{}}
    \hline
& \multirow{2}{*}{CMB}  & \multirow{2}{*}{CMB + $\xi(s)$} & \multirow{2}{*}{CMB + rBAO}& \multirow{2}{*}{CMB + SN} & CMB + $\xi(s)$ & CMB + $\xi(s)$ & \multirow{2}{*}{$\xi(s)$+ rBAO}\\
&     &                &                                  &                            &   + rBAO        &  + rBAO + SN &  \\  
\hline
$100\Theta$           & $ 1.0416_{-0.0023}^{+0.0023}$ & $1.0415_{-0.0022}^{+0.0022}$  & $ 1.0413_{-0.0022}^{+0.0022}$ & $ 1.0413_{-0.0022}^{+0.0022}$   & $1.0414_{-0.0022}^{+0.0022}$ &  $ 1.0412_{-0.0022}^{+0.0022}$ &$ 0.987_{ -0.069}^{+ 0.065}$\\[1.5mm]
$\omega_{\rm dm}$     & $ 0.1091_{-0.0053}^{+ 0.0053}$& $0.1076_{-0.0038}^{+0.0038}$  & $ 0.1108_{-0.0036}^{+0.0036}$ & $ 0.1125_{ -0.0038}^{+ 0.0038}$  & $ 0.1095_{-0.0032}^{+0.0032}$   &  $ 0.1110_{-0.0029}^{+0.0029}$&$ 0.091_{ -0.038}^{+ 0.040}$\\[1.5mm] 
$100\,\omega_{\rm b}$ & $ 2.282_{-0.051}^{+0.050}$    & $2.280_{-0.051}^{+0.050}$     & $ 2.277_{-0.050}^{+0.051}$    & $ 2.270_{-0.051}^{+ 0.051}$ &$2.276_{-0.049}^{+0.049}$   & $ 2.267_{-0.051}^{+0.050}$ &$3.0_{-1.7}^{+2.0}$\\[1.5mm]
$\tau $              & $ 0.088_{-0.016}^{+0.016}$    &  $0.090_{-0.017}^{+0.017}$   & $0.088_{-0.016}^{+0.016}$ &  $ 0.087_{-0.016}^{+0.016}$     &  $0.088_{-0.016}^{+0.016}$  & $ 0.086_{-0.016}^{+0.016}$& - \\[1mm] 
$n_{\rm s}$          &  $ 0.965_{ -0.013}^{+0.013}$  &   $0.965_{-0.012}^{+ 0.012}$  & $0.962_{-0.012}^{+0.012}$& $ 0.960_{-0.012}^{+0.012}$   &  $ 0.963_{-0.011}^{+0.011}$  & $ 0.960_{-0.011}^{+0.011}$&$ 1.06_{ -0.31}^{+0.34}$\\[1.5mm] 
$\ln(10^{10}A_{\rm s})$ & $ 3.066_{-0.037}^{+0.037}$   &  $3.059_{-0.038}^{+0.038}$   &$3.067_{-0.036}^{+0.036}$ & $ 3.070_{-0.036}^{+0.036}$     &  $3.062_{-0.036}^{+0.036}$   &$3.065_{-0.036}^{+0.036}$&-\\[1.5mm] 
$\Omega_{\rm DE}$       & $ 0.749_{-0.026}^{+0.026}$    & $ 0.755_{-0.018}^{+ 0.018}$    &  $0.740_{-0.017}^{+0.017}$  & $ 0.730_{-0.020}^{+ 0.019}$   &  $ 0.747_{-0.014}^{+0.014}$    & $ 0.739_{-0.013}^{+0.013}$& $0.782_{-0.041}^{+0.040}$\\[1.5mm]
$\Omega_{\rm m}$      &  $0.251_{-0.026}^{+0.026}$   &  $ 0.244_{-0.018}^{+ 0.018}$   & $ 0.260_{ -0.017}^{+ 0.017}$   & $ 0.270_{-0.019}^{+0.020}$    & $ 0.253_{-0.014}^{+0.014}$       & $ 0.261_{-0.013}^{+0.013}$& $ 0.218_{-0.040}^{+ 0.041}$\\[1.5mm]
$\sigma_{8}$          &  $ 0.795_{-0.029}^{+0.028}$   &  $ 0.787_{-0.024}^{+0.024}$  &  $ 0.802_{-0.024}^{+0.024}$  & $ 0.810_{ -0.023}^{+0.023}$  &  $ 0.795_{-0.022}^{+0.021}$    & $ 0.802_{-0.021}^{+0.021}$& -\\[1.5mm]
$t_{0}/{\rm Gyr}$     & $13.64_{-0.11}^{+0.11}$      &  $13.64_{-0.10}^{+0.10}$    &  $13.668_{-0.097}^{+0.099}$ & $13.69_{-0.10}^{+0.10}$ &  $13.660_{-0.096}^{+0.097}$   & $13.682_{-0.094}^{+0.095}$&$14.4_{-2.9}^{+3.2}$\\[1.5mm] 
$z_{\rm re}$          &  $10.5_{-1.3}^{+1.3}$        &   $10.5_{-1.3}^{+1.3}$      &  $10.4_{-1.3}^{+1.4}$  & $10.4_{-1.3}^{+1.3}$       &   $10.4_{-1.3}^{+1.3}$       & $10.4_{-1.3}^{+1.3}$ & - \\[1.5mm]
$h$                  &  $0.726_{-0.024}^{+0.025}$    &  $0.731_{-0.018}^{+0.018}$ & $0.718_{-0.016}^{+0.016}$ & $0.710_{-0.018}^{+0.017}$    &   $0.723_{-0.013}^{+0.013}$  & $0.716_{-0.012}^{+0.012}$& 
$0.73_{-0.14}^{+0.15}$\\
\hline
\end{tabular}
\label{tab:lcdm}
\end{minipage}
\end{table*}

Fig.~\ref{fig:scales} shows the one-dimensional marginalized constraints on a subset of 
the parameters space of Eq.~(\ref{eq:paramwde}) as a function of the minimum 
scale, $r_{\rm min}$, included in the analysis. 
For $r_{\rm min}<40\,h^{-1}\,{\rm Mpc}$ we see that the constraints start to vary with
$r_{\rm min}$. Coincidentally, in this regime the model of Eq.~(\ref{eq:xi_model}) 
starts to deviate from the N-body measurements of the correlation function discussed in Sec.~\ref{ssec:rpt}. 
On such scales, extra terms beyond the one-loop contribution $\sim \xi^{\prime}$ to $\xi_{\rm MC}$
are expected to become important, which implies the breakdown of our ansatz.

For $r_{\rm min}>40\,h^{-1}\,{\rm Mpc}$, Fig.~\ref{fig:scales} shows that the constraints 
on the values of the cosmological parameters are very stable. Furthermore, the mean values 
obtained for these parameters using CMB data plus the LRG correlation function are in 
complete agreement with the ones obtained from CMB information only.
The allowed regions of some parameters, like $\omega_{\rm b}$ or $n_{\rm s}$ which are 
tightly constrained by the CMB data alone 
show almost no change on varying  
$r_{\rm min}$. Other parameters, such as $\Omega_{\rm m}$ or $w_{\rm DE}$ show a 
substantial increase in their allowed regions as the data from small scales is gradually 
excluded from the analysis.
For $r_{\rm min}=42.5\,h^{-1}\,{\rm Mpc}$ we obtain a constraint on the dark energy 
equation of state of $w_{\rm DE}=-0.996_{-0.095}^{+0.097}$, while for 
$r_{\rm min}=82.5\,h^{-1}\,{\rm Mpc}$, that is including only scales close to the
acoustic peak in $\xi(s)$, we get $w_{\rm DE}=-0.99_{-0.19}^{+0.17}$. This approximately 
factor of two change in the error bar highlights the importance of the inclusion of 
information from the shape of $\xi(s)$ on intermediate scales. Based on this comparison  
and the results of Section~\ref{ssec:rpt}, we have chosen to set 
$r_{\rm min}=42.5\,h^{-1}\,{\rm Mpc}$ in the subsequent analysis.

We also studied the impact of the correction to the measured correlation function to take
into account changing the fiducial cosmology assumed when estimating $\xi(s)$. For this
test we obtained constraints in the same parameter space without applying the correction of
Eq.~(\ref{eq:cor_fact}).
The results obtained in this way are entirely consistent with
those obtained when this correction is applied. The mean values of all the parameters remain
practically identical with the exception of $\Omega_{\rm m}$ which shows a slight shift towards
higher values of approximately 0.2$\sigma$. The allowed regions for these parameters also show 
almost no variation with a slight increase in the confidence limits of $\Omega_{\rm m}$ and $w_{\rm DE}$ of
approximately 0.3$\sigma$.
A similar shift but towards smaller values of $\Omega_{\rm m}$ is obtained on setting $A_{\rm MC}=0$
in the second term of the right hand side of Eq.~(\ref{eq:xi_model}). 
This is in agreement with
the results of \citet{sanchez08b}, who found that the two most important parameters of this parametrization
required to obtain a good description of the shape of the two-point correlation function are
$k_{\star}$ and $b$ (which in this case is marginalized over). 

\begin{figure*}
\centering
\centerline{\includegraphics[width=0.85\textwidth]{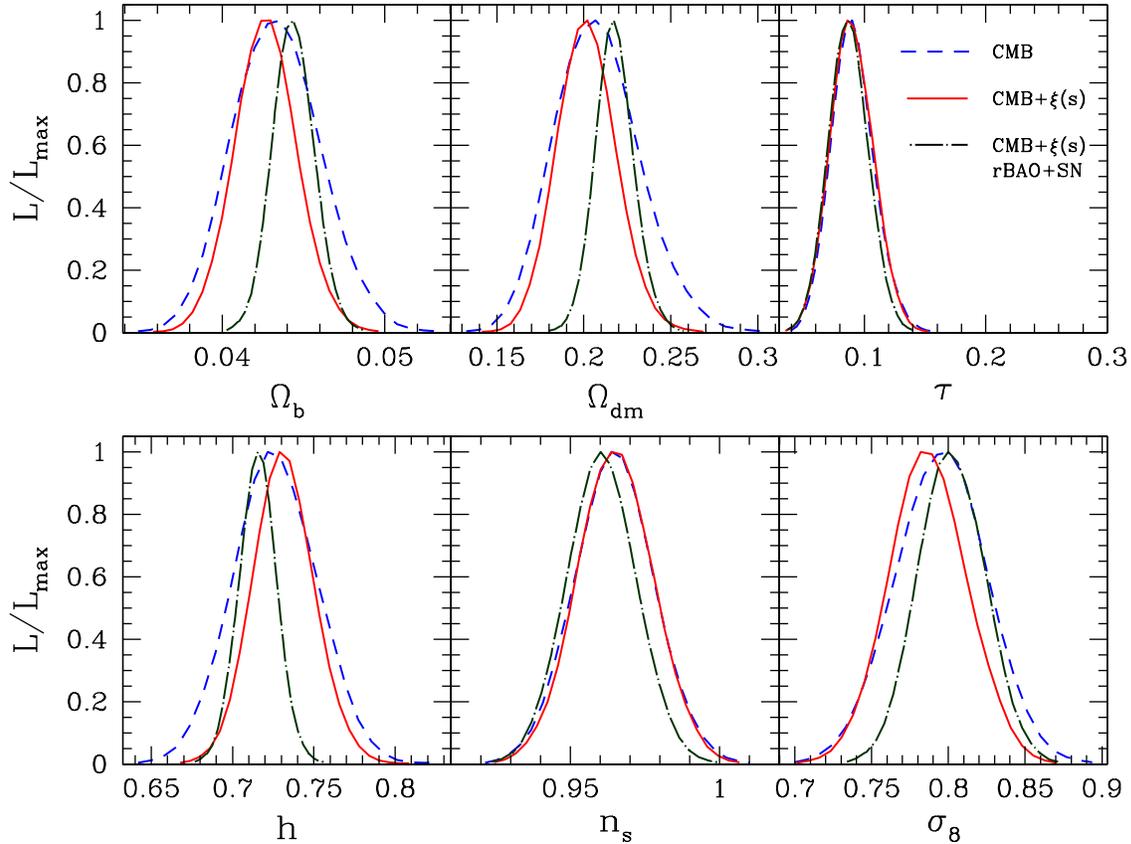}}
\caption{
The marginalized, one-dimensional posterior likelihood in the 
$\Lambda$CDM parameter space (Eq.~\ref{eq:param6}) obtained 
from CMB information only (dashed lines), CMB plus the shape 
of $\xi(s)$ (solid line) and the full constraints including also rBAO 
and SN data (dot-dashed lines). 
}
\label{fig:lcdm}
\end{figure*}

In Section~\ref{sec:results} we also present constraints obtained from the combination of the two large
scale structure datasets we use in our analysis, the LRG $\xi(s)$ and the position
of the radial acoustic peak, without including any CMB information. The fact that the correction for
the fiducial cosmology does not significantly change the obtained results when the LRG $\xi(s)$ is used in combination with the 
CMB, is not due to the later keeping the constraints sufficiently close to the fiducial cosmology. As we will
see later (see Section~\ref{ssec:wde}) the constraints in this parameter space from CMB data alone present a strong
degeneracy between $\Omega_{\rm m}$ and $w_{\rm DE}$, allowing for models that differ substantially from our fiducial
cosmology of $\Omega_{\rm m}=0.25$ and $w_{\rm DE}=-1$. Nonetheless, these models are ruled out by the $\xi(s)$ data
itself, (which on large scales is insensitive to this correction), since the position of the acoustic peak shows 
a strong variation in these models. The only region of the parameter space allowed by the data is that
where the effect of the correction of Eq.~(\ref{eq:cor_fact}) is small (although not completely
 negligible). This is the reason why the obtained constrains are not extremely sensitive to this correction. 
This also implies that this correction should not be too important even in the case 
in which the LRG $\xi(s)$ is combined with the radial acoustic peak whitout including any CMB data.
We have tested this explicitly and found that the resulting mean values of the cosmological parameters obtained
for this dataset combination when this correction is not applied show no variation with respect to the ones obtained when
it is used. The only effect of ignoring this correction is a small artificial increase of the allowed regions of 
$\Omega_{\rm m}$ and $w_{\rm DE}$ of the same order as in the previous case. Setting $A_{\rm MC}=0$
for this combination of datasets also produces a small shift towards smaller values of $\Omega_{\rm m}$, which in this 
case is much smaller than the allowed region for this parameter.
These tests show that our constraints are robust with respect to the details of our analysis technique.

\section{Constraints on cosmological parameters}
\label{sec:results}

In this section, we carry out a systematic study of the constraints placed on the values 
of the cosmological parameters for the different parameter spaces defined in 
Section~\ref{ssec:mechanics}. In Section~\ref{ssec:lcdm}, we present the results for 
the simple $\Lambda$CDM cosmological model with six free parameters. In Section~\ref{ssec:wde}, 
we consider an extension to this parameter set, allowing the dark energy equation of state 
$w_{\rm DE}$ to float (but without any redshift dependence). Section~\ref{ssec:wa} gives  
the constraints on models where the time variation of $w_{\rm DE}$ is parametrized according 
to Eq.~(\ref{eq:wa}). In Section~\ref{ssec:omk} we discuss our constraints on non-flat models, 
analysing the parameter space of Eq.~(\ref{eq:paramomk}). Finally, Section~\ref{ssec:wok} 
shows the constraints on the full parameter space of Eq.~(\ref{eq:param}), allowing both for 
non-flat models and more general dark energy models. Tables~\ref{tab:lcdm}-\ref{tab:wok} 
compare the constraints obtained in these parameter spaces using different combinations of 
the datasets described in Section~\ref{sec:data}. 

\subsection{The basic $\Lambda$CDM model}
\label{ssec:lcdm}

Due to the successful reproduction of a wide variety of observations, 
the $\Lambda$CDM model has emerged over the past decade as the new standard 
cosmological model. The recent results from five years of observations by 
the WMAP satellite have helped to reinforce this conclusion. The latest WMAP 
data give a much better estimation of the third acoustic peak in the CMB 
temperature power spectrum, as well as the low-$\ell$ polarization signal. 
Thanks to these improvements, the WMAP data alone has been able to provide 
much tighter constraints on this basic cosmological model than was possible 
when using earlier releases. 

Table~\ref{tab:lcdm} summarizes the constraints on the parameters of this 
simple model for different combinations of datasets. Fig.~\ref{fig:lcdm} 
shows the marginalized likelihoods for this parameter set obtained using 
CMB data alone (dashed lines), CMB plus the LRG $\xi(s)$ (solid lines) and 
the full combination of the datasets described in Section~\ref{sec:data} 
i.e. CMB+ LRG $\xi(s)$+rBAO+SN (dot-dashed line). Several 
parameters, such as $\omega_{\rm b}$, $\omega_{\rm c}$ and $\tau$, are tightly 
constrained by the CMB data alone, and show almost no variation when other 
datasets are included in the analysis. On the other hand, the constraints 
on the parameters of the energy budget do show a marked improvement on adding 
further datasets. 

\begin{table*} 
\centering
 \begin{minipage}{172mm}
  \caption{
    The marginalized 68\% interval constraints on cosmological parameters allowing for variations in the (redshift
    independent) dark energy equation of state (i.e. the parameter set defined by Eq.(\ref{eq:paramwde})),
    obtained using different combinations of the datasets described in Section~\ref{sec:data}, as stated
    in the column headings.}
    \begin{tabular}{@{}lccccccc@{}}
    \hline
& \multirow{2}{*}{CMB}  & \multirow{2}{*}{CMB + $\xi(s)$} & \multirow{2}{*}{CMB + rBAO}& \multirow{2}{*}{CMB + SN} & CMB + $\xi(s)$ & CMB + $\xi(s)$ & \multirow{2}{*}{$\xi(s)$+ rBAO}\\
&     &                &                                  &                            &   + rBAO        &  + rBAO + SN &  \\  
\hline
$w_{\rm DE}$        & $-0.73_{-0.30}^{+0.30}$       & $-0.988_{-0.088}^{+0.088}$     & $-0.92_{-0.15}^{+ 0.15}$       & $-0.950_{-0.055}^{+0.054}$    & $-0.999_{-0.091}^{+0.090}$ &  $-0.969_{-0.052}^{+0.052}$& $-1.05_{-0.15}^{+0.16}$\\[1.5mm] 
$100\Theta$         &$ 10413_{-0.0024}^{+0.0023}$   & $ 1.0415_{-0.0023}^{+0.0023}$  & $ 1.0414_{-0.0022}^{+ 0.0022}$ & $1.0413_{-0.0022}^{+0.0023}$  & $ 1.0413_{-0.0023}^{+0.0023}$ & $1.0414_{-0.0022}^{+0.0022}$& $0.994_{-0.075}^{+ 0.071}$\\[1.5mm] 
$\omega_{\rm dm}$   & $0.1105_{-0.0050}^{+0.0052}$  &  $ 0.1068_{-0.0048}^{+ 0.0049}$& $0.1095_{ -0.0045}^{+ 0.0045}$ & $ 0.10933_{ -0.0050}^{+ 0.0049}$ &  $0.1092_{-0.0042}^{+0.0042}$ &  $ 0.1088_{-0.0041}^{+0.0040}$& $0.099_{-0.045}^{+ 0.049}$\\[1.5mm] 
100\,$\omega_{\rm b}$ & $2.266_{-0.052}^{+0.053}$   &$ 2.280_{-0.052}^{+0.052}$     & $ 2.277_{-0.051}^{+0.051}$      & $ 2.272_{-0.052}^{+0.052}$    & $2.275_{-0.049}^{+0.00051}$ & $ 2.275_{-0.050}^{+0.051}$& $3.0_{-1.7}^{+1.8}$\\[1.5mm] 
$\tau $             & $0.089_{-0.017}^{+0.017}$     &  $0.090_{-0.017}^{+0.017}$    & $ 0.090_{ -0.017}^{+0.017}$     & $ 0.089_{ -0.017}^{+ 0.017}$  & $0.088_{-0.017}^{+0.017}$   & $ 0.088_{ -0.016}^{+ 0.016}$& -\\[1.5mm] 
$n_{\rm s}$         & $0.959_{-0.014}^{+0.014}$      &$ 0.965_{-0.013}^{+0.012}$     & $ 0.963_{-0.012}^{+0.012}$     & $ 0.961_{-0.013}^{+0.013}$    & $0.963_{-0.012}^{+0.011}$   & $ 0.963_{-0.012}^{+0.012}$&$ 1.03_{-0.32}^{+0.37}$\\[1.5mm] 
$\ln(10^{10}A_{\rm s})$ &$ 3.067_{ -0.037}^{+0.038}$ &  $ 3.057_{-0.037}^{+0.037}$   & $ 3.066_{-0.037}^{+0.037}$     & $ 3.064_{-0.037}^{+0.038}$    & $ 3.062_{-0.036}^{+0.035}$  & $ 3.060_{-0.036}^{+0.036}$& -\\[1.5mm] 
$\Omega_{\rm DE}$       &  $ 0.63_{-0.12}^{+0.12}$   &   $ 0.754_{-0.021}^{+0.020}$  & $ 0.723_{-0.034}^{+0.034}$     & $ 0.733_{-0.019}^{+0.018}$    & $ 0.746_{-0.017}^{+0.017}$  & $ 0.739_{-0.013}^{+0.013}$& $0.778_{-0.045}^{+0.045}$\\[1.5mm]
$\Omega_{\rm m}$      &$ 0.36_{-0.12}^{+ 0.12}$      & $0.245_{-0.020}^{+ 0.021}$    & $ 0.277_{-0.034}^{+0.034}$     & $ 0.267_{-0.018}^{+0.019}$    & $ 0.254_{-0.017}^{+0.017}$  & $ 0.261_{-0.013}^{+0.013}$&$0.222_{-0.045}^{+0.045}$\\[1.5mm]
$\sigma_{8}$          &  $ 0.724_{-0.085}^{+0.087}$  & $0.778_{-0.045}^{+0.045}$   & $ 0.774_{ -0.060}^{+0.060}$      & $ 0.780_{-0.037}^{+0.038}$    & $ 0.793_{-0.045}^{+0.043}$  & $ 0.781_{-0.034}^{+0.035}$& \\[1.5mm]
$t_{0}/{\rm Gyr}$     &  $14.01_{-0.38}^{+0.40}$     & $13.65_{ -0.11}^{+ 0.11}$     & $13.74_{-0.14}^{+0.14}$        & $13.71_{-0.10}^{+0.10}$       & $13.67_{-0.10}^{+0.10}$  & $13.689_{-0.095}^{+ 0.095}$ &$14.3_{-3.0}^{+3.1}$\\[1.5mm] 
$z_{\rm re}$          &  $10.7_{-1.4}^{+1.4}$        &  $10.5_{ -1.3}^{+ 1.3}$       & $10.6_{-1.3}^{+1.4}$           & $10.5_{-1.3}^{+1.3}$          & $10.4_{-1.4}^{+1.4}$        & $10.4_{ -1.3}^{+ 1.3}$& -\\[1.5mm]
$h$                  &    $0.63_{-0.10}^{+0.10}$     & $0.729_{-0.029}^{+0.027}$   & $0.695_{-0.046}^{+ 0.046}$       & $0.704_{-0.017}^{+0.017}$     & $0.722_{-0.025}^{+0.024}$   & $0.711_{-0.013}^{+0.014}$& $0.74_{-0.15}^{+0.16}$\\
\hline
\end{tabular}
\label{tab:wde}
\end{minipage}
\end{table*}

The CMB data is particularly sensitive to the value of $\omega_{\rm m}\equiv\Omega_{\rm m}h^2$.
This leads to a degeneracy between $\Omega_{\rm m}$ and $h$ which can be broken by combining 
the CMB measurements with other datasets \citep{percival02,spergel03,sanchez06,spergel07}. 
However, the improved estimation of the third acoustic peak in the WMAP5 temperature power 
spectrum helps to alleviate the degeneracy with respect to earlier data releases. The ratio 
of the amplitudes of the first and third acoustic peaks in the temperature power spectrum 
is sensitive to the ratio $\Omega_{\rm m}/\Omega_{\rm r}$.
This results in an improvement in the constraints 
on $\Omega_{\rm m}$ from CMB data alone, thereby reducing the degeneracy 
between $\Omega_{\rm m}$ and $h$ and improving the constraints on these parameters.
From the CMB data alone we get 
$\Omega_{\rm m}=0.251\pm0.026$ and $h=0.726_{-0.024}^{+0.025}$. Including the shape of 
$\xi(s)$, these constraints change to $\Omega_{\rm m}= 0.244 \pm 0.018$ and $h=0.731\pm0.018$, 
in complete agreement with the results from the CMB and with previous determinations based 
on the combination of CMB and large scale structure data \citep{sanchez06,spergel07,dunkley2008,
komatsu08}. The data from the radial BAO and SN prefer slightly higher values of 
$\Omega_{\rm m}$ than the CMB data and the LRG $\xi(s)$, but are consistent within 1$\sigma$.
Combining the information from all these datasets we get our tightest constraints, with 
$\Omega_{\rm m}= 0.261 \pm 0.013$ and $h=0.716\pm0.012$.

With the exception of the optical depth $\tau$ and the amplitude of density fluctuations $A_{\rm s}$,
it is also possible to obtain constraints on the same set of cosmological parameters from the combination of
the LRG $\xi(s)$ with the rBAO data without including any CMB information. For this we have included the bias parameter
$b$ explicitly in the analysis, but marginalized the results over a wide prior with $0.5<b<20$. 
This prior has minimal impact on the obtained constraints but it allowed us to effectively test if a
prior in $b$ can have implications on the obtained constraints. The results with 
this combination of datasets are shown in the last column of Table~\ref{tab:lcdm}. In this case we get 
$\Omega_{\rm m}=0.218_{-0.040}^{+0.041}$ and $h=0.73_{-0.14}^{+0.15}$. Although these constraints are weaker than those
obtained using CMB data, their importance lies in the fact that they are determined purely on the basis of large scale
structure information.

On combining the WMAP data with the BAO measurements from \citet{percival07c} and the same 
SN UNION data, \citet{komatsu08} found $\Omega_{\rm m}= 0.279 \pm 0.015$. This value is only 
marginally consistent with our results for CMB plus $\xi(s)$. This might indicate systematic 
problems introduced by the approximate treatment of the BAO measurements in previous analyses. 
We shall return to this point in Section~\ref{sec:conclusions}.

From the analysis of a compilation of CMB measurements with the final power spectrum of 
the 2dFGRS, \citet{sanchez06} found evidence for a departure from the scale invariant 
primordial power spectrum of scalar fluctuations, with the value $n_{\rm s}=1$ formally 
excluded at the 95\% level. This deviation was subsequently confirmed with higher 
significance with the availability of the three-year WMAP data \citep{spergel07}. 
By the full combination of the datasets of Section~\ref{sec:data} we find a constraint 
on the scalar spectral index of $n_{\rm s}= 0.963_{-0.011}^{+0.011}$, with the 
Harrison-Zel'dovich spectrum 3.6$\sigma$ away from the mean of the distribution.

The conclusion from this section is that the $\Lambda$CDM model gives a consistent and 
adequate description of all the datasets that we have included in our analysis. The 
precision and consistency of the constraints on the basic parameters on this model 
constitute a reassuring validation of the cosmological paradigm. In the following 
sections we will concentrate on two possible deviations from this model that can be 
better constrained by the shape of $\xi(s)$, namely alternative dark energy models and 
non-flat cosmologies.

\begin{figure}
\centering
\centerline{\includegraphics[width=\columnwidth]{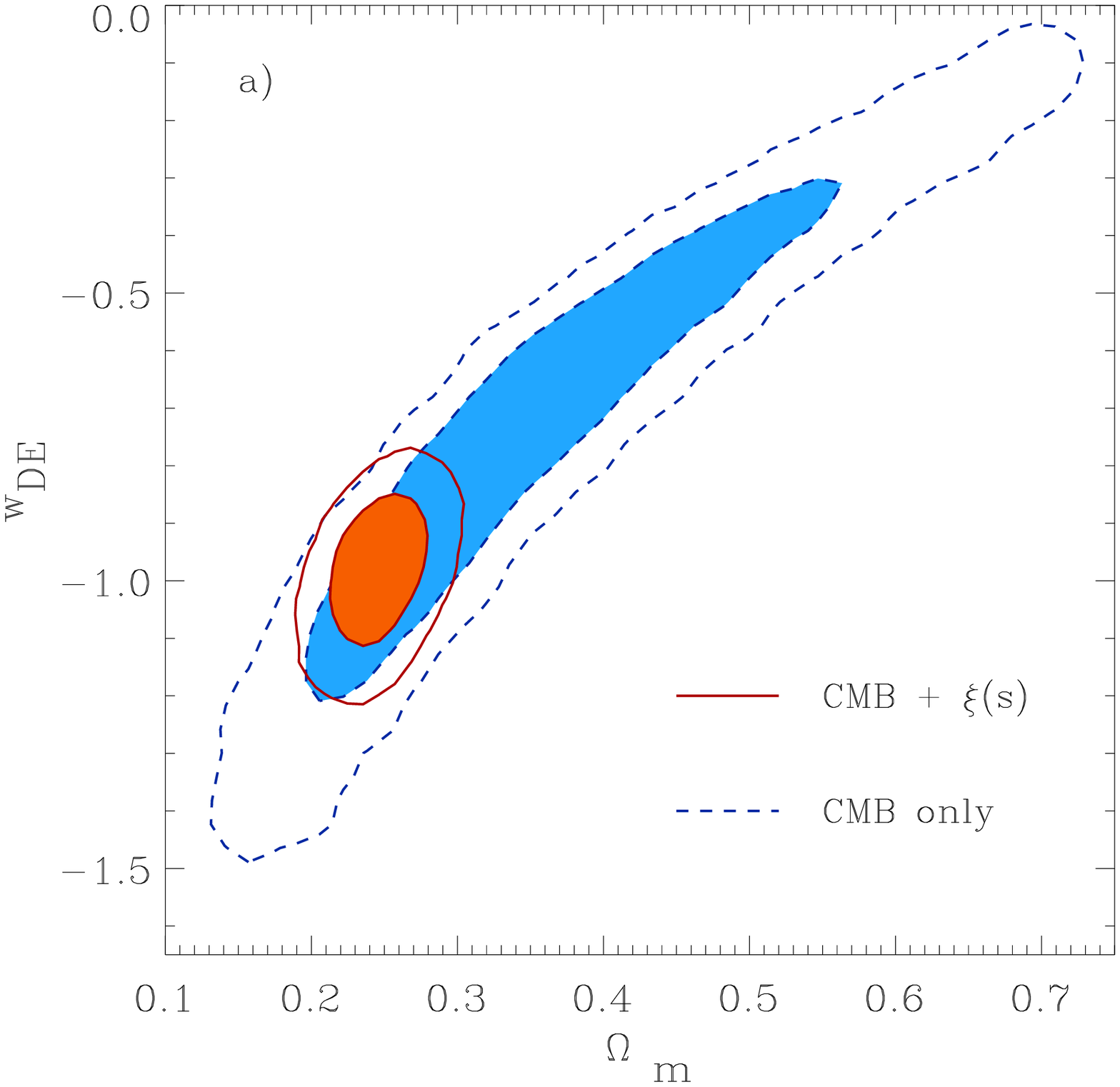}}
\centerline{\includegraphics[width=\columnwidth]{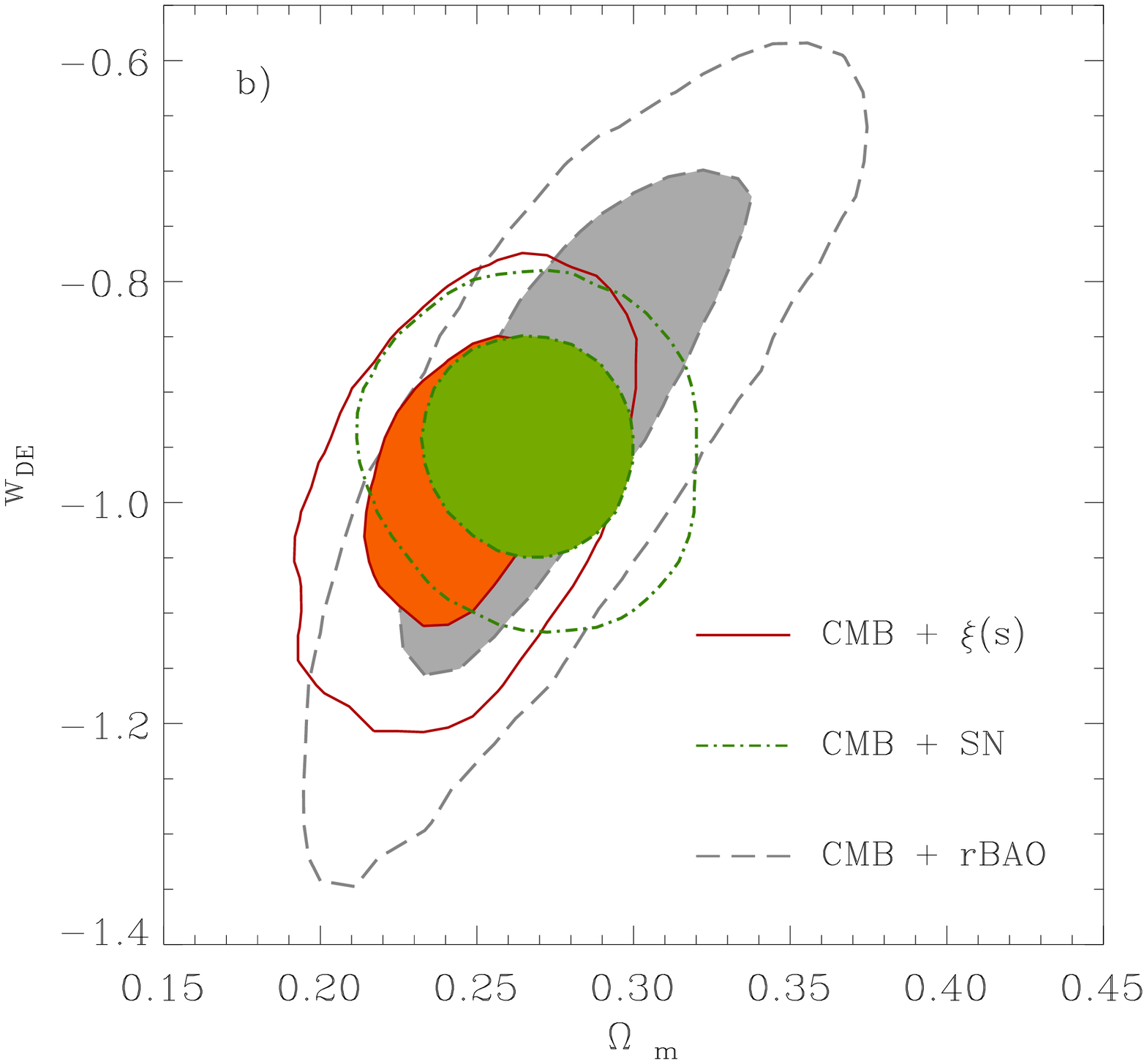}}
\caption{
Panel a): the marginalized posterior likelihood in the 
$\Omega_{\rm m}-w_{\rm DE}$ plane for the 
$\Lambda$CDM parameter set expanded by the 
addition of $w_{\rm DE}$ (Eq.~(\ref{eq:paramwde})).
The short-dashed lines show the 68 and 95 per
cent contours obtained using CMB information alone,
solid contours show CMB plus LRG $\xi(s)$ constraints. 
Panel b) Comparison of the marginalized posterior likelihood 
in the same parameter space obtained using CMB information plus LRG $\xi(s)$ 
(solid lines), CMB plus the radial BAO signal (long-dashed lines) and CMB+SN 
(dot-dashed lines), as indicated by the key. 
The filled contours correspond to the 68\% CL in each case.
}
\label{fig:bwde_om_wde}
\end{figure}

\subsection{The dark energy equation of state}
\label{ssec:wde}

When treated as standard candles, the apparent dimming of distant Type Ia supernovae 
surprisingly pointed towards an accelerating expansion of the Universe 
\citep{riess1998,perlmutter1999,riess2004}. This was the first piece 
of observational evidence in favour of the presence of a negative pressure component 
in the energy budget of the Universe. Independent support for this component, 
called dark energy, came from the combination of CMB measurements and large scale 
structure data \citep{efstathiou02,tegmark04}. Understanding the nature of dark 
energy has become one of the most important problems in physics today since it has strong
implications for our understanding of the fundamental physical laws of the Universe.
The simplest possibility is that the dark energy corresponds to the vacuum energy, 
in which case it behaves analogously to Einstein's cosmological constant with $w_{\rm DE}=-1$, 
but several alternative models have been proposed. One way to narrow down the 
wide range of possible models is to obtain constraints on the dark energy equation of 
state parameter $w_{\rm DE}$. In this section we extend the parameter space of the 
$\Lambda$CDM models to allow for variations in the (redshift-independent) value 
of $w_{\rm DE}$. In Section~\ref{ssec:wa} we drop this hypothesis to explore the 
possible redshift dependence of this parameter. 
Table~\ref{tab:wde} summarizes our 
constraints on this parameter set from different combinations of the datasets 
described in Section~\ref{sec:data}.

\begin{figure}
\centering
\centerline{\includegraphics[width=\columnwidth]{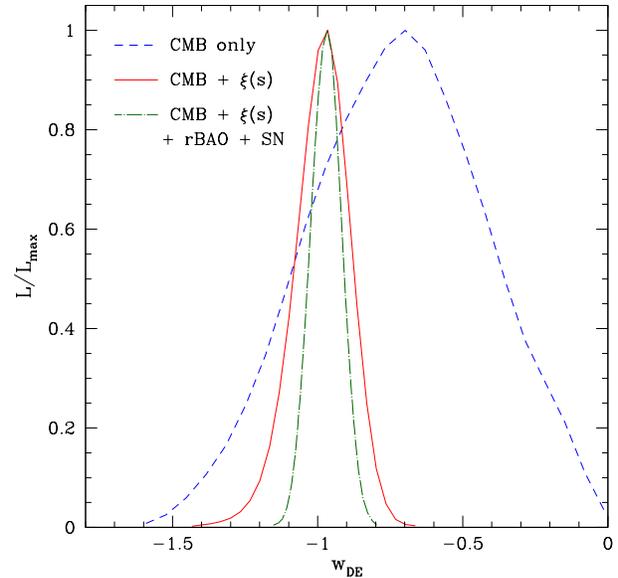}}
\caption{
The marginalized posterior likelihood for the dark energy equation of 
state parameter, $w_{\rm DE}$, in the case of the parameter set defined 
by Eq.~(\ref{eq:paramwde}). The dashed line shows the likelihood in the 
case of CMB data alone, the solid line shows the likelihood for CMB combined 
with the LRG $\xi(s)$ and the dot-dashed line shows the likelihood using 
all of the data sets described in Section~2.  
}
\label{fig:wde_1d}
\end{figure}

Fig.~\ref{fig:bwde_om_wde}a shows the two-dimensional marginalized constraints in 
the $\Omega_{\rm m}-w_{\rm DE}$ plane from CMB data alone (dashed lines) and CMB 
plus the LRG $\xi(s)$. There is a strong degeneracy between these parameters when 
only CMB data is included in the analysis which leads to poor one-dimensional 
marginalized constraints of $w_{\rm DE}=-0.73_{-0.30}^{+0.30}$ and 
$\Omega_{\rm m}=0.36_{-0.12}^{+0.12}$. Another view of this is given by 
Fig.~\ref{fig:wde_1d}, which shows the one-dimensional marginalized constraint on 
$w_{\rm DE}$. The CMB only case is again shown by the dashed line. 

The origin of this degeneracy is well understood. The position of the acoustic peaks 
in the CMB power spectrum depends on the size of the sound horizon at the decoupling 
epoch, $r_{\rm s}(z_{*})$, which is given by Eq.~(\ref{eq:soundh}).
The mapping of the physical scales of the acoustic peaks to angular scales on the 
sky depends on the comoving angular diameter distance, $D_{\rm A}(z_{*})$,
given by Eq.~(\ref{eq:dang}). 
Therefore the peak pattern in the CMB provides tight 
constraints on the ``acoustic scale'' given by 
\citep{bond1997,efstathiou1999,page2003,komatsu08} 
\begin{equation}
 \ell_{\rm A}=\frac{\pi D_{\rm A}(z_{*})}{r_{\rm s}(z_{*})}.
\label{eq:la}
\end{equation}
While $w_{\rm DE}$ is relevant for the calculation of $D_{\rm A}(z_{*})$, it has 
minimum impact on $r_{\rm s}(z_{*})$, since the dark energy is dynamically 
negligible at decoupling. For this reason, for fixed values of $\omega_{\rm b}$ and 
$\omega_{\rm dm}$, and given a value of $\Omega_{\rm m}$ (or $h$), it is always 
possible to find a value of $w_{\rm DE}$ such that the value of $\ell_{\rm A}$ remains 
constant. This gives rise to the degeneracy between these parameters seen in 
Fig.~\ref{fig:bwde_om_wde}a.

As shown in Fig.~\ref{fig:bwde_om_wde}a, the inclusion of the LRG correlation 
function breaks the degeneracy between $\Omega_{\rm m}$ and $w_{\rm DE}$ present in 
the CMB data. This is done in two ways; first the shape of $\xi(r)$ on intermediate 
scales tightens the constraints on $\Omega_{\rm m}$, which helps to break the degeneracy 
between this parameter and $w_{\rm DE}$. 
Second, through the position of the acoustic
peak it provides an independent estimation of the ratio $r_{\rm s}(z_{\rm d})/D_{\rm V}(z_{\rm m})$,
where $z_{\rm d}$ is the redshift of the drag epoch and $z_{\rm m}=0.35$ is the mean redshift of
the survey.

\begin{figure}
\centering
\centerline{\includegraphics[width=\columnwidth]{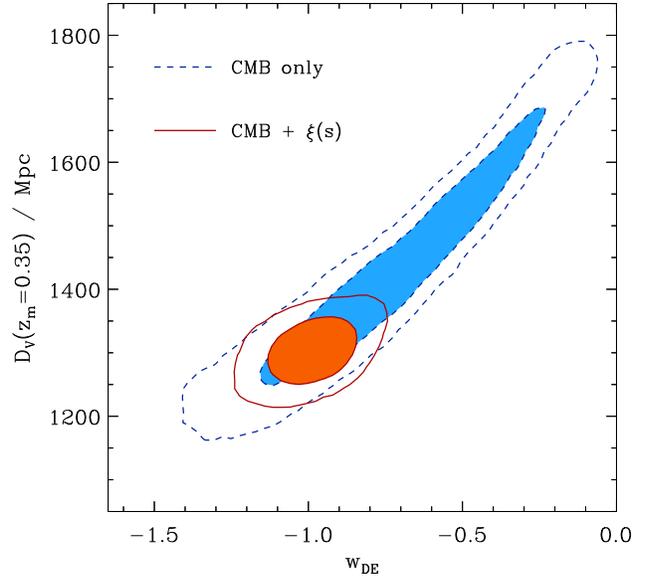}}
\caption{
The marginalized posterior likelihood in the $D_{\rm V}-w_{\rm DE}$ plane for the $\Lambda$CDM 
parameter set extended to allow $w_{\rm DE}$ to vary (Eq.~(\ref{eq:paramwde})). The dashed 
lines show the 68 and 95 per cent contours obtained using CMB information alone. The solid 
contours correspond to the results obtained from the combination of CMB data plus the shape 
of the LRG $\xi(s)$. The filled contours correspond 
to the 68\% CL in each case. 
}
\label{fig:bwde_wde_dv}
\end{figure}

When this information is combined with the constraint on $r_{\rm s}(z_{\rm d})$ provided by the 
CMB data, this provides an extra distance measurement, $D_{\rm V}(z_{\rm m}=0.35)$, which 
breaks the degeneracy in the CMB data. This can be seen more clearly in 
Fig.~\ref{fig:bwde_wde_dv}, which shows the two dimensional constraints in the 
plane $w_{\rm DE}-D_{\rm V}(z_{\rm m}=0.35)$. Varying $w_{\rm DE}$ to keep a constant 
value of $\ell_{\rm A}$ produces varying values of $D_{\rm V}(z_{\rm m}=0.35)$. 
The extra information from the shape of $\xi(s)$ fixes the value of $D_{\rm V}$, 
tightening the constraints on the dark energy equation of state. In this case we get 
$\Omega_{\rm m}=0.245_{-0.020}^{+ 0.021}$ and $w_{\rm DE}=-0.988_{-0.088}^{+0.088}$, 
in complete agreement with the cosmological constant. Again, Fig.~\ref{fig:wde_1d} 
shows the dramatic reduction in the width of the likelihood distribution for $w_{\rm DE}$ 
on combining the CMB data with the measurement of the LRG correlation function. 

Using WMAP data combined with the BAO measurement from \citet{percival07c}, 
\citet{komatsu08} found $w_{\rm DE}=-1.15_{-0.22}^{+0.21}$ (68\% CL). If we exclude the small 
scale CMB experiments and consider only WMAP measurements plus the LRG $\xi(s)$, 
we get $w_{\rm DE}=-0.996_{-0.095}^{+0.097}$, which corresponds to a reduction of 
almost a factor two in the allowed region for the equation of state parameter. This 
highlights the importance of the information contained in the shape of the correlation 
function. The method of \citet{percival07c} sacrifices the long-wavelength shape of 
$P(k)$, which is affected by scale-dependent effects, in order to obtain a purely 
geometrical test from the BAO oscillations.

It is also possible to obtain constraints in this parameter space from the combination
of the shape of the LRG $\xi(s)$ with the rBAO data without including any CMB information. 
In this case we get we get $w_{\rm DE}=-1.05_{-0.15}^{+0.16}$. This means that present 
day observations allow to obtain a competitive constraint on $w_{\rm DE}$ 
purely on the basis of large scale structure information, independently 
of CMB or SN observations.

The values of $w_{\rm DE}$ listed in Table~\ref{tab:wde} as being obtained by combining 
the CMB data with rBAO and SN information are completely consistent (see 
Section~\ref{sec:conclusions} and Fig.~\ref{fig:bwde_om_wde}b). Once again this 
shows the consistency between these datasets. Our results including rBAO data show an excellent 
agreement with those of \citet{gaztanaga08c}. When including systematic effects 
in the SN data as advocated by Kowalski et~al. (2008), the CMB plus SN result changes to 
$w_{\rm DE}=-0.900_{-0.078}^{+0.078}$, a 50\% increase in the allowed region and 
a shift in the mean value of $w_{\rm DE}$ of about one $\sigma$. This means that when the 
SN systematic errors are included in the analysis, the precision of the constraint on $w_{\rm DE}$
obtained from CMB plus SN is comparable to the one derived from CMB plus the shape of the LRG $\xi(s)$.
This highlights the importance of using the full shape of $\xi(s)$ as a cosmological probe.
This also shows the importance of a precise determination of the effects of systematic errors in
the SN data.

\subsection{The time evolution of $w_{\rm DE}$}
\label{ssec:wa}

In the previous section we analysed the possibility of extending the $\Lambda$CDM model 
with alternative dark energy models with a redshift-independent equation of state. 
From a theoretical perspective, if $w_{\rm DE}\neq-1$, there is no real reason why 
it should be constant. In this section, we analyse the constraints on the redshift 
dependence of this parameter,  using the popular linear parametrization of 
Eq.~(\ref{eq:wa}) \citep{chevalier2001,linder03}.

\begin{figure}
\centering
\centerline{\includegraphics[width=\columnwidth]{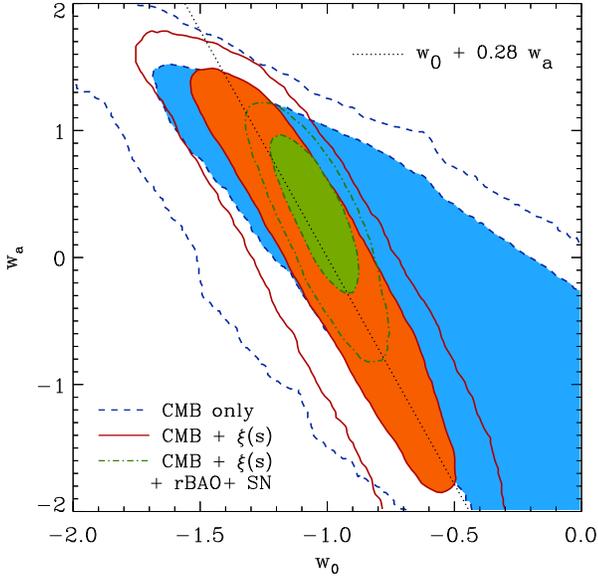}}
\caption{
The marginalized posterior likelihood in the $w_{0}-w_a$ plane for the $\Lambda$CDM 
parameter set extended with a redshift dependent dark energy equation of state 
parametrized according to Eq.~(\ref{eq:wa}). The dashed lines show the 68 and 95 
per cent contours obtained using CMB information alone. The solid contours 
correspond to the results obtained from the combination of CMB data plus the 
shape of the LRG $\xi(s)$. The dot-dashed contours show the constraints on using 
all datasets. The filled contours correspond 
to the 68\% CL in each case. The dotted straight line marks the degeneracy between 
$w_{0}$ and $w_{a}$. 
}
\label{fig:bwa_w0_wa}
\end{figure} 

\begin{figure}
\centering
\centerline{\includegraphics[width=\columnwidth]{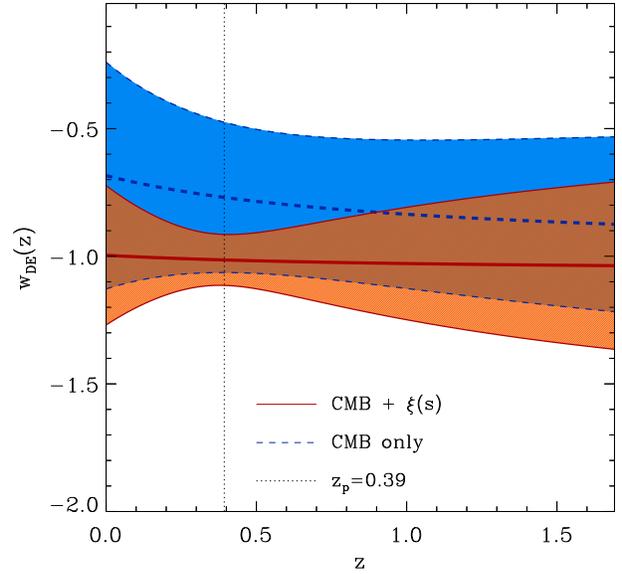}}
\caption{
The marginalized posterior likelihood in $w_{\rm DE}$ as a function of $z$
for the $\Lambda$CDM parameter set extended with a redshift dependent dark 
energy equation of state parametrized as in Eq.~(\ref{eq:wa}).
The dashed lines show the mean  value (thick line) and
the corresponding 68 per cent constraints (filled blue region between thin lines) obtained
using CMB information alone. The solid lines (and filled orange region) show the
corresponding results from the combination of CMB plus the shape of
the LRG $\xi(s)$. The dotted vertical line corresponds to the pivot redshift
$z_{\rm p}=0.39$.
}
\label{fig:bwa_z}
\end{figure}

This parametrization may lead to models in which the dark energy density has 
a dynamical impact at the epoch of Big Bang Nucleosynthesis (BBN), unlike 
the case with a cosmological constant. Such a scenario would affect the present 
day abundances of light elements, thereby violating the constraints on the 
possible variation in the Hubble 
parameter at the epoch of BBN as expressed through the 
ratio \citep{steigman2007}:
\begin{equation}
 S\equiv\frac{H'(a_{\rm BBN})}{H(a_{\rm BBN})}=0.942\pm0.030,
\label{eq:sprior}
\end{equation}
where $H(a_{\rm BBN})$ is the Hubble parameter in a standard $\Lambda$CDM model, 
in which case 
dark energy is completely negligible at the expansion factor corresponding 
to the epoch of Big Bang nucleosynthesis, $a_{\rm BBN}$, and $H'(a_{\rm BBN})$ 
is the Hubble parameter at the same epoch in an alternative model in which 
dark energy does play a role at this early epoch.
Here we follow \citet{wright2007} and \citet{komatsu08} and impose a Gaussian 
prior given by Eq.~(\ref{eq:sprior}) on $S$ which can be written as
\begin{equation}
 S=\sqrt{ 
1 +
\frac{\Omega_{\rm DE}\,a_{\rm BBN}^{-3(1+w_{\rm eff}(z_{\rm BBN}))}}
{\Omega_{\rm m}\,a_{\rm BBN}^{-3}+
\Omega_{\rm r}\,a_{\rm BBN}^{-4}+\Omega_{k}\,a_{\rm BBN}^{-2}}
},
\label{eq:priorbbn}
\end{equation}
where $a_{\rm BBN}=10^{-9}$ and $w_{\rm eff}$ is the effective dark energy 
equation of state, defined as
\begin{equation}
 w_{\rm eff}(a)\equiv \frac{1}{\ln(a)}\int_0^{\ln(a)}w_{\rm DE}(a')\,{\rm d}\ln a'.
\label{eq:weff}
\end{equation}
With the parametrization of Eq.~(\ref{eq:wa}) this becomes
\begin{equation}
 w_{\rm eff}(a)=w_0+w_a\left(1+\frac{1-a}{\ln(a)}\right).
\label{eq:weff2}
\end{equation}
Table~\ref{tab:wa} summarizes the constraints obtained in this parameter space 
using different combinations of datasets.

\begin{table*} 
\centering
 \begin{minipage}{172mm}
  \caption{
The marginalized 68\% interval constraints on cosmological parameters allowing for an evolving
dark energy equation of state (i.e. the parameter set defined by Eq.(\ref{eq:paramwa})), 
obtained using different combinations of the datasets described 
in Section~\ref{sec:data}, as stated in the column headings. 
}
    \begin{tabular}{@{}lccccccc@{}}
    \hline
& \multirow{2}{*}{CMB}  & \multirow{2}{*}{CMB + $\xi(s)$} & \multirow{2}{*}{CMB + rBAO}& \multirow{2}{*}{CMB + SN} & CMB + $\xi(s)$ & CMB + $\xi(s)$ & \multirow{2}{*}{$\xi(s)$+ rBAO}\\
&     &                &                                  &                            &   + rBAO        &  + rBAO + SN &  \\  
\hline
$w_{\rm 0}$          &  $-0.68_{-0.48}^{+0.48}$   &  $-1.00_{-0.29}^{+0.29}$ &   $-0.85_{ -0.49}^{+0.49}$&  $-1.07_{ -0.15}^{+ 0.16}$  & 
                        $-1.10_{ -0.19}^{+ 0.19}$     & $-1.03_{ -0.10}^{+ 0.10}$ &$-1.06_{-0.18}^{+0.19}$\\[1.5mm] 
$w_{a}$              & $-0.30_{-0.96}^{+0.96}$     &$-0.065_{-1.02}^{+0.99}$  &  $-0.2_{ -1.0}^{+1.0}$ &   $ 0.45_{-0.61}^{+0.59}$    & 
                        $ 0.31_{ -0.59}^{+ 0.60}$   & $ 0.30_{-0.41}^{+0.41}$ &$-0.0_{-1.1}^{+0.9}$\\[1.5mm] 
$100\Theta$             &  $ 1.0413_{ -0.0023}^{+ 0.0023}$  & $ 1.0416_{-0.0023}^{+0.0023}$ & $ 1.0415_{ -0.0022}^{+0.0022}$ &
                        $ 1.0413_{ -0.0022}^{+0.0022}$    & $ 1.0414_{ -0.0022}^{+ 0.0022}$ &  $ 1.0416_{-0.0022}^{+ 0.0022}$ &$ 0.997_{-0.078}^{+0.073}$\\[1.5mm] 
$\omega_{\rm dm}$    &  $ 0.1098_{ -0.0054}^{+ 0.0054}$ & $ 0.1076_{-0.0050}^{+ 0.0051}$ &  $ 0.1098_{-0.0044}^{+0.0043}$ &   
                        $ 0.1107_{ -0.0052}^{+ 0.0053}$    & $ 0.1087_{-0.0047}^{+0.0047}$   & $ 0.1077_{-0.0045}^{+0.0045}$&$ 0.096_{-0.045}^{+0.048}$\\[1.5mm] 
$100\,\omega_{\rm b}$     & $ 2.270_{-0.052}^{+0.051}$   & $ 2.284_{-0.052}^{+0.053}$ &$ 2.279_{-0.051}^{+0.051}$ &
                        $ 2.271_{-0.053}^{+ 0.053}$   & $ 2.278_{-0.051}^{+ 0.051}$ & $ 2.280_{-0.051}^{+0.050}$& $2.6_{-1.5}^{+1.6}$ \\[1.5mm] 
$\tau $              & $ 0.090_{ -0.017}^{+ 0.017}$    & $ 0.090_{ -0.017}^{+ 0.017}$   &  $ 0.090_{-0.017}^{+0.017}$  & 
                        $ 0.089_{ -0.017}^{+ 0.017}$    & $ 0.089_{-0.017}^{+ 0.017}$   &$ 0.091_{ -0.017}^{+ 0.016}$& - \\[1.5mm]
$n_{\rm s}$            &  $ 0.961_{ -0.013}^{+ 0.013}$ & $ 0.966_{-0.013}^{+0.013}$  & $ 0.964_{ -0.012}^{+ 0.012}$ &  
                         $ 0.960_{ -0.013}^{+ 0.013}$    & $ 0.963_{ -0.012}^{+ 0.012}$   & $ 0.963_{-0.012}^{+ 0.012}$& $ 1.04_{-0.32}^{+0.32}$\\[1.5mm] 
$\ln(10^{10}A_{\rm s})$ & $ 3.065_{-0.038}^{+ 0.038}$    &$ 3.059_{-0.037}^{+0.038}$    & $ 3.068_{-0.038}^{+0.038}$    & 
                         $ 3.068_{-0.037}^{+ 0.037}$    & $ 3.061_{ -0.037}^{+ 0.037}$   & $ 3.062_{-0.037}^{+0.036}$& -\\[1.5mm] 
$\Omega_{\rm DE}$      & $ 0.66_{-0.11}^{+ 0.11}$   & $ 0.755_{-0.023}^{+0.023}$ & $ 0.715_{-0.062}^{+0.062}$   &   
                        $ 0.718_{ -0.024}^{+ 0.024}$       &   $ 0.749_{ -0.018}^{+ 0.018}$   & $ 0.738_{-0.013}^{+ 0.013}$& $0.780_{-0.047}^{+0.045}$\\[1.5mm]
$\Omega_{\rm m}$       & $ 0.34_{-0.11}^{+ 0.11}$   &  $ 0.245_{ -0.023}^{+0.023}$ & $ 0.285_{-0.062}^{+0.062}$    &   
                        $ 0.282_{ -0.024}^{+ 0.024}$      &  $ 0.251_{ -0.018}^{+ 0.018}$     &$ 0.262_{ -0.013}^{+ 0.013}$& $ 0.220_{-0.045}^{+0.047}$ \\[1.5mm]
$\sigma_{8}$           & $ 0.733_{-0.082}^{+0.084}$   & $ 0.786_{-0.045}^{+0.045}$ & $ 0.773_{-0.070}^{+0.071}$  &  
                        $ 0.7790_{ -0.041}^{+ 0.041}$    &  $ 0.791_{-0.048}^{+0.047}$    & $ 0.769_{ -0.040}^{+ 0.040}$&-\\[1.5mm]
$t_{0}/{\rm Gyr}$      & $13.89_{-0.31}^{+0.32}$  & $13.65_{ -0.18}^{+ 0.19}$    &  $13.72_{-0.14}^{+0.14}$  &   
                          $13.83_{ -0.17}^{+0.18}$      &  $13.72_{-0.13}^{+0.13}$    & $13.76_{ -0.12}^{+ 0.12}$& $14.6_{-3.0}^{+3.3}$\\[1.5mm] 
$z_{\rm re}$           & $10.6_{-1.4}^{+1.4}$   & $10.5_{-1.3}^{+1.4}$   &  $10.5_{-1.4}^{+1.4}$   &  
                        $10.6_{ -1.4}^{+ 1.4}$        & $10.5_{ -1.4}^{+ 1.4}$ &  $10.6_{ -1.3}^{+ 1.3}$& -\\[1.5mm]
$h$                  & $0.65_{ -0.10}^{+0.10}$  &  $0.732_{-0.031}^{+0.028}$ &  $0.692_{-0.075}^{+0.078}$  & 
                       $0.689_{-0.023}^{+ 0.023}$     & $0.726_{-0.027}^{+0.025}$   & $0.706_{-0.015}^{+0.015}$ & $73_{-0.14}^{+0.16}$\\
\hline
\end{tabular}
\label{tab:wa}
\end{minipage}
\end{table*}

The dashed lines in Fig.~\ref{fig:bwa_w0_wa} show the two dimensional marginalized 
constraints in the $w_0-w_a$ plane from CMB data alone. As discussed in the previous 
section, the CMB data follows a degeneracy of constant $\ell_{\rm A}$. Allowing the 
dark energy to evolve with redshift, this degeneracy gains an extra degree of freedom 
leading to poor constraints on $w_0$ and $w_a$. 

These constraints can be transformed into constraints on the value of $w_{\rm DE}(a)$ which will be given by
\begin{equation}
\langle \delta w_{\rm DE}(a)^2 \rangle = \langle(\delta w_{\rm 0}+(1-a)\delta w_{\rm a})^2 \rangle.
\label{eq:delta_wa}
\end{equation}

The dashed lines in Fig.~\ref{fig:bwa_z} show the mean value of $w_{\rm DE}(z)$ as a function of $z$ (thick line), as well as the 68
per cent confidence limits (thin lines) obtained from CMB data alone.
These constraints can accommodate large variations of $w_{\rm DE}(z)$. The solid lines show the correspondent results when the LRG
correlation function is included in the analysis. Although considerable deviations from the simple cosmological constant case are
still allowed, these are much strongly constrained by the data.

   The pivot scale factor $a_{\rm p}$ is defined as the point where Eq.~(\ref{eq:delta_wa}) is minimized
\citep{huterer01,hu04,albretch06}. That is
\begin{equation}
 a_{\rm p}=1+\frac{\langle \delta w_{\rm 0}\delta w_{\rm a} \rangle}{\langle\delta w_{\rm a}^2 \rangle}.
\label{eq:pivot}
\end{equation}
For the combination of CMB plus the shape of $\xi(s)$ the corresponding pivot redshift is given by $z_{\rm p}=0.39$,
which is shown by the dotted line in Fig.~\ref{fig:bwa_z}. At this redshift we get our tightest constraint on the
dark energy equation of state, with $w_{\rm DE}(z_{\rm p}=0.39)=-1.01\pm 0.10$, entirely consistent with a cosmological constant.
In this case also, adding the information from the shape of $\xi(s)$ allows to obtain a tight constraint on 
$\Omega_{\rm m}=0.245_{-0.022}^{+0.023}$.

The solid lines in Fig.~\ref{fig:bwa_w0_wa} 
show the two dimensional marginalized constraints in the $w_0-w_a$ plane for CMB plus the 
shape of $\xi(s)$. The constraint on $w_{\rm DE}(z_{\rm p}=0.39)$ corresponds to a degeneracy 
between these parameters approximately given by $w_{\rm 0}+0.28w_{\rm a}=-1$, which is shown by 
the dotted line in Fig.~\ref{fig:bwa_w0_wa}.

Adding information from rBAO or SN gives completely consistent results, which 
are shown by the dot-dashed lines in Fig.~\ref{fig:bwa_w0_wa}. The combination 
of all the datasets gives the tightest constraints, with 
$w_0=-1.03\pm0.10$, $w_a=0.30_{-0.41}^{+0.41}$
and $\Omega_{\rm m}=0.262\pm0.013$.
These values correspond to a lower pivot redshift $z_{\rm p}=0.28$ for which we get
$w_{\rm DE}(z_{\rm p}=0.28)=-0.969\pm0.049$.
Even allowing for dynamic dark energy models current observations can constrain 
the present value of $w_{\rm DE}$ at the 10\% level.

Combining the shape of the LRG correlation function with the rBAO measurements we obtain the constraints
$w_0=-1.06_{-0.18}^{+0.19}$ and $w_a=0.0_{-1.1}^{+ 0.9}$. These constraints give a much lower pivot redshift,
with $z_{\rm p}=0.11$, for which we find $w_{\rm DE}(z_{\rm p}=0.11)=-1.06\pm0.19$. This shows that current
large scale structure data allows for the present value of a redshift-dependent dark energy equation of state
to be determined at the 20$\%$ level.

\subsection{Non-flat models}
\label{ssec:omk}

\begin{figure}
\centering
\centerline{\includegraphics[width=\columnwidth]{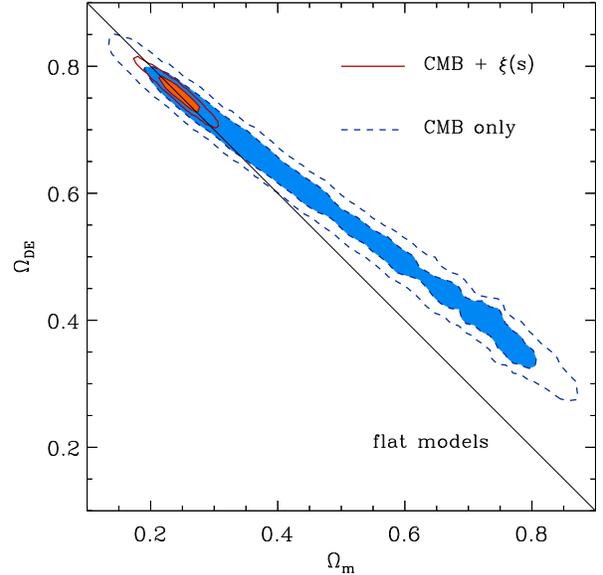}}
\caption{
The marginalized posterior likelihood in the $\Omega_{\rm m}-\Omega_{\rm DE}$ 
plane for the $\Lambda$CDM parameter set extended with $\Omega_k$ 
(Eq.~(\ref{eq:paramomk})). The dashed lines show the 68 and 95 per cent contours
obtained using CMB information alone. The solid contours correspond to the results 
obtained from the combination of CMB with the shape of the LRG $\xi(s)$. The thick 
solid line corresponds to flat models, where $\Omega_k=0$. The filled contours correspond 
to the 68\% CL in each case.
}
\label{fig:bomk_om_ode}
\end{figure}

\begin{table*} 
\centering
 \begin{minipage}{172mm}
  \caption{
The marginalized 68\% interval constraints on cosmological parameters for non-flat $\Lambda$CDM models
(i.e. the parameter set defined by Eq.(\ref{eq:paramomk})), obtained using different combinations of the datasets 
described in Section~\ref{sec:data}, as stated in the column headings.}
    \begin{tabular}{@{}lccccccc@{}}
    \hline
& \multirow{2}{*}{CMB}  & \multirow{2}{*}{CMB + $\xi(s)$} & \multirow{2}{*}{CMB + rBAO}& \multirow{2}{*}{CMB + SN} & CMB + $\xi(s)$ & CMB + $\xi(s)$ & \multirow{2}{*}{$\xi(s)$+ rBAO}\\
&     &                &                                  &                            &   + rBAO        &  + rBAO + SN &  \\  
\hline
$\Omega_{k}$        &  $-0.064_{-0.063}^{+0.056}$  &  $-0.0016_{-0.0074}^{+0.0070}$  &  $-0.0036_{-0.0053}^{+0.0053}$  &  
                   $-0.012_{-0.010}^{+0.010}$   &   $-0.0034_{ -0.0051}^{+ 0.0051}$ &  $-0.0035_{-0.0051}^{+0.0050}$  & $-0.12_{-0.088}^{+0.089}$\\[1.5mm] 
$100\,\Theta$             & $ 1.0412_{ -0.0023}^{+0.0022}$ &$ 1.0415_{-0.0022}^{+0.0022}$&  $ 1.0415_{-0.0022}^{+0.0022}$&
                     $ 1.0414_{-0.0023}^{+0.0022}$   & $ 1.0415_{-0.0022}^{+0.0022}$ &  $ 1.0415_{-0.0023}^{+0.0023}$&  $1.40_{-0.34}^{+0.34}$\\[1.5mm] 
$\omega_{\rm dm}$        & $ 0.1104_{-0.0050}^{+0.0050}$ & $ 0.1064_{-0.0048}^{+0.0048}$  & $ 0.1086_{-0.0046}^{+0.0047}$ & 
                      $ 0.1080_{-0.0048}^{+0.0047}$     &  $ 0.1072_{-0.0047}^{+0.0046}$ &  $ 0.1084_{-0.0044}^{+0.0044}$& $0.112_{-0.049}^{+0.051}$\\[1.5mm] 
$100\,\omega_{\rm b}$      & $ 2.253_{-0.051}^{+0.049}$ &$ 2.275_{-0.051}^{+0.051}$   & $ 2.272_{-0.051}^{+0.051}$ &
                    $ 2.266_{-0.051}^{+0.050}$  & $ 2.274_{ -0.050}^{+0.050}$ &  $ 2.267_{-0.050}^{+0.050}$ & $ 3.1_{-1.6}^{+1.7}$  \\[1.5mm] 
$\tau $              & $ 0.087_{-0.017}^{+0.017}$    &  $ 0.089_{-0.017}^{+0.017}$   &   $ 0.090_{-0.018}^{+0.017}$  &
                        $ 0.090_{-0.017}^{+0.017}$    & $ 0.089_{ -0.017}^{+ 0.017}$   &  $ 0.088_{-0.016}^{+0.016}$& - \\[1.5mm] 
$n_{\rm s}$              & $ 0.956_{-0.013}^{+0.012}$    & $ 0.964_{-0.012}^{+0.012}$   &  $ 0.963_{-0.012}^{+0.012}$  &
                      $ 0.960_{-0.012}^{+0.012}$        &  $ 0.963_{ -0.012}^{+0.012}$   & $ 0.961_{-0.012}^{+0.012}$& $0.99_{-0.31}^{+0.32}$ \\[1.5mm] 
$\ln(10^{10}A_{\rm s})$ &$ 3.061_{ -0.037}^{+ 0.037}$ &$ 3.052_{-0.037}^{+ 0.038}$   & $ 3.062_{-0.039}^{+0.038}$    &
                       $ 3.058_{-0.038}^{+ 0.038}$     & $ 3.055_{ -0.038}^{+ 0.037}$ & $ 3.055_{-0.035}^{+0.035}$ & - \\[1.5mm] 
$\Omega_{\rm DE}$       &   $ 0.56_{-0.17}^{+0.16}$& $ 0.756_{-0.021}^{+0.021}$  &  $ 0.740_{-0.018}^{+0.018}$ &
                       $ 0.717_{-0.023}^{+0.023}$       &  $ 0.748_{-0.015}^{+ 0.015}$      & $0.741_{-0.013}^{+0.013}$ &  $0.894_{-0.090}^{+0.091}$\\[1.5mm]
$\Omega_{\rm m}$           & $ 0.498_{-0.22}^{+0.23}$ &  $ 0.245_{-0.024}^{+0.025}$  &   $ 0.263_{-0.018}^{+0.019}$ &
                      $ 0.295_{-0.030}^{+0.030}$        &  $ 0.256_{-0.015}^{+0.016}$     & $ 0.262_{-0.013}^{+0.013}$ & $ 0.226_{-0.047}^{+0.049}$\\[1.5mm]
$\sigma_{8}$           &   $ 0.765_{-0.032}^{+0.032}$  & $ 0.778_{-0.028}^{+0.028}$   &  $ 0.791_{-0.028}^{+0.028}$  &     
                       $0.782_{-0.028}^{+0.028}$    &  $ 0.782_{-0.027}^{+ 0.027}$      &  $ 0.787_{-0.026}^{+0.026}$ & - \\[1.5mm]
$t_{0}/{\rm Gyr}$          &   $15.7_{-1.6}^{+1.6}$&  $13.71_{-0.38}^{+0.39}$     & $13.85_{-0.26}^{+0.26}$  &
                        $14.27_{-0.46}^{+0.46}$        & $13.83_{-0.25}^{+0.25}$   &  $13.85_{-0.25}^{+0.25}$ &  $14.2_{-2.6}^{+2.9}$\\[1.5mm] 
$z_{\rm re}$             & $10.3_{-1.4}^{+1.4}$   &  $10.4_{-1.3}^{+1.3}$        &  $10.5_{-1.4}^{+1.4}$  &
                        $10.49_{-1.4}^{+1.3}$        &   $10.4_{-1.4}^{+1.3}$     & $10.3_{-1.3}^{+1.3}$  & - \\[1.5mm]
$h$                  &   $0.55_{-0.15}^{+0.13}$ & $0.730_{-0.037}^{+0.034}$    &  $0.707_{-0.021}^{+0.021}$  &
                        $0.668_{-0.037}^{+0.037}$      & $0.713_{-0.019}^{+0.019}$   & $0.707_{-0.017}^{+0.017}$ & $0.78_{-0.14}^{+0.15}$ \\
\hline
\end{tabular}
\label{tab:omk}
\end{minipage}
\end{table*}

The flatness of the Universe is one of the generic predictions of most common 
models of inflation. A detection of a non-negligible curvature 
would have profound implications for our understanding of the mechanism thought to 
be responsible for seeding density fluctuations in the Universe.
The flatness hypothesis is also important because it has a strong impact on 
the constraints on the remaining cosmological parameters, since many of them 
are degenerate with $\Omega_k$. In this section, we analyse non-flat models and 
include $\Omega_k$ in our parameter space, assuming that the dark energy is 
given by vacuum energy with $w_{\rm DE}=-1$.

\begin{figure}
\centering
\centerline{\includegraphics[width=\columnwidth]{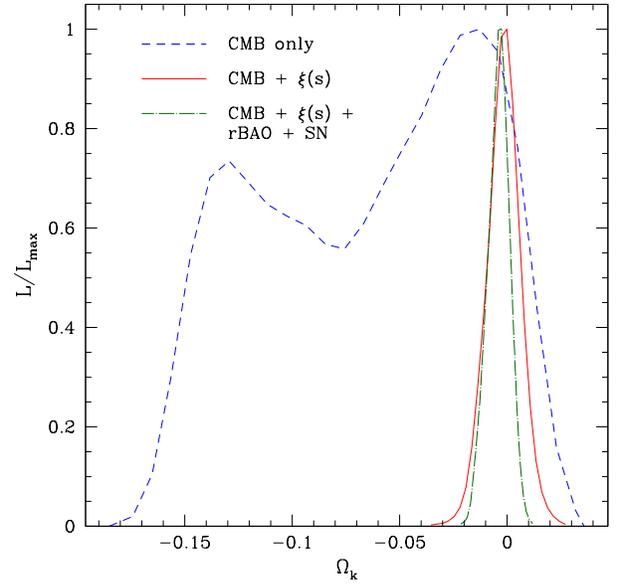}}
\caption{
The marginalized posterior likelihood in $\Omega_{\rm k}$ for the $\Lambda$CDM 
parameter set extended with $\Omega_k$ (Eq.~(\ref{eq:paramomk})). The dashed line 
shows the likelihood obtained using CMB information alone. The solid line corresponds 
to the result obtained from the combination of CMB data with the shape of the 
LRG $\xi(s)$. The dot-dashed line shows the likelihood when all datasets are considered. 
}
\label{fig:omk_1d}
\end{figure}

When $\Omega_k$ is allowed to float, the CMB data alone are unable to constrain 
all the parameters at the same time, giving rise to the well known 
geometrical degeneracy, which is completely analogous to the one between 
$w_{\rm DE}$ and $\Omega_{\rm m}$ described in Section~\ref{ssec:wde}. For each 
choice of $\omega_{\rm b}$, $\omega_{\rm dm}$ and $\Omega_{\rm m}$, it is possible 
to find a value of $\Omega_{k}$ (or $\Omega_{\rm DE}$) which will give the same value of
$\ell_{\rm A}$. This strong degeneracy can be seen in the dashed lines of 
Fig.~\ref{fig:bomk_om_ode}, which shows the two dimensional marginalized constraints 
in the plane $\Omega_{\rm m}-\Omega_{\rm DE}$. We plot the marginalized constraints 
on $\Omega_{k}$ in Fig.~\ref{fig:omk_1d}. The dashed curve shows the results for 
CMB data alone. This narrow degeneracy produces poor marginalized constraints on 
the curvature of the Universe, with $\Omega_{k}=-0.064_{-0.063}^{+0.056}$.

Adding the independent constraint from the shape of the LRG $\xi(s)$ helps to break 
this degeneracy. The solid lines in Fig.~\ref{fig:bomk_om_ode} show how with this 
extra information the two dimensional constraints in the plane 
$\Omega_{\rm m}-\Omega_{\rm DE}$ close up over the locus of the flat models 
(shown by the thick solid line). From CMB data plus the LRG $\xi(s)$ we get 
$\Omega_{k}=-0.0016_{-0.0074}^{+0.0070}$. Combining the CMB data with the other 
external datasets yields similar results, with $\Omega_k=-0.012_{-0.010}^{+0.010}$ 
for CMB plus SN and $\Omega_{k}=-0.0036_{-0.0053}^{+0.0053}$ from CMB plus rBAO. 

\begin{figure}
\centering
\centerline{\includegraphics[width=\columnwidth]{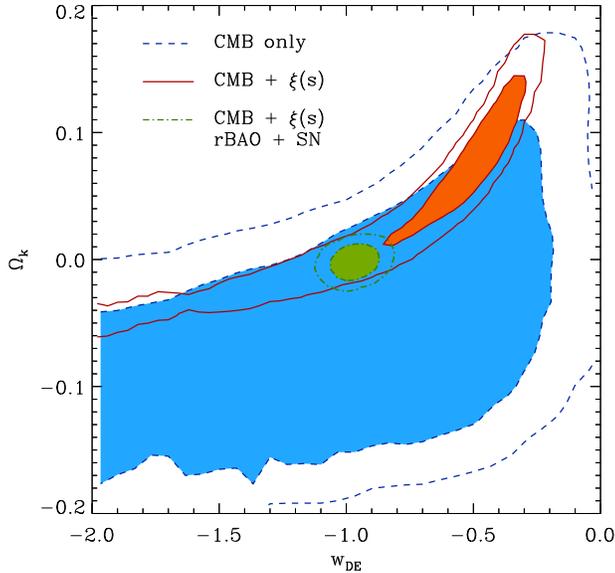}}
\caption{
The marginalized posterior likelihood in the $\Omega_{k}-w_{\rm DE}$ plane for 
the $\Lambda$CDM parameter set extended with $\Omega_k$ and $w_{\rm DE}$ 
(Eq.~(\ref{eq:param})). The dashed lines show the 68 and 95 per cent
contours obtained using CMB information alone. The solid contours correspond 
to the results obtained from the combination of CMB data with the shape of 
the LRG $\xi(s)$. The dot-dashed lines show the constraints obtained using 
all data sets. The filled contours correspond 
to the 68\% CL in each case.
}
\label{fig:bwok_wde_ok}
\end{figure}

The combination of the shape of the LRG $\xi(s)$ and the rBAO data is not able to give
meaningful constraints on $\Omega_{k}$. This data combination allows for a wide range of 
closed models giving a very poor one dimensional marginalized constraint of
$\Omega_{k}=-0.12_{-0.088}^{+0.089}$. On the other hand, these datasets can give a constraint 
on the matter density of $\Omega_{\rm m}= 0.226_{-0.047}^{+0.049}$, with a similar precision to 
the one obtained in the parameter spaces analysed in the previous sections.

Our tightest constraint comes from the combination of the four datasets, which 
gives $\Omega_k=-0.0035_{-0.0050}^{+0.0051}$, 
which is dominated by the combination 
of CMB and rBAO data. This result is in full agreement with that found by \citet{komatsu08} from the 
combination of WMAP-5yr data with the BAO measurements from Percival et al. of 
$\Omega_{k}=-0.0050_{-0.0060}^{+0.0061}$. This means that, with the hypothesis 
that vacuum energy is the source of the accelerated expansion of the Universe, 
current data allow us to probe spatial curvature up to 
$\Omega_{\rm k}\simeq5\times10^{-3}$. As we shall see in Section~\ref{ssec:wok}, 
this limit does not change when the assumption of $w_{\rm DE}=-1$ is relaxed.

\begin{table*} 
\centering
 \begin{minipage}{172mm}
  \caption{
The marginalized 68\% interval constraints on the cosmological parameters allowing for non-flat models and 
variations in the (redshift independent) dark energy equation of state parameter
(i.e. the parameter space defined by Eq.~(\ref{eq:param})), obtained using different combinations of the datasets 
described in Section~\ref{sec:data}, as stated in the column headings.}
    \begin{tabular}{@{}lccccccc@{}}
    \hline
& \multirow{2}{*}{CMB}  & \multirow{2}{*}{CMB + $\xi(s)$} & \multirow{2}{*}{CMB + rBAO}& \multirow{2}{*}{CMB + SN} & CMB + $\xi(s)$ & CMB + $\xi(s)$ & \multirow{2}{*}{$\xi(s)$+ rBAO}\\
&     &                &                                  &                            &   + rBAO        &  + rBAO + SN &  \\  
\hline
$\Omega_{k}$             & $-0.045_{-0.073}^{+0.067}$   & $ 0.040_{-0.060}^{+ 0.057}$  & $0.0089_{ -0.016}^{+ 0.014}$ & 
                          $-0.046_{ -0.032}^{+0.037}$   & $-0.0034_{-0.0065}^{+0.0065}$ & $-0.0018_{-0.0054}^{+0.0054}$&$-0.169_{-0.092}^{+0.099}$\\[1.5mm] 
$w_{\rm DE}$             & $-0.98_{ -0.67}^{+0.57}$     & $-0.81_{-0.48}^{+0.39}$      & $-0.83_{-0.29}^{+0.29}$         &
                          $-1.31_{-0.34}^{+0.37}$       & $-1.03_{-0.12}^{+0.12}$       & $-0.965_{-0.081}^{+0.078}$&$-0.91_{-0.14}^{+0.13}$\\[1.5mm] 
$100\,\Theta$            & $1.0411_{-0.0022}^{+0.0023}$ & $ 1.0412_{-0.0022}^{+0.0022}$& $ 1.0415_{-0.0022}^{+0.0022}$   & 
                          $ 1.0413_{-0.0023}^{+0.0022}$ & $ 1.0416_{-0.0022}^{+0.0022}$ & $ 1.0415_{-0.0022}^{+0.0022}$&$1.58_{-0.42}^{+0.39}$\\[1.5mm] 
$\omega_{\rm dm}$        & $ 0.1104_{-0.0051}^{+0.0049}$& $ 0.1057_{-0.0049}^{+0.0050}$& $ 0.1085_{-0.0048}^{+0.0048}$   &
                          $ 0.1094_{-0.0047}^{+0.0048}$ & $ 0.1073_{-0.0046}^{+0.0046}$ & $ 0.1078_{-0.0046}^{+0.0047}$& $ 0.106_{-0.043}^{+0.045}$\\[1.5mm] 
$100\,\omega_{\rm b}$    & $ 2.254_{-0.054}^{+0.052}$   & $ 2.259_{-0.048}^{+0.049}$   & $ 2.270_{-0.051}^{+0.050}$      & 
                          $ 2.257_{-0.052}^{+0.050}$    & $ 2.274_{-0.050}^{+0.050}$    & $ 2.270_{-0.046}^{+0.046}$&$ 0.030_{-0.016}^{+0.018}$\\[1.5mm] 
$\tau $                  & $ 0.087_{-0.017}^{+0.017}$   & $ 0.089_{-0.017}^{+0.017}$   & $ 0.090_{-0.017}^{+0.017}$      & 
                          $ 0.087_{-0.017}^{+0.017}$    &$ 0.089_{-0.017}^{+0.017}$     & $ 0.088_{-0.017}^{+0.017}$&-\\[1.5mm] 
$n_{\rm s}$              & $ 0.956_{ -0.013}^{+0.013}$  & $ 0.963_{-0.013}^{+0.013}$   & $ 0.962_{-0.012}^{+0.012}$      & 
                          $ 0.957_{-0.012}^{+0.012}$    &$ 0.963_{-0.012}^{+0.012}$     & $ 0.962_{-0.012}^{+0.012}$&$ 0.99_{-0.30}^{+0.31}$\\[1.5mm] 
$\ln(10^{10}A_{\rm s})$  & $ 3.059_{-0.038}^{+0.037}$   & $ 3.047_{-0.038}^{+0.037}$   &$ 3.061_{-0.038}^{+ 0.037}$      & 
                          $ 3.057_{-0.037}^{+0.039}$     &$ 3.055_{-0.039}^{+0.038}$     & $ 3.055_{-0.037}^{+ 0.036}$&-\\[1.5mm] 
$\Omega_{\rm DE}$        & $ 0.54_{-0.16}^{+ 0.15}$     & $ 0.726_{-0.034}^{+0.033}$   & $ 0.697_{-0.065}^{+0.066}$      & 
                          $ 0.649_{-0.054}^{+0.058}$    &  $ 0.750_{-0.020}^{+0.019}$   & $ 0.740_{-0.013}^{+0.013}$& $ 0.94_{ -0.10}^{+0.09}$\\[1.5mm]
$\Omega_{\rm m}$         & $ 0.50_{-0.20}^{+0.22}$      & $ 0.233_{-0.041}^{+0.047}$   & $ 0.294_{-0.052}^{+0.052}$      & 
                          $ 0.397_{-0.093}^{+0.087}$    & $ 0.253_{-0.017}^{+0.017}$    & $ 0.260_{-0.013}^{+0.013}$&$ 0.23_{-0.044}^{+0.044}$\\[1.5mm]
$\sigma_{8}$             & $ 0.73_{-0.10}^{+0.10}$      & $ 0.667_{-0.15}^{+0.17}$     & $ 0.735_{-0.091}^{+0.090}$      &
                           $ 0.805_{-0.045}^{+0.047}$     & $ 0.789_{-0.046}^{+0.046}$    & $ 0.775_{-0.033}^{+0.034}$&-\\[1.5mm]
$t_{0}/{\rm Gyr}$        & $15.4_{-1.9}^{+1.9}$         & $12.6_{-1.8}^{+2.1}$         & $13.67_{-0.38}^{+0.38}$         &
                           $15.6_{-1.4}^{+1.2}$          & $13.85_{-0.31}^{+0.30}$       & $13.78_{-0.26}^{+0.26}$&$14.3_{-2.6}^{+2.7}$\\[1.5mm] 
$z_{\rm re}$             & $10.4_{-1.5}^{+1.5}$         & $10.9_{-1.6}^{+1.6}$         & $10.7_{-1.4}^{+1.4}$          &
                           $10.2_{-1.4}^{+ 1.4}$          & $10.4_{-1.4}^{+1.4}$          & $10.4_{-1.3}^{+1.3}$&-\\[1.5mm]
$h$                      & $0.54_{-0.14}^{+0.12}$       & $0.750_{-0.068}^{+0.061}$    & $0.676_{-0.060}^{+ 0.060}$      &
                           $0.589_{ -0.066}^{+0.070}$     &$0.718_{-0.025}^{+0.024}$      & $0.706_{-0.017}^{+0.017}$&$0.76_{-0.13}^{+0.14}$\\
\hline
\end{tabular}
\label{tab:wok}
\end{minipage}
\end{table*}

\subsection{Dark energy and curvature}
\label{ssec:wok}

The assumption of a flat Universe has important implications for the constraints 
on the dark energy equation of state. Neither the curvature nor the dark energy 
play a role in determining the physical scales of the acoustic peaks in the 
CMB. These quantities do, however, have an impact on the angular diameter distance 
to the decoupling epoch. Furthermore, their contribution is degenerate.
An increment in the value of $\Omega_k$, which reduces the value of $D_{\rm A}(z_*)$, 
can be compensated for by an increase in  $w_{\rm DE}$, which produces the opposite effect.
This means that, instead of a one dimensional degeneracy as in the cases analysed 
in Sections~\ref{ssec:wde} and \ref{ssec:omk}, this parameter space has an extra 
degree of freedom for every choice of $\omega_{\rm b}$, $\omega_{\rm dm}$ and 
$\Omega_{\rm m}$. Fig.~\ref{fig:bwok_wde_ok} shows the marginalized posterior 
likelihood in the $w_{\rm DE}-\Omega_{k}$ plane. Due to this degeneracy, the 
region allowed by the CMB data, shown by the dashed lines, covers a wide range of this
parameter space. When the information from the shape of the correlation function 
is included in the analysis, the extra distance measurement reduces this degeneracy 
to an (approximately) one dimensional region in the $w_{\rm DE}-\Omega_{k}$ plane 
shown by the solid lines, but is also unable to place meaningful constraints on 
these parameters. 

An extra piece of information, coming from rBAO or SN data can break the remaining 
degeneracy. The dot-dashed lines in Fig.~\ref{fig:bwok_wde_ok} correspond to the 
results obtained by combining the CMB measurements, the SDSS $\xi(s)$, rBAO and SN data.
Combining all these datasets, the contours close-up over the canonical $\Lambda$CDM values, 
with $\Omega_k=-0.0018_{-0.0054}^{+0.0054}$ and $w_{\rm DE}=-0.965_{-0.053}^{+0.054}$
It is important to note that the precision of these constraints is similar to those 
obtained with the assumptions of a flat Universe (Section~\ref{ssec:wde}) or a 
cosmological constant (Section~\ref{ssec:omk}), showing the robustness of our results.

As in Sec.~\ref{ssec:omk}, the combination of the LRG $\xi(s)$ and the rBAO measurements
is not able to give useful constraints in this parameter space. In this case, there is a
wide range of closed models allowed by the data, which leads to a poor marginalized constraint of
$\Omega_k=-0.169_{-0.092}^{+0.099}$. On the other hand, the large scale structure data is able to 
constrain the dark energy equation of state to $w_{\rm DE}=-0.91\pm0.14$, with the same level
of accuracy obtained with the assumption of a flat Universe.

\section{Distance constraints}
\label{sec:distances}

In this section we explore the distance measurements that can be obtained from the shape of the 
LRG redshift space correlation function and its combination with other datasets. For the results of this section we 
have imposed the conservative constraint on the LRG bias factor of $b<4$.

\begin{figure}
\centering
\centerline{\includegraphics[width=\columnwidth]{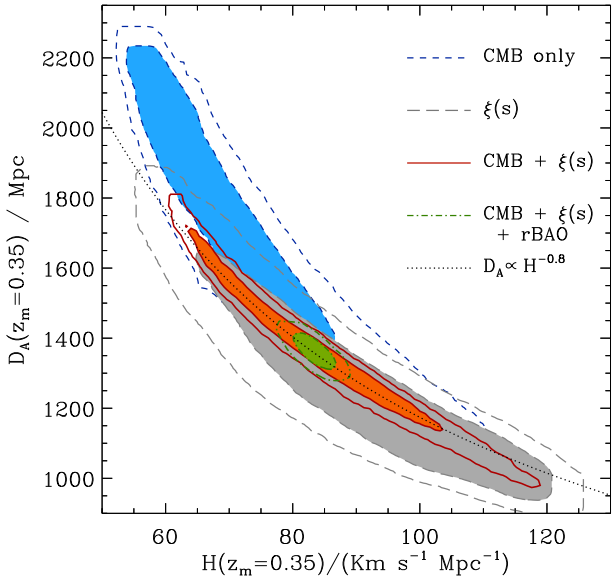}}
\caption{
The marginalized posterior likelihood in the $D_{\rm A}(z_{\rm m}=0.35)-H(z_{\rm m}=0.35)$ plane for 
the $\Lambda$CDM parameter set extended with $\Omega_k$ and $w_{\rm DE}$ 
(Eq.~(\ref{eq:param})). 
The short-dashed and long-dashed lines show the 68 and 95 per 
cent contours obtained individually from CMB information and the LRG $\xi(s)$ respectively.
The solid contours correspond to the results obtained from the combination of CMB data plus the shape 
of the LRG $\xi(s)$, which follow approximately a line of constant $G(z_{\rm m}=0.35)$ (dotted line)
defined by Eq.~(\ref{eq:g_z}). The dot-dashed lines show the contours obtained including also rBAO
information. The shading fills the 68\% CL contours.
}
\label{fig:bwok_dist}
\end{figure}

\citet{eisenstein05} used the correlation function of a sample of 
around 46,000 LRGs from the SDSS to obtain constraints on $\Omega_{\rm m}h^2$
and the combinations of distance measurements $D_{\rm V}(z)$ (defined in Eq.~\ref{eq:dv})
and $A(z)$, defined as
\begin{equation}
 A(z)\equiv D_{\rm V}(z)\frac{\sqrt{\Omega_{\rm m}H_0^2}}{zc},
\label{eq:aa}
\end{equation}
at the mean redshift of the survey $z_{\rm m}=0.35$ and found $D_{\rm V}(z_{\rm m}=0.35)=1370\pm64\,{\rm Mpc}$ and 
$A(z_{\rm m}=0.35)=0.469\pm 0.017$. In their analysis, 
\citet{eisenstein05} fixed the values of $w_{\rm b}=0.0223$ and $n_{s}=0.98$, as well as 
adopting a fixed BAO damping scale $k_{\star}=0.11\,h\,{\rm Mpc}^{-1}$ and varied only $\Omega_{\rm m}h^2$ and
$D_{\rm V}(z_{\rm m}=0.35)$. Fixing the values of these parameters has implications 
for the obtained constraints. They also implemented a different model for the non-linear distortion  
of the shape of the correlation function and included a wider range of scales 
in their analysis ($10\,h^{-1}{\rm Mpc}<r<177\,h^{-1}{\rm Mpc}$) than we consider. 
\citet{eisenstein05} used a fixed template for the nonlinear distortion of the 
linear perturbation theory correlation function, ignoring changes expected on 
varying the cosmological parameters (e.g. $\sigma_{8}$). Also, no attempt was 
made to model scale dependent effects such as redshift space distortions or 
bias, which could become important for pair separations approaching $10 h^{-1}\,$Mpc. 
In our analysis, we only consider pair separations given by $r > 42.5\,h^{-1}{\rm Mpc}$, 
for which our theoretical model is in excellent agreement with the results of N-body 
simulations (see Fig.~\ref{fig:modelRZspace}).
For the wide priors listed in Table~1, and using only 
information from the shape of $\xi(s)$ over the range of pair separations 
$42.5\,h^{-1}{\rm Mpc}<r<150\,h^{-1}{\rm Mpc}$, we find weaker constraints,
$D_{\rm V}(z_{\rm m}=0.35)=1230\pm220\,{\rm Mpc}$ and $A(z_{\rm m}=0.35)=0.424\pm 0.064$.
To achieve an accuracy comparable to that of Eisenstein et al. 
it is necessary to combine the $\xi(s)$ constraints with
CMB information. In this case we find $D_{\rm V}(z_{\rm m}=0.35)=1300\pm31\,{\rm Mpc}$ and $A(z_{\rm m}=0.35)=0.447\pm 0.015$.

In the analysis of the fifth year data of the WMAP satellite, \citet{komatsu08} produced a set of
distance priors that contain most of the information in the WMAP power spectrum. This set contains
({\it i}) the physical baryon density $w_{\rm b}$, ({\it ii}) the redshift to the decoupling epoch $z_{*}$,
({\it iii}) the ``acoustic scale'' of Eq.~(\ref{eq:la}) and 
({\it iv}) the ``shift parameter'' $R$ given by 
\begin{equation}
 R(z_{*})\equiv D_{\rm A}(z_{*})\frac{\sqrt{\Omega_{\rm m}H_0^2}}{c},
\label{eq:shift}
\end{equation}
with their respective covariance matrix. Here, we provide an extended set of distance priors combining the
information from the clustering of the LRGs with CMB data. For this, we expand the set of parameters provided by \citet{komatsu08} by adding an extra constraint on the quantity
\begin{equation}
G(z_{\rm m}) \equiv D_{\rm A}(z_{\rm m})\times h(z_{\rm m})^{0.8},
\label{eq:g_z}
\end{equation}
where $h(z_{\rm m})=H(z_{\rm m})/100\,{\rm km\,s^{-1}\,Mpc^{-1}}$. Now we show that this parameter is well constrained
by the combination of the CMB measurements and the LRG $\xi(s)$.

Fig.~\ref{fig:bwok_dist} shows the marginalized posterior likelihood in the $D_{\rm A}(z_{\rm m})-H(z_{\rm m})$ plane for 
the full parameter space of Eq.~(\ref{eq:param}). The constraints obtained individually from CMB information (short-dashed
lines) and the LRG $\xi(s)$ (long-dashed lines) show strong degeneracies between these parameters. The combination of the
two datasets reduces the allowed region for these parameters to a narrow degeneracy (solid lines) which follows approximately
a line of constant $G$ (shown by the dotted line). For this combination we find the constraint $G(z_{\rm m}=0.35)=1175_{-17}^{+18}\,{\rm Mpc}$ 
(a 1.4\% error) which we include in our set of extended distance priors. 
The degeneracy between $D_{\rm A}$ and $H$ 
can be broken by including an extra piece of information. In this case it is possible
to obtain separate constraints on these parameters. Including the rBAO measurements (as shown by the dot-dashed lines in 
Fig.~\ref{fig:bwok_dist}) we obtain $D_{\rm A}(z_{\rm m}=0.35)=1363\pm34$ Mpc and 
$H(z_{\rm m}=0.35)=83.3\pm2.2 \,{\rm km}\,{\rm s}^{-1}\,{\rm Mpc}^{-1}$. By using the SN data instead we obtain 
$D_{\rm A}(z_{\rm m}=0.35)=1336_{-84}^{+80}$ Mpc and $H(z_{\rm m}=0.35)=86.4\pm5.6 \,{\rm km}\,{\rm s}^{-1}\,{\rm Mpc}^{-1}$.

\begin{figure}
\centering
\centerline{\includegraphics[width=\columnwidth]{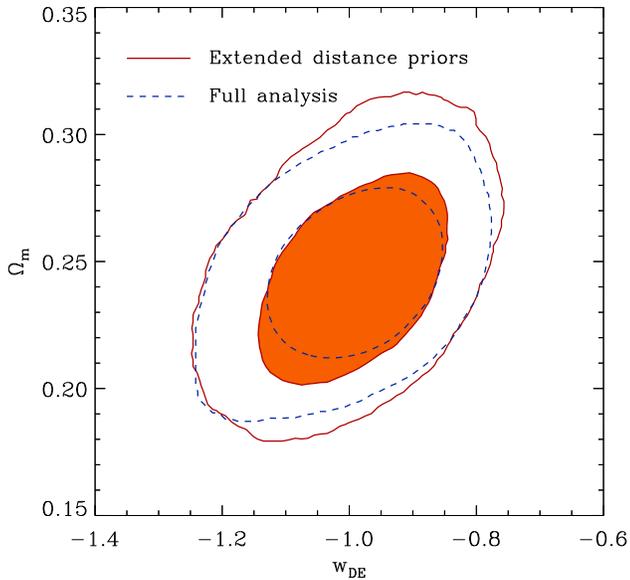}}
\caption{
The marginalized posterior likelihood in the $w_{\rm DE}-\Omega_{\rm m}$ plane for 
the $\Lambda$CDM parameter set extended with $w_{\rm DE}$ (Eq.~(\ref{eq:paramwde})). 
The dashed lines show the 68 and 95 per cent contours obtained using the full combination of CMB data
plus the shape of the LRG $\xi(s)$. The solid contours correspond to the results obtained using the set of
extended distance priors of Tables~B1 and B2 of Appendix B. The shading fills the 68\% CL contours. 
}
\label{fig:bwde_edp}
\end{figure}

The mean values and covariance matrix of the set of extended distance priors obtained from the
combination of CMB information and the shape of the LRG $\xi(s)$ are listed in Tables~B1 and B2 of Appendix B. 
Fig.~\ref{fig:bwde_edp} shows a comparison of the marginalized constraints 
in the plane $\Omega_{\rm m}-w_{\rm DE}$
for the parameter space of Eq.~(\ref{eq:paramwde}) obtained by using the set of extended distance priors
(solid lines) and the ones obtained using the full CMB and $\xi(s)$ data (dashed lines). This shows that this
set of constraints contains the most relevant information from the combination of these datasets and can be
used to replace them to obtain constraints on cosmological parameters in combination with other cosmological
observations like SN data or weak lensing, simplifying and accelerating the required numerical procedure.

\section{Conclusions}
\label{sec:conclusions}

In this paper, we have applied a new model for the shape of the correlation function at 
large pair separations to the measurement of LRG clustering from the SDSS made by \citet{cabre2008}.
Using a full Monte-Carlo markov chain analysis and combining the LRG measurements with
the latest compilations of CMB and SNe data, we have presented a comprehensive set of
constraints on cosmological parameters for different combinations of datasets and for
different parameter spaces. 

Large scale structure measurements and modelling have now reached a 
level of precision where they provide constraints on cosmological 
parameters which are competitive with those obtained from other datasets. 
Using only the LRG correlation function and the measurement of the radial BAO
peak, it is possible to determine the dark energy equation of state parameter to 
$w_{\rm DE}=-1.05^{+0.16}_{-0.15}$. This provides further confirmation of the
dark energy scenario, independent of CMB or SN observations. 

The availability of several high quality datasets gives us the opportunity to scrutinize the 
derived constraints and to isolate possible systematic effects in either the observations 
themselves or the theoretical model used to describe them.  In general, different datasets 
are sensitive to different parameter combinations and the constraints are optimized when 
datasets are combined. However, in the case in which datasets are responsive to similar 
parameter combinations, it is important to check that consistent results are obtained 
before combining the datasets. It would be meaningless to combine two such datasets if a 
tension existed between the parameters values returned from each one individually. 

The tables of parameter constraints given in this paper show that, to first 
order at least, the ``Union'' supernova dataset of \citet{kowalski08} and 
the LRG clustering from the SDSS yield very similar constraints on cosmological 
parameters when analysed in combination with CMB data, for all of the parameter 
spaces considered. As shown in Section~\ref{ssec:wde}, the recovered value of the dark 
energy equation of state in the CMB plus $\xi(s)$ case of $w_{\rm DE}=-0.996\pm0.090$ 
shows a remarkable agreement with that obtained from the combination of CMB with the 
SN data of $w_{\rm DE}=-0.950\pm0.055$. The consistency of the constraints we obtain 
from different combinations of datasets is a reassuring validation of our analysis 
technique.

\citet{percival07c}, however, found a tension at the 2.4$\sigma$ level between the 
constraints on $D_{\rm V}(z=0.35)/D_{\rm V}(z=0.20)$ coming from the BAO signal measured 
from galaxy samples drawn from the 2dFGRS and the SDSS (including a smaller LRG sample than 
the one considered here), and the SN data from \citet{astier06}. When folded with the 
CMB data, this discrepancy leads to different preferred values for the dark energy equation 
of state, with the SN data preferring $w_{\rm DE}\simeq -1$ and the BAO data pointing 
to $w_{\rm DE}<-1$ with a significance of 1.4$\sigma$ (see their figures 12 and 13; 
see also \cite{lazkoz08}). 
Our analysis shows that the LRG clustering and SNe data give consistent constraints 
on the distance scale $D_{\rm V}(z=0.35)$. The tension reported by \citet{percival07c} 
could result from the BAO signal at $z=0.2$ (which corresponds to the 
main 2dFGRS and SDSS galaxy samples). This is unlikely to be the answer, however, since 
the rBAO measurements from \citet{gaztanaga08b}, which separate the signal at $z=0.24$ 
and $z=0.43$, also give a consistent answer with $w_{\rm DE}=-0.92\pm0.15$. The treatment 
of the BAO feature is quite different in the two analyses. \citet{percival07c} effectively 
take a BAO template generated for a given set of cosmological parameters and then fit this 
to the measured BAO features for different parameters. Furthermore, the technique of Percival 
et~al.  requires a reference spectrum to be defined. Finally, as \citet{percival07c} speculate 
themselves, the problem could lie in the way in which the BAO are damped; these authors 
assume a fixed damping scale, whereas we treat the damping scale as a parameter. 

\begin{figure}
\centering
%\centerline{\includegraphics[width=\columnwidth]{figs/DR6systematicslin.ps}}
\centerline{\includegraphics[width=\columnwidth]{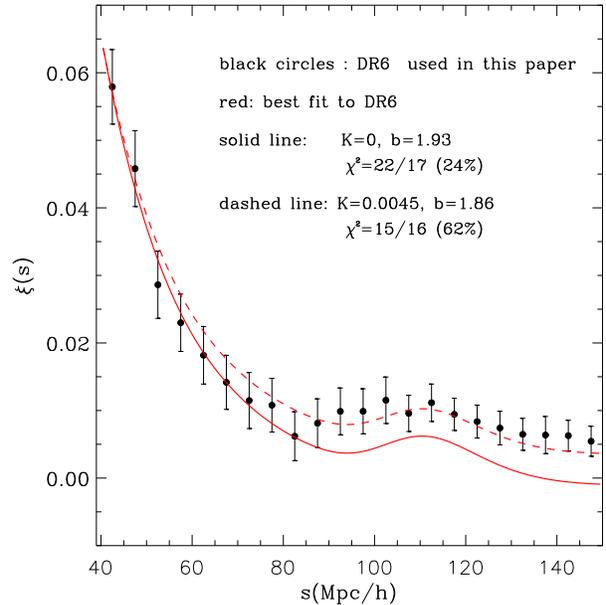}}
\caption{
Circles with error bars show the redshift space correlation function 
used in this paper \citep[from][]{cabre2008}. The solid line 
corresponds to our best-fit model within the framework of the basic 
$\Lambda$CDM parameter space described in Sec.~\ref{ssec:lcdm}, while the 
dashed line includes in addition a systematic K-shift with $K=0.0045$. 
This shift may originate from unknown systematic effects as described in 
Sec.~\ref{sec:Kshift}. Even though the visual impact seems important, 
this constant shift (note the bias factor also changes) does not 
compromise the goodness of fit of our best fitting model or the range 
of cosmological parameters derived in this paper. Our best fitting 
model (ie using $K=0$) has a probability of 24 \% to be in 
agreement with the data, as compared with 62 \% for the model with $K=0.0045$. 
Thus there is not a significant improvement on adding a constant shift to the model.
}
\label{fig:corfunc2}
\end{figure}

One criticism that could be levelled at our analysis is the apparent level of agreement 
between our new theoretical model and the measured LRG clustering. We have shown that the  
model is remarkably accurate when compared to the correlation function measured for the dark 
matter or for samples of haloes drawn from N-body simulations \citep[see also][]{crocce08,sanchez08b}.
However, in the case of the observed LRG clustering, 
the relative height of the BAO feature is greater than predicted by the model.
This is illustrated in Fig.~\ref{fig:corfunc2}. This behaviour  
is {\it not} peculiar to the measurement by \citet{cabre2008}. The earlier measurement 
of LRG clustering by \citet{eisenstein05} had a similar form \citep[see also][]{okumura08}.   
A key result from this paper is the dependence of the parameter constraints on the minimum 
pair separation used in the analysis (Fig.~5). The mean parameter value returned is insensitive 
to the choice of the minimum pair separation; the only noticeable difference is an increase 
in the $1 \sigma$ range on selected parameters as fewer data points are included. Similar 
conclusions were reached using a less challenging range of scales by \citet{okumura08}. 
This confirms that the information we extract from the overall shape of the correlation 
function is consistent with that contained in the BAO feature.

Fig.~\ref{fig:corfunc2} also considers an extension to our best fitting model 
which involves allowing for a constant shift in amplitude (Sec 2.1.1). 
We have tested that 
the marginalization over a constant additive term has a negligible
impact on the cosmological constraints that we derive, which are driven by relative changes
in the goodness of fit. However, the addition of this extra nuisance parameter can improve the appearance 
of the fit and therefore reduce the absolute value of $\chi^2$. 
If we use an absolute  $\chi^2$ fit with the full covariance matrix, 
we find a probability of about 25\% that our best fitting model describes 
the measured correlation function (the exact value varies between 20\% and 33\%
depending on which parameters we consider as degrees of freedom). If we allow for a constant systematic
shift, as argued in section \ref{sec:Kshift}, the best fit model
for the cosmological parameters remains the same, but the probability of the 
model describing the data improves slightly to 60\% in the range $K=0.003-0.006$.
This improvement is not significant enough to justify the extra parameter, 
despite the visual impact of this shift in Fig.~\ref{fig:corfunc2}. 

We used the combined information of the CMB measurements and the 
shape of the LRG $\xi(s)$ to obtain a set of distance priors 
that combine the most relevant information from these data sets 
to place constraints on dark energy models and spatial curvature.
For this, we expanded the set of distance priors provided by 
\citet{komatsu08} by adding a constraint on the quantity $G(z_{\rm m})$, 
defined in Eq.~(\ref{eq:g_z}), which is well constrained by the 
combination of these datasets (with an error of less than $2\%$).
The mean values and covariance matrix of this extended set of distance 
priors are listed in Tables~B1 and B2 of Appendix B.
These can be used instead of the combination of CMB and $\xi(s)$ 
data used in our analysis to obtain constraints on cosmological
parameters in combination with other cosmological observations, 
making the required calculations faster and simpler.

Our analysis shows that current large-scale structure datasets (using the shape of the 
correlation function at large pair separations, along with the form of the BAO peak) give 
constraints on the basic cosmological parameters that are competitive with those from 
the best available SNe datasets. Moreover, possible systematic distortions in 
the clustering signal have been modelled extensively. The theoretical model applied 
in this paper has been shown to provide an accurate description of the clustering for 
the level of error expected in volumes up to two orders of magnitude larger than the 
current LRG datasets. This bodes well for experiments planned for ten year's time, which 
will produce surveys covering volumes in the region of $100 h^{-3}{\rm Gpc}^{3}$ 
\citep[e.g.][]{cimatti2008}. Nevertheless, it will be prudent to pursue as many approaches as 
possible to measure dark energy to allow comparisons to be made between the results obtained 
with different datasets, as carried out in this paper. 

\section*{Acknowledgements}

We thank Francesco Montesano and Emilio Donoso for their 
assistance in preparing preliminary versions of many of our 
plots. We thank Eiichiro Komatsu, Andr\'es Balaguera, Emiliano Sefusatti and Pablo Fosalba for
useful comments and suggestions, and Roman Scoccimarro for providing the simulation data.
MC, AC and EG acknowledge support from Spanish Ministerio de Ciencia y Tecnologia (MEC),
project AYA2006-06341, Consolider-Ingenio CSD2007-00060, research project 2005SGR00728
from Generalitat de Catalunya and the Juan de la Cierva MEC program.
We acknowledge the use of the Legacy Archive for Microwave Background 
Data Analysis (LAMBDA). Support for LAMBDA is provided by the NASA
Office of Space Science. This work was supported in part by the STFC.

\appendix

\section{The theoretical motivation for the model for the shape of $\xi(r)$} 
\label{ssec:shape}

In recent years substantial progress has been made in understanding the 
non-linear gravitational evolution of density fluctuations \citep{crocce06,
crocce08,matarrese2007,matsubara08,pietroni2008,taruya08}. Within the theoretical 
framework of Renormalized Perturbation Theory (RPT), \citet{crocce06,crocce06b} 
showed that the evolution of the power spectrum from its initial value in the 
linear regime, $P_{\rm L}(k)$, can be described as the sum of two contributions
\begin{equation}
P_{\rm NL}(k,z)=P_{\rm L}(k)G^2(k,z)+P_{\rm MC}(k,z),
 \label{eq:p_rpt}
\end{equation}
where $G$ is the nonlinear propagator which weights how much power can be directly 
linked to the linear epoch (e.g. damping the higher BAO harmonics), and $P_{\rm MC}$ 
represents the new variance generated by mode coupling. 

In configuration 
space, the first term in Eq.~(\ref{eq:p_rpt}) leads to a convolution of the 
linear theory correlation function with a nearly Gaussian kernel 
(since $G\sim {\rm e}^{-k^2}$, see \cite{crocce06b}), which causes the acoustic peak to broaden and shift to smaller values \citep{smith08,crocce08}. 
In turn, the contribution of the mode-coupling term can be computed to  leading order at large pair separation, yielding $\xi_{\rm MC}(r) \propto \xi'_{\rm L}(r)\,\xi^{(1)}_{\rm L}(r)$ \citep{crocce08}, 
where $\xi'_{\rm L}(r)$ is the first derivative of the linear theory correlation 
function and $\xi^{(1)}_{\rm L}(r)$ is defined by Eq~(\ref{eq:xi1}). 

In fact, Eq.(~\ref{eq:p_rpt}) is also valid for the nonlinear spectra of density ($\delta$)/velocity divergence ($\theta$) fields, $P_{\delta\theta}$ and $P_{\theta\theta}$,
\begin{equation}
P_{NL, ab}=P_L G_a G_b + P_{MC,ab},
\label{eq:pall_rpt}
\end{equation}
with $a=\delta , \theta$.

To study the impact of redshift distortions and bias we put together the discussion above and the ansatz given in \cite{scoccimarro04} (see also \cite{percival08}):
\begin{eqnarray}
P_s({\bf k})= {\rm e}^{-k^2 f^2 \mu^2 \sigma_v^2}
\left[b^2 P_{\delta\delta}+ 2 f \mu^2 b P_{\delta\theta} 
+ f^2 \mu^4 P_{\theta\theta} \right],
\label{roman}
\end{eqnarray}
where $P_{\delta\delta}, P_{\delta\theta}$ and $P_{\theta\theta}$ were introduced above and a local linear bias 
relation between the density fluctuations in the distribution of galaxies 
and dark matter is assumed (but with unbiased galaxy velocities). 
The damping factor in  Eq.~(\ref{roman}) arises after assuming a Gaussian PDF 
for the pairwise velocities with {\it rms} (if evaluated in the linear regime), 
\begin{equation}
\sigma_v = \frac{1}{3}\int \frac{P_{\rm L}(q)}{q^2} {\rm d}^3 q.
\end{equation}
In Eq.~(\ref{roman}) $f$ is the logarithmic 
derivative of the linear growth rate with respect to the scale factor, 
$f = {\rm d}\,{\rm ln} D /{\rm d}\,{\rm ln} a$, and $\mu=k_z/k$ is the cosine of 
the line-of-sight angle. Note that the small scale redshift space distortions, commonly referred to 
as ``fingers of God'' (FOG) due to the elongation of virialised structures, are difficult 
to model accurately, partly due to the non-Gaussianity of the PDF, which persists 
even to large scales \citep{scoccimarro04}.  In addition, the damping factor depends on galaxy type (or mean halo mass and satellite population)

We are interested in the monopole of the correlation function, i.e. the angle average of the Fourier transform of $P_s({\bf k})$ in Eq.~(\ref{roman}). 
Since the time required to evaluate the different power spectra is not well suited to 
the exploration of large parameter spaces, we would like to motivate 
a parametric description of the problem.

\cite{crocce06b} showed that both $G_{\theta}$ and $G_{\delta}$ are of approximately Gaussian form, and with
similar smoothing length. Therefore the main contribution to the correlation function at large separations can be modeled, combining Eqs.~(\ref{roman},\ref{eq:pall_rpt}), 
as $\propto \xi_L \otimes {\rm e}^{-(k_{\star} r)^2}$.  The FOG prefactor in Eq.(\ref{roman}) leads to an extra suppression, which is subdominant to that of nonlinear gravity
for low redshift data (where $f \sim 0.5$). However, it can be accounted for to some degree by smaller values of $k_{\star}$.

The mode coupling spectra in Eq.~(\ref{eq:pall_rpt}) do differ at the several $\%$ level \citep{scoccimarro04}, but they all give contributions scaling as $\sim \xi^{\prime}$ 
(the derivative of the correlation function) 
in configuration space at large separations. The convolution with the FOG suppression leaves this
contribution unaltered as the smoothing length introduced by the FOG is smaller than features present in $\xi^{\prime}$, even at BAO scales.

Therefore we are left with the following parametrization,
\begin{equation}
 \xi_{\rm NL}(r) = b^2 \{ \xi_{\rm L}(r)\otimes {\rm e}^{-(k_{\star}r)^2} 
+ A_{\rm MC} \,\xi'_{\rm L}(r)\,\xi^{(1)}_{\rm L}(r) \}, 
\label{eq:last}
\end{equation}
for the correlation function at large pair separations in redshift space, where $b$ and
$A_{\rm MC}$ are nuisance parameters to account for bias and also the enhanced correlation amplitude due to redshift distortions \citep{kaiser87}. 

The assumptions behind the model in Eq.~(\ref{eq:last}), one could argue, are over-simplifications, particularly for biased tracers such as LRGs (Zehavi et~al. 2005; Blake et~al. 2007), or for deeper surveys where a more systematic derivation of redshift distortions may be needed. Nonetheless this form is well motivated given the size of statistical errors in the SDSS-DR6 LRG sample, and based on results from Sec.~\ref{ssec:rpt} (see also \cite{sanchez08b}) should be accurate also for future larger surveys.

\section{Extended distance priors} 
\label{ssec:edp}

Here we list the constraints on the set of extended distance priors
described in Section~\ref{sec:distances}. Table B1 lists the mean values and
variances of the parameters in this set and Table B2 gives the corresponding
covariance matrix. This information can be used to replace the CMB and $\xi(s)$ data
in multiparametric analysis since it contains most of the information of the combination
of these two datasets.
With this simplification, the likelihood of a given cosmological model can be easily computed
as $-2\ln{\mathcal L}\propto({\mathbf D}-{\mathbf T})^{\rm t}\textsf{ \mathbfss{ C}}^{-1}({\mathbf D}-{\mathbf T})^{\rm t}$
where ${\mathbf D}$ is a vector containing the constraints of Table B1, ${\mathbfss T}$ contains the
% CMB sp.
corresponding 
prediction for a given cosmological model and ${\textsf {\textbf C}}^{-1}$ is the inverse of
the covariance matrix of Table B2.
This procedure considerably reduces the computing time required to obtain
constraints on cosmological parameters 
compared with the full analysis of the data and can be used 
in combination with other datasets like weak lensing or the Hubble diagram of type Ia supernovae.

\begin{table}
\begin{center}
\caption{ 
The mean values and variances of the parameters in the set of extended distance
priors described in Section~\ref{sec:distances},
using the compilation of CMB data in combination with the LRG correlation function.
}
\end{center}
\begin{center}
\begin{tabular}[t]{cc}
\hline\hline
Parameter                 & value \\ \hline\hline
$100\omega_{\rm b}$          & $2.28 \pm 0.55$  \\ 
$z_{*}$                   & $1090.12 \pm 0.93$ \\
$\ell_{\rm A}(z_{*})$     & $301.58  \pm 0.67$ \\
$R(z_{*})$                & $1.701   \pm 0.018$ \\
$G(z_{\rm m})$            & $1175   \pm 21$ \\
\hline\hline
\end{tabular}
\end{center}
\label{tab:edp_values}
\end{table}

\begin{table*}
\begin{center}
\caption{ The covariance matrix of the parameters in the set of extended distance
priors described in Section~\ref{sec:distances}.
}
\end{center}
\begin{center}
\begin{tabular}[t]{cccccc}
\hline\hline
                      &$100\omega_{\rm b}$     & $z_{*}$     & $\ell_{\rm A}(z_{*})$  &  $R(z_{*})$   &  $G(z_{\rm m})$   \\
$100\omega_{\rm b}$      & $2.99\times10^{-7}$ & $-4.2030\times10^{-4}$ &     $-1.9988\times10^{-4}$       &   $-3.4393\times10^{-6}$ &  $-1.9978\times10^{-3}$ \\
$z_{*}$               &                      &  $8.6812\times10^{-1}$ &   0.2557      &   $1.1999\times10^{-1}$   & 3.2355\\
$\ell_{\rm A}(z_{*})$ &                    &              &  0.4558    &   $3.1265\times10^{-2}$   & 1.8247\\
$R(z_{*})$            &                    &              &            &   $3.1460\times10^{-4}$   &  $3.9649\times10^{-2}$\\
$G(z_{\rm m})$        &                    &              &            &                &  $4.3784\times10^{2}$\\
\hline\hline
\end{tabular}
\end{center}
\label{tab:edp_cov}
\end{table*}


\begin{thebibliography}{}

\bibitem[Ade et~al.~(2008)]{ade08}
Ade P., ock J., Bowden M., Brown M.~L., Cahill G., Carlstrom J.~E., Castro P.~G., Church, S.,
2008, ApJ, 674, 22

\bibitem[Albretch et~al.~(2006)]{albretch06}  
Albrecht, A., et al., 2006, preprint astro-ph/0609591

\bibitem[Angulo et~al.~(2005)]{angulo05}
Angulo R. E., Baugh C. M., Frenk C. S., Bower R.G., Jenkins A., Morris S. L., 
2005, MNRAS, 362, L25

\bibitem[Angulo et~al.~(2008)]{angulo08} 
Angulo R. E., Baugh C. M., Frenk C. S., Lacey C. G., 
2008, MNRAS, 383, 755

\bibitem[Astier et~al.~(2006)]{astier06}  
Astier P. et al.,
2006,  A\&A, 447, 31

\bibitem[Barris et~al.~(2004)]{barris2004}  
Barris B. J. et al.,
2004, ApJ, 602, 571

\bibitem[Bennet et~al.~(2003)]{bennet03}  
Bennett C. L. et al.,
2003, ApJS, 148, 1

\bibitem[Blake \& Glazebrook~(2003)]{blake03}  
Blake C., Glazebrook K., 
2003, ApJ, 594, 665

\bibitem[Blanton et~al.~(2005)]{blanton2005}  
Blanton M. R. et al.,
2005, AJ, 129, 2562  

%\bibitem[Bond, Efstathiou \& Tegmark~(1997)]{bond1997}     %% 3 authors 
\bibitem[Bond et~al.~(1997)]{bond1997}   
Bond J. R., Efstathiou G., Tegmark M., 
1997, MNRAS, 291, L33

\bibitem[Bridle et~al.~(2003)]{bridle2003} 
Bridle S. L., Lahav O., Ostriker J. P., Steinhardt P. J., 
2003, Science, 299, 1532

\bibitem[Cabre \& Gazta\~naga~(2009a)]{cabre2008}  
Cabre A., Gazta\~naga E., 
2009a, MNRAS, 393, 1183, eprint arXiv:0807.2460

\bibitem[Cabre \& Gazta\~naga~(2009b)]{cabre2008b}  
Cabre A., Gazta\~naga E., 
2009b, MNRAS, 396, 1119, eprint arXiv:0807.2461

\bibitem[Chevallier \& Polarski~(2001)]{chevalier2001} 
Chevallier M., Polarski D., 
2001, Int. J. Mod. Phys. D., 10, 213

\bibitem[Cimatti et~al.~(2008)]{cimatti2008} 
Cimatti A. et al.,
2008, Experimental Astronomy, in press, eprint arXiv:0804.4433

\bibitem[Cole et~al.~(2005)]{cole05}   
Cole S. et al. (The 2dFGRS Team), 
2005, MNRAS, 362, 505     %P(k) final

\bibitem[Colless et~al.~(2001)]{colless01}    
Colless M. et~al. (The 2dFGRS Team), 
2001, MNRAS, 328, 1039

\bibitem[Colless et~al.~(2003)]{colless03}   
Colless M. et~al. (The 2dFGRS Team), 
2003, eprint arXiv:0306581

\bibitem[Cresswell \& Percival~(2009)]{cresswell09} 
Cresswell J.G., Percival W.J.,
2009, MNRAS, 392, 682

\bibitem[Crocce \& Scoccimarro~(2006a)]{crocce06} 
Crocce M., Scoccimarro R., 
2006a, PRD, 73, 063519

\bibitem[Crocce \& Scoccimarro~(2006b)]{crocce06b} 
Crocce M., Scoccimarro R., 
2006b, PRD, 73, 063520

\bibitem[Crocce \& Scoccimarro~(2008)]{crocce08}   
Crocce M., Scoccimarro R., 
2008, PRD, 77, 023533

\bibitem[Crocce et~al.~(2009)]{crocce09}   
Crocce M., Fosalba P., Castander F, Gaztanaga E.,
2009, eprint arXiv:0907.0019

\bibitem[Desjacques~(2008)]{desjacques08}
Desjacques V.,
2008, PRD, 78, 103503

\bibitem[Dunkley et~al.~(2009)]{dunkley2008}  
Dunkley J. et al.,
2009, ApJS, 180, 306

\bibitem[Efstathiou \& Bond~(1999)]{efstathiou1999}  
Efstathiou G., Bond J. R., 
1999, MNRAS, 304, 75

\bibitem[Efstathiou et~al.~(2002)]{efstathiou02} 
Efstathiou G. et al.,
2002, MNRAS, 330, L29

\bibitem[Eisenstein \& Hu~(1998)]{EH98}  
Eisenstein D. J., Hu W., 
1998, ApJ. 496, 605

\bibitem[Eisenstein \& Hu~(1999)]{EH99}  
Eisenstein D. J., Hu W., 
1998, ApJ. 511, 5

\bibitem[Eisenstein et al.~(2001)]{eisenstein01} 
Eisenstein D. J. et~al., 
2001, AJ, 122, 2267

\bibitem[Eisenstein et~al.~(2005)]{eisenstein05}   
Eisenstein D. J. et~al., 
2005, ApJ, 633, 560

\bibitem[Eisenstein et~al.~(2006)]{eisenstein06}  
Eisenstein D. J., Seo H., Sirko E., Spergel D. N., 
2006, ApJ, 664, 675

\bibitem[Ellis et~al.~(2008)]{ellis08}   
Ellis R. S. et al.,
2008, ApJ, 674, 51 

\bibitem[Estrada et~al.~(2009)]{estrada08}  
%\bibitem[Estrada, Sefusatti \& Frieman~(2008)]{estrada08}  %% 3 authors 
Estrada J., Sefusatti E., Frieman J.~A., 
2009, ApJ, 692, 265

\bibitem[Fang et~al.~(2008)]{fang2008}
%\bibitem[Fang, Hu \& Lewis~(2008)]{fang2008}  %% 3 authors 
Fang W., Hu W., Lewis A., 
2008, PRD, 78, 087303

\bibitem[Ferramacho et~al.~(2009)]{ferramacho08}
Ferramacho L. D., Blanchard A., Zolnierowski Y.,
2009, A\&A, 499, 21

\bibitem[Fosalba et~al.~(2008)]{fosalba08}  
Fosalba P., Gazta\~naga, E., Castander F.~J., Manera M.,
2008, MNRAS, 391, 435

\bibitem[Gallagher et~al.~(2005)]{gallagher05}  
Gallagher J.~S., Garnavich P.~M., Berlind P., Challis P., Jha S.,  Kirshner R.~P.,
2005, ApJ, 634, 210

\bibitem[Gazta\~naga et al.(2008a)]{gaztanaga08a} 
Gaztanaga E., Cabre A., Castander F., Crocce M., Fosalba P., 
2008a, MNRAS in press, eprint arXiv:0807.2448

\bibitem[Gazta\~naga et~al.~(2008b)]{gaztanaga08b}  
%\bibitem[Gazta\~naga, Cabre \& Hui~(2008b)]{gaztanaga08b}  %% 3 authors
Gazta\~naga E., Cabre A., Hui L., 
2008b, MNRAS in press, eprint arXiv:0807.3551

\bibitem[Gazta\~naga et~al.~(2008c)]{gaztanaga08c}  
%\bibitem[Gazta\~naga, Miquel \& Sanchez~(2008c)]{gaztanaga08c}   %% 3 authors
Gazta\~naga E., Miquel R., S\'anchez E., 
2008c, eprint arXiv:0808.1921

\bibitem[Gelman \& Rubin~(1992)]{GR92} 
Gelman A., Rubin D., 
1992, Stat. Sci., 7, 457

\bibitem[Gorski et~al.~(1999)]{gorski99}
Gorski K. M., Wandelt B. D., Hansen F. K., Hivon E., Banday A. J.,
1999, eprint arXiv:9905275

\bibitem[Glazebrook et~al.~(2007)]{glazebrook2007}  
Glazebrook K. et al.
2007, ASPC, 379, 72

\bibitem[Guzik et~al.~(2007)]{guzik2007}      
Guzik J., Bernstein G., Smith R. E., 
2007, MNRAS, 375, 1329

\bibitem[Guy et~al.~(2005)]{guy2005}  
Guy J., Astier P., Nobili S., Regnault N., Pain R., 
2005, A\&A, 443, 781 

\bibitem[Hamann et~al.~(2008)]{hamann08}
Hamann J., Hannestad S., Melchiorri A., Wong Y. Y. Y., 
2008, JCAP, 07, 017 

\bibitem[Hicken et~al.~(2009)]{hicken09}
Hicken M., Wood-Vasey W.M., Blondin S., Challis P., Jha S., Kelly P.L., Rest A., Kirshner R.P.,
2009, eprint arXiv:0901.4804


\bibitem[Hinshaw et~al.~(2003)]{hinshaw03} 
Hinshaw G. et al.,
2003, ApJS, 148, 135

\bibitem[Hinshaw et~al.~(2007)]{hinshaw07}  
Hinshaw G. et al.,
2007, ApJS, 170, 288 

\bibitem[Hinshaw et~al.~(2009)]{hinshaw08}  
Hinshaw G. et al.,
2009, ApJS, 180, 225

\bibitem[Hogg et~al.(2005)]{hogg2005}
Hogg D.W., Eisenstein D.J., Blanton M.R., Bahcall N.A., Brinkmann J., Gunn J.E., Schneider D.P.,
2005, ApJ, 624, 54

\bibitem[Howell et~al.(2009)]{howell08}
Howell D. A. et al.,
2009, ApJ, 691, 661

\bibitem[Hu \& Haiman~(2003)]{hu03}
Hu W., Haiman Z., 
2003, PRD, 68, 063004

\bibitem[Hu \& Jain~(2004)]{hu04}
Hu W., Jain B., 
2004, PRD, 70, 043009

\bibitem[Huff et~al.~(2007)]{huff07}
Huff E., Schulz A.~E., White M., Schlegel D.~J., Warren M.~S., 
2007, Astroparticle Physics, 26, 351

\bibitem[Huterer \& Turner~(2001)]{huterer01}
Huterer D., Turner M.S.,
2001, PRD, 64, 123527

\bibitem[H\"utsi~(2006)]{hutsi06}
H\"utsi G., 
2006, A\&A, 449, 891

\bibitem[H\"utsi~(2007)]{hutsi07}
H\"utsi G., 
2007, MNRAS, submitted, eprint arXiv:0705.1843


\bibitem[Jones et~ al.~(2006)]{jones2006}
Jones W. C. et al.,
2006, ApJ, 647, 823

% CMB Kiaser effect ref 
\bibitem[Kaiser~(1987)]{kaiser87}
Kaiser N., 1987, MNRAS, 227, 1

\bibitem[Komatsu et~al.~(2009)]{komatsu08}
Komatsu E. et al.,
2009, ApJS, 180, 330

\bibitem[Kowalski et~al.~(2008)]{kowalski08}
Kowalski M. et al.,
2008, ApJ, 686, 749

\bibitem[Kuo et~al.~(2007)]{kuo2007}
Kuo C. L. et al.,
2007, ApJ, 664, 687

\bibitem[Lazkoz et~al.~(2008)]{lazkoz08}
%\bibitem[Lazkoz, Nesseris \& Perivolaropoulos (2008)]{lazkoz08}  %% 3 authors
Lazkoz R., Nesseris S., Perivolaropoulos L., 
2008, JCAP, 07, 012

\bibitem[Lee et~al.~(2001)]{lee2001}
Lee A. T. et al.,
2001, ApJ, 561, L1 

\bibitem[Lewis \& Bridle~(2002)]{cosmomc}
Lewis A., Bridle, S.,
2002, PRD, 66, 103511   %COSMO-MC

\bibitem[Lewis et al.~(2000)]{camb}
%\bibitem[Lewis, Challinor \& Lasenby (2000)]{camb}  %% 3 authors %%
Lewis A., Challinor A., Lasenby A., 
2000, ApJ, 538, 473

\bibitem[Linder et~al.~(2003)]{linder03}
Linder E. V., 
2003, PRL, 90, 091301

\bibitem[Martinez et~al.~(2009)]{martinez2008}  %%new
Martinez V.~J., Arnalte-Mur P., Saar E., de la Cruz P., Pons-Borderia M.~J., Paredes S., Fernandez-Soto A., Tempel E.,
2009, ApJ, 696L, 93

\bibitem[Matarrese \& Pietroni~(2007)]{matarrese2007}
Matarrese S., Pietroni M., 
2007, JCAP, 06, 026

\bibitem[Matsubara~(2004)]{matsubara04}
Matsubara T., 
2004, ApJ, 615, 573

\bibitem[Matsubara~(2008)]{matsubara08}
Matsubara T., 
2008, PRD, 77, 063530 

\bibitem[Meiksin et~al.~(1999)]{meiksin1999}
Meiksin A., White M., Peacock J.~A., 
1999, MNRAS, 304, 851

\bibitem[Miknaitis et~al.~(2007)]{miknaitis2007}
Miknaitis G. et al.,
2007, ApJ, 666, 674

\bibitem[Montroy et~al.~(2006)]{montroy2006}
Montroy T. E. et al.,
2006, ApJ, 647, 813 

\bibitem[Nolta et~al.~(2009)]{nolta2008}           
Nolta M. R. et al.,
2009, ApJS, 180, 296

\bibitem[Norberg et~al.~(2009)]{norberg08}
Norberg P., Baugh C.~M., Gazta\~naga E., Croton D.~J., 
2009, MNRAS, 396, 19

\bibitem[Okumura et~al.~(2008)]{okumura08}     
Okumura T., Matsubara T., Eisenstein D.~J., Kayo I., 
Hikage C., Szalay A.~S., Schneider D.~P., 
2008, ApJ, 676, 889

\bibitem[Padmanabhan et~al.~(2007)]{padmanabhan07}
Padmanabhan N., et~al., 
2007, MNRAS, 378, 852

\bibitem[Page et~al.~(2003)]{page2003}      %% this paper does not exist with that order of authors%
%Page L., Hinshaw G., Komatsu E., Nolta M.~R., 
%Spergel D.~N., Bennett C.~L., Barnes C., Bean R., et~al. 
Page L. et al.,
2003, ApJS, 148, 233

\bibitem[Percival et~al.~(2002)]{percival02}
Percival W.~J. et al.,
2002, MNRAS, 337, 1068

\bibitem[Percival et~al.~(2007a)]{percival07a} % SDSS P(k) 
Percival W.~J. et~al., 
2007a, ApJ, 657, 51

\bibitem[Percival et~al.~(2007b)]{percival07b} % SDSS P(k) 
Percival W.~J. et~al., 
2007b, ApJ, 657, 645

\bibitem[Percival et~al.~(2007c)]{percival07c} % BAO en 2dF-SDSS P(k)      
Percival W.J., Cole S., Eisenstein D. J., Nichol R. C., Peacock J. A., Pope A. C., Szalay A. S., 
2007, MNRAS, 381, 1053

\bibitem[Percival \& White~(2009)]{percival08}   
Percival W.~J., White M., 
2009, MNRAS, 393, 297

\bibitem[Perlmutter et~al.~(1999)]{perlmutter1999}
Perlmutter S. et al.,
1999, ApJ, 517, 565

\bibitem[Piacentini et~al.~(2006)]{piacentini2006}
Piacentini F. et al.,
2006, ApJ, 647, 833

\bibitem[Pietroni~(2008)]{pietroni2008}
Pietroni M., 
2008, JCAP, 10, 036

%\bibitem[Pryke et al.(2008)]{pryke2008}
%Pryke et al., 2008

\bibitem[Readhead et~al.~(2004)]{readhead2004}
Readhead A.~C. et al.,
2004, ApJ, 609, 498

\bibitem[Reichardt et~al.~(2009)]{reichardt2008}   
Reichardt C.~L. et al.,
2009, ApJ, 694, 1200

\bibitem[Reid et~al.~(2008)]{reid08}         
%\bibitem[Reid, Spergel \& Bode (2008)]{reid08}   %% 3 authors
Reid B.~A., Spergel D.~N., Bode P., 
2008, ApJ, submitted, eprint arXiv:0811.1025

\bibitem[Riess et~al.~(1998)]{riess1998}
Riess A. G. et al.,
1998, AJ, 116, 1009

\bibitem[Riess et~al.~(2004)]{riess2004}        %% this paper does not exist with these authors%%
%Riess A.~G., Li W., Stetson P.~B., Filippenko A.~V., 
%Jha S., Kirshner R.~P., Garnavich P.~M., Chornock R., 
%Challis P., 
Riess A. G. et al.,
2004, ApJ, 607, 665

\bibitem[Riess et~al.~(2007)]{riess2007}
Riess A.~G. et al.,
2007, ApJ, 659, 98 

\bibitem[Ruhl et~al.~(2003)]{ruhl2003}
Ruhl J. E. et al.,
2003, ApJ, 599, 786 

\bibitem[S\'anchez et~al.~(2006)]{sanchez06}
S\'anchez A.~G., Baugh C.~M., Percival W.~J., Peacock J.~A., 
Padilla N.~D., Cole S., Frenk C.~S., Norberg P.,  
2006, MNRAS, 366, 189  % 2dF P(k) + CMB

\bibitem[S\'anchez \& Cole~(2008)]{sanchez08}
S\'anchez A.G., Cole S., 
2008, MNRAS, 385, 830 % 2dF vs SDSS

%\bibitem[S\'anchez, Baugh \& Angulo (2008)]{sanchez08b}     %% 3 authors %%
\bibitem[S\'anchez et al.~(2008)]{sanchez08b}
S\'anchez A.G., Baugh C.M., Angulo R., 
2008, MNRAS, 390, 1470 % BAO with xi(r)

\bibitem[Scoccimarro~(2004)]{scoccimarro04}
Scoccimarro R., 
2004, PRD, 70, 083007

\bibitem[Seljak et~al.~(2005)]{seljak05}
Seljak U. et al.,
2005, PRD, 71, 103515

\bibitem[Seljak et~al.~(2006)]{seljak06}
Seljak U., Slosar A., McDonald P., 
2006, JCAP, 10, 014

\bibitem[Seo \& Eisenstein~(2003)]{seo03}
Seo H., Eisenstein D. J.,
2003, ApJ, 598, 720

\bibitem[Seo \& Eisenstein~(2007)]{seo2007}
Seo H., Eisenstein D. J., 
2007, ApJ, 665, 14

\bibitem[Seo et~al.~(2008)]{seo2008}  %%HERE%%
Seo H., Siegel E. R., Eisenstein D. J., White M., 
2008, ApJ, 686, 13

\bibitem[Smith et~al.~(2003)]{smith03}
Smith R.~E., Peacock J.~A., Jenkins A., White, S.~D.~M., Frenk C.~S., 
Pearce F.~R., Thomas P.~A., Efstathiou G., Couchman H.~M.~P., 
2008, MNRAS, 341, 1311

%\bibitem[Smith, Scoccimarro \& Sheth (2007)]{smith07}       %% 3 names %%
\bibitem[Smith et~al.~(2007)]{smith07}
Smith R.~E., Scoccimarro R., Sheth R.~K., 
2007, PRD, 75, 063512

%\bibitem[Smith, Scoccimarro \& Sheth (2008)]{smith08}       %% 3 names %%
\bibitem[Smith et~al.~(2008)]{smith08}
Smith R.~E., Scoccimarro R., Sheth R.~K., 
2008, PRD, 77, 043525

\bibitem[Spergel et~al.~(2003)]{spergel03}
Spergel D.~N. et al.,
2003, ApJS, 148, 175

\bibitem[Spergel et~al.~(2007)]{spergel07}
Spergel D. N. et al.,
2007, ApJS, 170, 377

\bibitem[Steigman~(2007)]{steigman2007}
Steigman G., 2007, on SUSY06: The 14th International Conference on 
Supersymmetry and the Unification of Fundamental Interactions Edited 
by Feng J.L., p.40

\bibitem[Sugiyama~(1995)]{sugiyama95}
Sugiyama N., 
1995, ApJS, 100, 281

\bibitem[Sullivan et~al.~(2003)]{sullivan03}          
Sullivan M. et al., 
2003, MNRAS, 340, 1057

\bibitem[Swanson et~al.~(2008)]{swanson08}
Swanson M.E.C., Tegmark M., Blanton M., Zehavi I., 
2008, MNRAS, 385, 1635

\bibitem[Szapudi~(2004)]{szapudi04}
Szapudi I.,
2004, ApJ, 614, 51

\bibitem[Takahasi et~al.~(2008)]{takahashi08}     
Takahashi R. et al.,
2008, MNRAS, 389, 1675

\bibitem[Taruya \& Hiramatsu~(2008)]{taruya08}
Taruya A., Hiramatsu T., 
2008, ApJ, 674, 617

\bibitem[Tegmark et~al.~(1994)]{tegmark1994}     %%HERE %%
Tegmark M., Silk, J., Blanchard A., 1994, ApJ, 420, 484

\bibitem[Tegmark et~al.~(2004)]{tegmark04}
Tegmark M. et~al., 
2004, ApJ 606, 702

\bibitem[Tegmark et~al.~(2006)]{tegmark2006}
Tegmark M. et al,
2006, PRD, 74, 123507

\bibitem[Tonry et~al.~(2003)]{tonry2003}
Tonry J. L. et~al.,
2003, ApJ, 594, 1

\bibitem[Wood-Vasey et~al.~(2007)]{wood-vasey07}
Wood-Vasey W. M.et al.,
2007, ApJ, 666, 694

\bibitem[Wang~(2006)]{wang06}
Wang Y., 
2006, ApJ, 647, 1

\bibitem[Wang \& Mukherjee~(2006)]{wang06b}
Wang Y., Mukherjee P., 
2006, ApJ, 650, 1

\bibitem[Wood-Vasey et~al.~(2007)]{wv07}
Wood-Vasey M., et al. 
2007, ApJ, 666, 694

\bibitem[Wright~(2007)]{wright2007}
Wright E.L., 
2007, ApJ, 664, 633

\bibitem[Xia et~al.~(2008)]{xia08}
Xia J., Li H., Zhao G., Zhang X., 
2008, PRD, 78, 0835524

\end{thebibliography}
\end{document}